\documentclass[a4paper,11pt]{amsart}
\usepackage[a4paper,margin=2cm]{geometry}
\usepackage{amsmath,amssymb,amsthm}
\usepackage{mathtools,stmaryrd}
\usepackage{url}
\usepackage{color}
\usepackage{enumitem}
\usepackage{comment}
\usepackage{tikz,graphicx}
\usetikzlibrary{arrows.meta,calc,decorations.pathreplacing,shapes.geometric}
\definecolor{proofInk}{HTML}{182F41}
\definecolor{proofBlue}{HTML}{365875}
\definecolor{proofGreen}{HTML}{216559}
\definecolor{proofTint}{HTML}{F2F7F5}
\usepackage{todonotes}

\usepackage[colorlinks,linkcolor=blue, citecolor=blue]{hyperref}

\usepackage[backend=bibtex,style=alphabetic,url=false,doi=true,giveninits=true,isbn=false,maxbibnames=10]{biblatex}
\AtEveryBibitem{\clearlist{language}}
\theoremstyle{plain}
\newtheorem{theorem}{Theorem}[section]

\newtheorem{corollary}[theorem]{Corollary}
\newtheorem{lemma}[theorem]{Lemma}

\newtheorem{proposition}[theorem]{Proposition}

\theoremstyle{definition}

\theoremstyle{remark}
\newtheorem{remark}[theorem]{Remark}
\newtheorem{convention}[theorem]{Convention}

\def\I{\mathsf{i}}

\allowdisplaybreaks[2]

\title[Figure eight knot and topological recursion]{Colored Jones polynomial of figure eight knot and non-perturbative topological recursion}

\author{Alexander Hock}

\address{University of Geneva}

\email{Alexander.Hock@unige.ch}

\author{Sergey Shadrin}

\address{University of Amsterdam}

\email{S.Shadrin@uva.nl}

\begin{document}

\begin{abstract}
The asymptotic expansion of the colored Jones polynomial of the figure eight knot is proved to be connected with non-perturbative topological recursion on the underlying elliptic spectral curve, following the proposal of Borot and Eynard. 
\end{abstract}

\maketitle

\tableofcontents

\section{Introduction}

Knots are fundamental objects in mathematics. Many familiar invariants fail to distinguish every pair of knots, so finding simple, useful complete invariants remains a challenge. Knot invariants are defined in many different ways and connect knot theory to representation theory, algebraic geometry, other areas of low-dimensional topology, and mathematical physics. Sometimes, connections between different invariants defined in very different ways reveal unexpected connections between these fields.

For example, an important invariant is the Jones polynomial \cite{Jones1985}, which associates to each oriented knot a Laurent polynomial in $q$. It can be computed by a skein relation which, by successively changing and smoothing crossings, generates simpler knots and links, which finally reduces the computation to unknots, which is normalized to be $1$. The first three-dimensional interpretation of the Jones polynomial was given by Witten \cite{Witten1989}, who showed that it can be derived from theoretical physics. He showed that, for knots in $S^3$, so-called Wilson loop expectation values along such a knot in the fundamental representation of $\mathrm{SU}(2)$ Chern--Simons theory, normalized by the value of the unknot, give the Jones polynomial evaluated at roots of unity, thus connecting three-dimensional quantum field theory to knot invariants. The Jones polynomial was further refined in mathematics using representations of quantum groups \cite{ReshetikhinTuraev1990}, and in physics by taking Wilson loops in the $N$-dimensional irreducible representation of $\mathrm{SU}(2)$ \cite{Witten1989}. On the quantum group side, taking the $N$-dimensional irreducible representation of $U_q(\mathfrak{sl}_2)$ gave rise to the definition of the \emph{colored Jones polynomial} $J_N(q)$. The colored Jones polynomial has a generalization to the colored HOMFLY-PT invariant \cite{LinZheng2010},
whose specialization at rank $M$ is obtained by coloring the knot
with symmetric representation of $U_q(\mathfrak{sl}_M)$.

Kashaev observed in several examples \cite{Kashaev1997} that the
large-$N$ growth of the so-called Kashaev invariant recovers the hyperbolic
volume of the knot complement in $S^3$. Later, in
\cite{MurakamiMurakami2001} this Kashaev invariant was identified with
the normalized $N$-colored Jones polynomial evaluated at the root of unity
$q=e^{2\pi\I/N}$.
Nowadays one formulates the volume conjecture immediately for the colored Jones polynomial. The volume conjecture has also been generalized by considering the
colored Jones polynomial at $q_N=\exp((2\pi\I+s)/N)$, with $s$
fixed as $N\to\infty$. Gukov proposed
that the full asymptotic expansion is governed by perturbative
$SL_2(\mathbb C)$ Chern--Simons theory on the knot complement
\cite{Gukov2005}. In that formulation, the
leading term is given by the classical Chern--Simons action, while
the subleading terms are given by the one-loop and higher-loop
contributions \cite{GukovMurakami2008}.

A different invariant associated to a knot is the so-called
$A$-polynomial, which is actually described by its vanishing
locus. The $A$-polynomial $A$ is obtained from the
$\mathrm{SL}_2(\mathbb{C})$-character variety of the knot complement
in $S^3$ \cite{CCGLS1994}. One takes representations of the knot
group $\pi_1(S^3\setminus K)$ into $\mathrm{SL}_2(\mathbb{C})$ and
obtains their characters by taking traces.
For knots the $A$-polynomial has two variables and thus gives rise to a complex curve.
For links with several components, one gets a higher-dimensional variety.
There are also refinements
of the $A$-polynomial to more general varieties. On the mathematical
side, knot contact homology gives rise to augmentation varieties
\cite{Ng2008}. On the physics side, Aganagic and Vafa
\cite{AganagicVafa2012} proposed a $Q$-deformation of the
$A$-polynomial via the large $N$ transition to the
resolved conifold.

The AJ conjecture gives a precise connection between the colored
Jones polynomial and the $A$-polynomial, two invariants defined in
very different ways. Its mathematical formulation goes back to
Garoufalidis \cite{Garoufalidis2004}. Around the same time, Gukov
\cite{Gukov2005} proposed a physical interpretation in terms of the
quantization of the $A$-polynomial.
The starting point is the theorem of Garoufalidis and L\^e
\cite{GaroufalidisLe2005} that, for every knot, the sequence of
colored Jones polynomials $J=(J_N(q))_{N\geq 1}$ is $q$-holonomic.
This means that it satisfies a nontrivial linear $q$-difference
equation in the color $N$. Thus there exists an operator
$\hat{A}$, with coefficients depending on $q$ and $q^N$, such that
\begin{equation}
    \hat{A}J=0.
\end{equation}
Choosing $\hat{A}$ in some minimal way, the AJ conjecture states that its
semiclassical limit, obtained by setting $q=1$ with $X=q^N$ held
fixed, agrees with the classical $A$-polynomial up to a
rational factor depending only on $X$.

What makes this conjecture surprising is that the two invariants are
defined in very different ways. The colored Jones polynomial can,
for instance, be computed by successively smoothing the crossings of the knot.
The classical $A$-polynomial is defined through representations of
the knot group and their restriction to the boundary torus. A
smoothing can turn a knot into a two-component link. Its exterior
then has two boundary tori, and the corresponding character variety
can have two-dimensional components.

Thus one goes from the sequence of all colored Jones polynomials
$(J_N(q))_{N\geq 1}$ to their recurrence operator $\hat{A}$, and
then, by taking its semiclassical limit, from $\hat{A}$ to the
classical $A$-polynomial, as predicted by the AJ conjecture:
\begin{equation}
    (J_N(q))_{N\geq 1}
    \xrightarrow{\text{recurrence}} \hat{A}
    \xrightarrow{q\to 1} A.
\end{equation}
There is also a generalization of this picture for the
colored HOMFLY-PT polynomials. Here one again passes from the
polynomials to their recurrence operator, whose semiclassical limit,
with the HOMFLY-PT parameter $Q$ held fixed, is conjectured to define
the augmentation curve of the knot
\cite{AganagicEkholmNgVafa2014,EkholmNg2020}.\\

Let us now introduce the second half of the title. Topological recursion
(TR) is a universal procedure \cite{EynOra} that originated in the study
of loop equations for correlators of matrix resolvents.
In TR, an infinite family of meromorphic multidifferentials
$\omega^{(g)}_{n}$ is recursively defined from initial data called a
spectral curve. Depending on the spectral curve, the corresponding
multidifferentials have different interpretations.

Nowadays, TR appears in several areas and connects them. In enumerative
geometry, examples include the stationary Gromov--Witten invariants of
$\mathbb{P}^1$ \cite{NorburyScott2014,DOSS2014} and the Gromov--Witten
invariants of toric Calabi--Yau threefolds \cite{EynardOrantin2015}.
There is also a connection to topological string theory through the
remodeling of the $B$-model, where TR is applied to the mirror curve
\cite{BKMP2009}. In the theory of integrable systems, TR appears, for
instance, in the study of Painlev\'e equations
\cite{Iwaki2020,MarchalOrantin}, the KP hierarchy \cite{ABDKS2026},
and the KdV hierarchy \cite{EynOra}. This gives just a glimpse of the
connections that TR provides.

The starting point is the spectral curve, whose data can be specified
in different ways depending on the problem one wants to study with TR.
A simple example is given by an irreducible polynomial in two variables
$P(x,y)$. The compact curve $P(x,y)=0$ is a Riemann
surface equipped with the two meromorphic functions $x$ and $y$.
A choice of symplectic basis of its first homology determines the
normalized bidifferential used in TR. If the spectral curve has positive
genus, a non-perturbative completion of TR gives a KP-integrable system
\cite{ABDKS-NP}, building on the constructions in
\cite{EynardMarino2011,BE-integrable}. This is also related to the
construction of quantum curves
\cite{Gukov:2011qp,Iwaki2020,MarchalOrantin,EGFMO2024}.\\

Motivated by the connection of TR to topological string theory through
the remodeling of the $B$-model, and by the connection of topological
string theory to knot theory through Chern--Simons theory, a conjectural relation between TR and knot
invariants was studied by Dijkgraaf, Fuji and Manabe \cite{DijkgraafFujiManabe}. They took the irreducible component
of the $A$-polynomial curve as the spectral curve and expected the
integrated multidifferentials to give the all-order asymptotic expansion
of the colored Jones polynomial. This provided a conjectural path from
the $A$-polynomial to the colored Jones polynomial via TR. A related
approach to the quantization of the $A$-polynomial was developed already in \cite{Gukov:2011qp}.

For the figure eight knot, \cite{DijkgraafFujiManabe} observed
discrepancies in the first few orders and adjusted certain constants
by hand. Borot and Eynard resolved these adjustments by including
the non-perturbative contributions \cite{BE2015}. They formulated an
explicit conjecture relating the all-order asymptotic expansion of the
colored Jones polynomial to non-perturbative TR on the non-abelian (geometric) irreducible
component of the $A$-polynomial curve. They tested this conjecture explicitly
at the first few orders for the figure eight knot, and this was further extended through order $\hbar^7$ in~\cite{Potter2023}.

In this sense, the conjecture of Borot and Eynard provides a way to
construct the all-order asymptotic expansion of the colored Jones
polynomial \emph{from} the $A$-polynomial via non-perturbative TR.
This can be written schematically as
\begin{equation}
     A=0
    \xrightarrow{\text{non-pert TR}}
    \operatorname*{Asymp}_{N\to\infty} J_N(e^{x/N})
    \qquad ( x\ \text{fixed}).
\end{equation}
In this article, we prove this relation for the figure eight knot. We prove that the logarithmic differentials of the
two wave functions agree to all stable orders.

\subsection{Main statement} There are three different wave functions that can be associated to the figure eight knot in the context of the study of its colored Jones polynomial:
\begin{itemize}
    \item $\Psi(x,\hbar)$ controls the all-orders asymptotic expansion of the colored Jones polynomial for $N\to \infty$ (where $N$ is the dimension of the irreducible representation of $\mathfrak{sl}_2$ used in the construction of the colored Jones polynomial).  
    \item  $\Psi^{\mathsf{rec}}(x,\hbar)$ is the formal WKB expansion of the recursion for the colored Jones polynomial given by an operator  $\widehat A$, the quantum $A$-polynomial. 
    \item $\Psi^{\mathsf{np}}(x,\hbar)$ is the Borot-Eynard wave function constructed from the non-perturbative topological recursion on the spectral curve obtained from the semi-classical limit of $\widehat A$, the $A$-polynomial.
\end{itemize}
In the definition of $\Psi^{\mathsf{rec}}(x,\hbar)$ we choose the initial conditions (the disk term and the amplitude of the wave function, in physics jargon) such that they match exactly the ones of $\Psi(x,\hbar)$, and we also define $\widetilde{\Psi}^{\mathsf{np}}(x,\hbar)$ by manually replacing in $\Psi^{\mathsf{np}}(x,\hbar)$ the disk term and the amplitude by the ones that match $\Psi(x,\hbar)$. The main goal of this paper is to prove the following theorem:

\begin{theorem} \label{thm:main} We have 
\begin{align}
    d\log \Psi(x,\hbar) = d\log \Psi^{\mathsf{rec}}(x,\hbar) \qquad \text{and}\qquad d\log \Psi^{\mathsf{rec}}(x,\hbar) = d \log \widetilde{\Psi}^{\mathsf{np}}(x,\hbar)
\end{align}
(the differential $d$ here is taken with respect to $x$, assuming $\hbar$ fixed).
\end{theorem}

In this formulation, we suppress important differences on the domains of definitions of the involved wave functions, and the proper statements require a detailed analysis of the analytic behavior of the involved wave functions to make the comparison mathematically precise. 

Indeed, $\Psi(x,\hbar)$ is defined as an asymptotic expansion of the Jones polynomial, and its existence requires a proof. What we obtain is an expansion in $\hbar$ with coefficients analytically depending on $x$ for $x\in 2\pi\I + I$, where $I$ is a compact interval $I\subset (0,\log \frac{3+\sqrt{5}}{2})$. It can be analytically extended in $x$ to an open domain $2\pi\I + I\subset U \subset \mathbb{C}$. The wave function $d\log \Psi^{\mathsf{rec}}(x,\hbar)$ is defined by recursion, pointwise, and by taking the initial disk and amplitude terms as in $\Psi(x,\hbar)$, it can be also considered as an analytic function on $U$. So, the first comparison is some uniqueness argument that concerns $\hbar$-expansions with coefficients in the holomorphic differentials on $U$. 

By contrast, $\widetilde{\Psi}^{\mathsf{np}}(x,\hbar)$ is rather a global object defined on a compact elliptic curve $\Sigma$. The domain $U$ can be identified with a branch of $\Sigma$, and we \emph{restrict $\widetilde{\Psi}^{\mathsf{np}}(x,\hbar)$ to this branch for comparison with $\Psi^{\mathsf{rec}}(x,\hbar)$ in the second equality of the theorem}. But in fact, we prove that the coefficients of $\hbar$-expansion of  $\log \Psi^{\mathsf{rec}}(x,\hbar)$ extend (in the stable range, that is, starting from $\hbar^1$) to global meromorphic functions of $\Sigma$. Then, the second equality of the theorem can also be seen as a comparison of meromorphic differentials on $\Sigma$.

\subsection{Main idea of the construction}
In the conjecture of Borot-Eynard, how the asymptotic of the colored Jones can be constructed from non-perturbative TR on the $A$-polynomial, the so-called $2|2$-spinor kernel appears.
We now explain why the $2|2$-spinor kernel appears in the construction
of $\widetilde{\Psi}^{\mathsf{np}}(x,\hbar)$. Borot and Eynard
\cite{BE2015} take the square root of the normalized $2|2$-spinor
kernel of non-perturbative TR, evaluated at the two pairs of points
$(p,\iota(p))$ and $(o,\iota(o))$. Our proof gives
an explanation for this choice of kernel in the case of the
figure eight knot.

The importance of the $2|2$-spinor kernel is that the two integration
paths, from $o$ to $p$ and from $\iota(o)$ to $\iota(p)$, are related
by the involution $\iota$. Thus, in each variable of a connected
non-perturbative differential, the two integrals combine as
\begin{align}
    \int_o^p \alpha+\int_{\iota(o)}^{\iota(p)}\alpha
    =\int_o^p\bigl(\alpha+\iota^*\alpha\bigr).
\end{align}
The resulting differential is $\iota$-invariant in each variable and
therefore descends to the quotient curve $\mathsf C=\Sigma/\iota$. 

For the figure eight knot, the original spectral curve $\Sigma$ has
genus one, and the quotient $\mathsf C$ has genus zero. With the choice of
homology basis, the even theta characteristic $(0,\tfrac12)$ gives a
zero of the kernel that makes the extra term in the projected quadratic
loop equation vanish. The projected differentials therefore satisfy the
ordinary loop equations on $\mathsf C$. Their stable terms have poles only at
the two ramification points $z=\pm1$ of
$x_\mathsf C=\tfrac12(z+z^{-1})$. Since on a genus zero curve the principal
parts determine a meromorphic differential, these equations identify
the projected differentials with those of ordinary TR on $\mathsf C$. This
recursion is local in the sense that $y_\mathsf C$ is needed only near
$z=\pm1$.

The final identification takes place on the quotient curve. The
higher-order logarithmic differentials obtained from the colored Jones
recurrence also descend to $\mathsf C$. We compare the two families order by
order. Regularity of their local second-order differential equations
at $z=\pm1$ implies that the first possible difference is holomorphic
there. Since these are its only possible poles and $\mathsf C$ has genus zero,
the difference vanishes. Pulling the equality back to $\Sigma$ gives
\begin{align}
    d\log\widetilde{\Psi}^{\mathsf{np}}(x,\hbar)
    =d\log\Psi^{\mathsf{rec}}(x,\hbar)
\end{align}
to all orders on the chosen branch on $\Sigma$.

\subsection{Outline of the article}
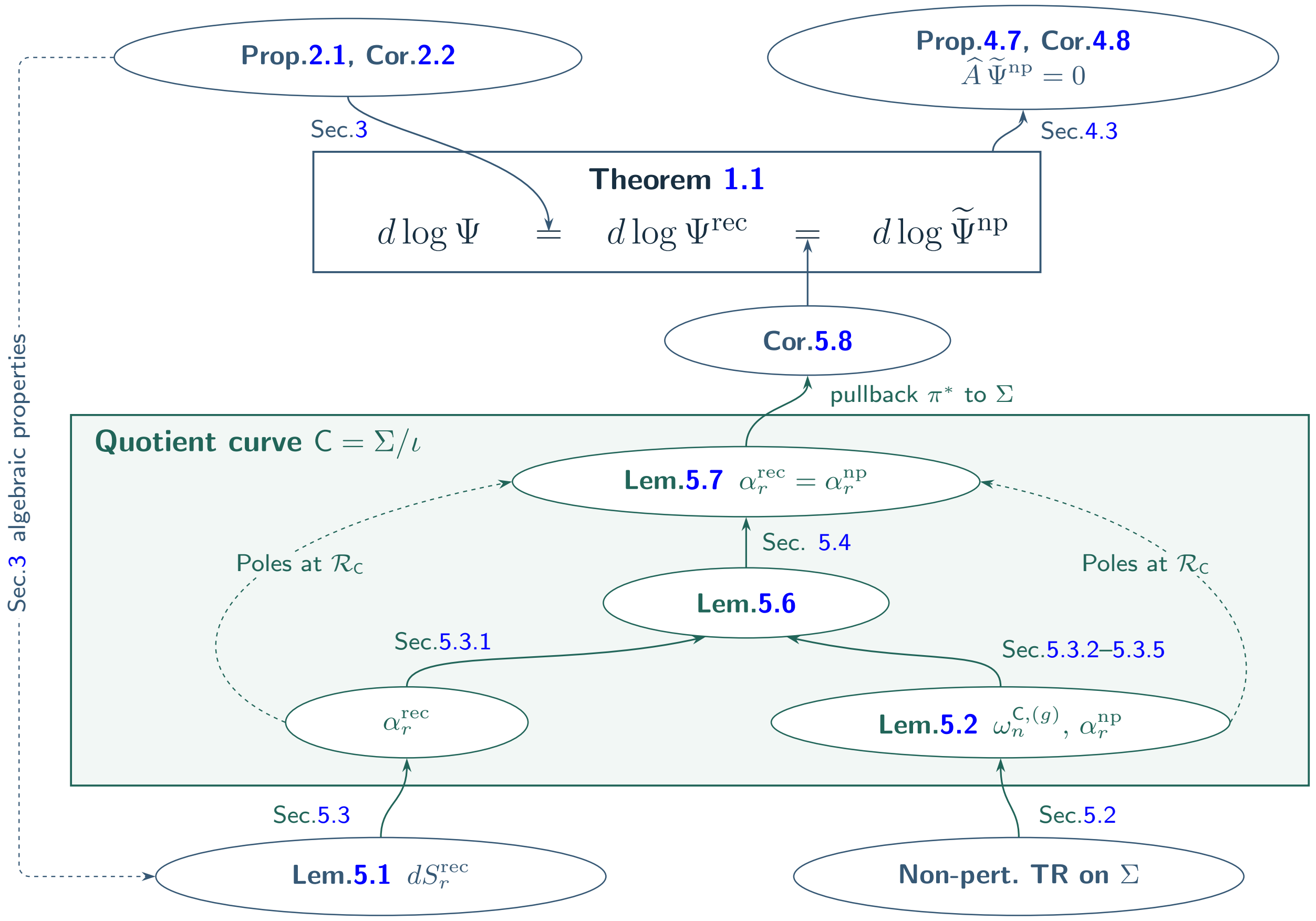
\begin{figure}[tbp]
\centering
\resizebox{\linewidth}{!}{%
\begin{tikzpicture}[
  x=1mm, y=-1mm,
  result/.style={
    ellipse, draw, fill=white, align=center,
    inner sep=0pt, outer sep=.1pt, line width=.45pt,
    font=\sffamily\bfseries\fontsize{9}{10.5}\selectfont
  },
  upstairs/.style={result, draw=proofBlue, text=proofBlue},
  downstairs/.style={result, draw=proofGreen, text=proofGreen},
  arr/.style={
    -{Stealth[length=1.6mm,width=1.05mm]}, line width=.55pt
  },
  supporting/.style={
    arr, line width=.4pt,
    dash pattern=on 1.35pt off 1.45pt, rounded corners=1.2mm
  },
  formula/.style={
    anchor=base, inner sep=0pt, outer sep=0pt,
    text=proofInk, font=\fontsize{12.8}{15}\selectfont
  },
  tag/.style={
    inner xsep=.65mm, inner ysep=.2mm,
    font=\sffamily\fontsize{7.7}{8.7}\selectfont
  }
]
\path[use as bounding box] (0,0) rectangle (156,109);

\node[upstairs, minimum width=54mm, minimum height=9mm]
  (jones) at (42,7.7) {Prop.\ref{prop:all-order-existence}, Cor.\ref{cor:psipsirec}};

\node[upstairs, minimum width=59mm, minimum height=12mm]
  (consequences) at (120,7.7)
  {Prop.\ref{prop:equation-for-tildepsi}, Cor.\ref{cor:A-hat-Psi-np}\\[-.2pt]
   $\widehat A\,\widetilde\Psi^{\mathrm{np}}=0$};

\draw[proofBlue, line width=.65pt]
  (38,18.6) rectangle (122,32.5);
\node[text=proofInk, font=\sffamily\bfseries\small]
  at (80,21.7) {Theorem \ref{thm:main}};

\node[formula]       at (51.4,29.2)  {$d\log\Psi$};
\node[formula] (eq1) at (65.2,29.2)  {$=$};
\node[formula]       at (80.1,29.2)  {$d\log\Psi^{\mathrm{rec}}$};
\node[formula] (eq2) at (95.1,29.2)  {$=$};
\node[formula]       at (110.5,29.2) {$d\log\widetilde\Psi^{\mathrm{np}}$};

\node[upstairs, minimum width=33mm, minimum height=8mm]
  (pullback) at (95.1,40.4) {Cor.\ref{cor:dsnpdsrec}};

\begin{scope}[color=proofBlue]
  \draw[arr] (jones.south)
    .. controls (42,16.1) and (65.2,18.6)
    .. ([yshift=-.45pt]eq1.north);

  \draw[arr] (pullback.north)
    -- ([yshift=1.55pt]eq2.south);

  \draw[arr] (116.5,18.6)
    .. controls (116.5,16.5) and (120,16.2)
    .. (consequences.south);

  \node[tag] at (41,15.9)    {Sec.\ref{sec:analytic-comparison}};
  \node[tag] at (126.5,16.1) {Sec.\ref{sec:eqwave}};
\end{scope}

\filldraw[fill=proofTint, draw=proofGreen, line width=.65pt]
  (10,49) rectangle (153,91.8);

\node[anchor=west, inner sep=0pt, text=proofGreen,
      font=\sffamily\bfseries\small]
  at (12.6,52.3) {Quotient curve $\mathsf{C}=\Sigma/\iota$};

\node[downstairs, minimum width=54mm, minimum height=8.1mm]
  (comparison) at (88,56.7)
  {Lem.\ref{lem:quotient-comparison}\enspace $\alpha_r^{\mathrm{rec}}=\alpha_r^{\mathrm{np}}$};

\node[downstairs, minimum width=33mm, minimum height=8.1mm]
  (regularity) at (88,70.7) {Lem.\ref{lem:rec-equation}};

\node[downstairs, minimum width=28mm, minimum height=8.2mm]
  (recforms) at (48.8,84.5) {$\alpha_r^{\mathrm{rec}}$};

\node[downstairs, minimum width=53mm, minimum height=8.2mm]
  (trforms) at (117.4,84.5)
  {Lem.\ref{lem:descent}\enspace $\omega_n^{\mathsf C,(g)},\,\alpha_r^{\mathrm{np}}$};

\begin{scope}[color=proofGreen]
  \draw[arr] (comparison.north)
    .. controls (88,48) and (95.1,48) .. (pullback.south);
  \node[tag] at (108.3,46.8) {pullback $\pi^*$ to $\Sigma$};

  \draw[arr] (regularity.north) -- (comparison.south);
  \node[tag] at (95,63.6) {Sec. \ref{subsec:comparison}};

  \draw[arr] (recforms.north)
    .. controls (48.8,75.7) and (67,78.3) .. (regularity.220);
  \draw[arr] (trforms.north)
    .. controls (117.4,75.6) and (108.4,78.1) .. (regularity.320);
  \node[tag] at (53,75.1) {Sec.\ref{sec:recdif}};
  \node[tag] at (127,76)  {Sec.\ref{sec:difeqnp}--\ref{sec:difeqnpend}};

  \draw[supporting] (recforms.west)
    .. controls (20,78.5) and (24,63.8) .. (comparison.west);
  \draw[supporting] (trforms.east)
    .. controls (151,73) and (137,61) .. (comparison.east);

  \node[tag, fill=proofTint] at (36.5,66.2)  {Poles at $\mathcal{R}_{\mathsf C}$};
  \node[tag, fill=proofTint] at (134.2,66.2) {Poles at $\mathcal{R}_{\mathsf C}$};
\end{scope}

\node[upstairs, minimum width=52mm, minimum height=9mm]
  (recSigma) at (45.8,102.3)
  {Lem.\ref{lem:rec-meromorphy}\enspace $dS_r^{\mathrm{rec}}$};

\node[upstairs, minimum width=52mm, minimum height=9mm]
  (trSigma) at (119.5,102.3) {Non-pert.\ TR on $\Sigma$};

\begin{scope}[color=proofGreen]
  \draw[arr] (recSigma.north)
    .. controls (45.8,93.9) and (48.8,93.7) .. (recforms.south);
  \draw[arr] (trSigma.north)
    .. controls (119.5,94.2) and (117.4,94) .. (trforms.south);

  \node[tag] at (37.8,95.1)  {Sec.\ref{subsec:alpha-equations}};
  \node[tag] at (126.3,95.1) {Sec.\ref{subsec:rational-reduction}};
\end{scope}

\draw[supporting, proofBlue] (jones.west)
  -- (4,7.7) -- (4,102.3) -- (recSigma.west);
\node[tag, rotate=90, fill=white, text=proofBlue]
  at (4,55.7) {Sec.\ref{sec:analytic-comparison}\enspace algebraic properties};

\end{tikzpicture}%
}
\caption{Outline how the different parts of the article fit together and how their lemmata, propositions and corollaries finally contribute to the proof of Theorem \ref{thm:main}}
\label{fig:figure-eight-proof}
\end{figure}

We outline the main steps in the proof of Theorem~\ref{thm:main}.
In Sec.~\ref{sec:knot}, we start with the colored Jones polynomial
$J_N(q)$ of the figure-eight knot. Prop.~\ref{prop:all-order-existence}
establishes its all-orders asymptotic expansion for $q=e^{x/N}$ as
$N\to\infty$, locally uniformly for $x$ in a complex neighbourhood
of $2\pi i+I$. We then explain how the exact recurrence for finite $N$
passes to a homogeneous difference equation at the level of the
asymptotic expansion. The resulting wave function $\Psi$ agrees with
the formal WKB solution $\Psi^{\mathrm{rec}}$ of this equation up to
a constant factor, as stated in Cor.~\ref{cor:psipsirec}.
This proves the first equality in Theorem~\ref{thm:main}.

The proofs of Prop.~\ref{prop:all-order-existence} and
Cor.~\ref{cor:psipsirec} are given in Sec.~\ref{sec:analytic-comparison}.
We extend the saddle point analysis of Murakami
\cite{Murakami2013,Murakami2024} to a small complex neighbourhood
of the interval considered there, using the methods of Ohtsuki
\cite{Ohtsuki2016}. The local uniformity of the expansion is crucial
for applying it to the recurrence.
The Gaussian saddle point expansion also gives the algebraic structure of the higher-order
coefficients on an elliptic curve $\Sigma$ which turns out to be birationally equivalent to the geometric component of the $A$-polynomial.

In Sec.~\ref{sec:A-polynomial}, we recall the construction of the 
Borot--Eynard wave function $\Psi^{\mathrm{np}}$ from the 
$2|2$-spinor kernel \cite{BE2015}. We use the same elliptic curve $\Sigma$ as the spectral curve for
topological recursion. The differentials $dx=d\log X$ and
$dy=d\log Y$ are obtained by pulling back the coordinates
$X$ and $Y$ from the geometric component of the $A$-polynomial curve
to $\Sigma$ via the birational transformation. The non-perturbative construction uses a specific choice of homology
basis and an even theta characteristic. We describe the resulting
wave function on $\Sigma$, and define $\widetilde{\Psi}^{\mathrm{np}}$
by matching its disk term and amplitude with those of $\Psi$.
As a consequence of Theorem~\ref{thm:main},
Cor.~\ref{cor:A-hat-Psi-np} shows that
$\widehat A\,\widetilde{\Psi}^{\mathrm{np}}=0$ as a formal identity.

Sec.~\ref{sec:quantum-curve-proof} is devoted to the second equality
in Theorem~\ref{thm:main}. The main step is to identify the stable
coefficients of $d\log\Psi^{\mathrm{rec}}$ and
$d\log\widetilde{\Psi}^{\mathrm{np}}$ on the genus-zero quotient curve
$\mathsf C=\Sigma/\iota$. On the recurrence side, the stable
differentials are invariant under $\iota$ and hence descend to $\alpha^{\text{rec}}_r$ on 
$\mathsf C$. On the Borot--Eynard side, the chosen theta characteristic
gives the vanishing needed for the projected differentials to satisfy
the loop equations of ordinary local topological recursion on
$\mathsf C$. This provides us with the differentials $\omega^{\mathsf C,(g)}_n$ and $\alpha^{\text{np}}_r$ on $\mathsf C$.

Lemma~\ref{lem:rec-equation} shows that, near each ramification point,
the two branches of each quotient wave function satisfy a second-order
differential equation with regular coefficients. On the recurrence
side, we obtain this equation by passing locally from the second-order
difference equation to the associated second-order differential equation that infinitesimally generates the $\hbar$-shift. On the topological recursion side, we use the interpretation
of the wave function as a deformed Baker--Akhiezer kernel, together
with determinantal identities from KP integrability and the loop
equations on $\mathsf C$. These techniques are closely related to the known techniques of deriving quantum curves from topological recursion \cite{MarchalOrantin,EGFMO2024}.

The two local differential equations may a priori be different.
In Lemma~\ref{lem:quotient-comparison}, we use their regularity to
prove by induction in $\hbar$ that the stable differentials
$\alpha_r^{\mathrm{rec}}$ and $\alpha_r^{\mathrm{np}}$ agree.
At each order, their difference can have poles only at the
ramification points, and the induction shows that it is holomorphic
there as well. Since $\mathsf C$ has genus zero, every holomorphic one-form vanishes. Pulling the equality back to $\Sigma$ gives
Cor.~\ref{cor:dsnpdsrec} and completes the proof of
Theorem~\ref{thm:main}. Fig.~\ref{fig:figure-eight-proof} illustrates
how the different steps fit together.

\subsection{Beyond the figure eight knot and the colored Jones polynomial}

The figure eight knot is often the first example used to test a
conjecture or a new relation in knot theory. It is the hyperbolic
knot with the fewest crossings, yet it already has enough structure
to give nontrivial results. Our result makes it natural to ask
whether non-perturbative TR applied to the $A$-polynomials of other
hyperbolic knots computes the all-orders asymptotic expansion of the
colored Jones polynomial, as proposed by Borot and Eynard
\cite{BE2015}.

A natural infinite family to consider next is that of hyperbolic
twist knots. Their $A$-polynomials are known explicitly
\cite{HosteShanahan2004,Mathews2014,Mathews2014Erratum}. The
geometric genus of the corresponding curve can already be higher
than in the figure eight case. For instance, the curve for the
knot $5_2$ has genus two, which makes the application
of non-perturbative TR more involved. However, these curves have
an important property. If $\Sigma$ is the non-abelian factor from the
$A$-polynomial, then its quotient by the involution
$\iota(X,Y)=(X^{-1},Y^{-1})$ has genus zero, see for instance
\cite{GaroufalidisSun2006}.
Thus $\Sigma$ is a double cover of a rational curve and is
hyperelliptic. It is therefore
natural to ask whether the same reduction to local TR on the
quotient curve can be carried out for these knots. 
In particular, a natural question is under what conditions on the spectral curve one
can choose a theta characteristic and a symplectic homology basis so
that the normalized $2|2$-spinor kernel descends to the quotient curve
and the projected differentials satisfy the ordinary loop equations
and are given by topological recursion on the quotient curve.
We are exploring this question for some examples of hyperelliptic
curves, taking the quotient by the hyperelliptic
involution~\cite{HS-in-preparation}. When such a choice exists, one
can further ask whether the theta characteristic is unique up to
symplectic equivalence.
This allows to put forward the proposal of Borot and Eynard
\cite{BE2015}.

Another natural question is how to go beyond the colored Jones
polynomial. The proposed generalization of the AJ conjecture
relates the augmentation curve to the colored
HOMFLY-PT polynomials.
It is therefore natural to ask whether TR can also be applied to
the augmentation curve. Torus knots provide an example. Although
they are non-hyperbolic, the volume conjecture in terms of simplicial
volume applies to them \cite{MurakamiMurakami2001} and was proved
for torus knots by Kashaev and Tirkkonen
\cite{KashaevTirkkonen2003}. Their classical $A$-polynomial is
particularly simple. On each non-abelian component, the logarithmic
projection $x=\log X$ has no ramification points at which to apply
the usual TR.

Motivated by large $N$ duality, Brini, Eynard, and Mari\~no
\cite{BriniEynardMarino2012} obtained an auxiliary spectral curve
by applying an $\mathrm{SL}_2(\mathbb Z)$ transformation to the
mirror curve of the resolved conifold. This construction gives
a fractional framing, and TR on the resulting curve recovers the
colored HOMFLY-PT invariants of the torus knot after
symmetrization. The recursion also produces additional terms.
These were identified with the coefficients of the extended
Ooguri--Vafa partition function, from which the ordinary
Ooguri--Vafa partition function is recovered by the same
symmetrization \cite{DKPSS2022}.

To recover the physical amplitudes directly, Gu, Jockers, Klemm,
and Soroush \cite{GuJockersKlemmSoroush2015} proposed a modified
TR on a curve of generally higher genus, using a calibrated annulus
kernel. They conjectured that this curve coincides with the
augmentation curve of the torus knot, in accordance with the
large $N$ picture of Aganagic and Vafa \cite{AganagicVafa2012}.
This identification and the corresponding recursion for the
colored HOMFLY-PT correlation differentials were recently proved
for torus knots \cite{HockShadrin2026}.

These examples lead to a more general question. How should one
choose the initial data on the augmentation curve or on the
$A$-polynomial curve so that TR, with a non-perturbative
completion where needed, recovers the colored HOMFLY-PT invariants
or the asymptotic expansion of the colored Jones polynomial,
respectively? Answering this question would help explain the
relations between these differently defined knot invariants and
the limits connecting them.

\subsection{Acknowledgement} We thank Rinat Kashaev for useful discussions. A.H. was funded by the Swiss National Science Foundation (SNSF) through the
Ambizione project “TRuality: Topological Recursion, Duality and Applications”
under the grant agreement PZ00P2\_\allowbreak223297 and partially by the DFG -- project ID 551478549.
S.S. was supported by the Dutch Research Council, grant no. OCENW.M.21.233.

AI agents were used substantially in the initial development of proof ideas. Some of the key ideas underlying the proofs in Sections 3 and 5 were suggested by AI agents, often in a rather unclear and difficult-to-follow form. Though, following the general philosophy of these suggestions, the authors independently developed, verified, and completely (re)wrote all proofs in the form presented in this paper. AI tools were also used to improve the language, grammar, and clarity of the manuscript. The authors take full responsibility for the mathematical content and correctness of the results.

\section{Colored Jones polynomial}\label{sec:knot}

\subsection{Colored Jones polynomial expansion}
Let $4_1$ be the figure eight knot in $S^3$. As an explicit definition one can use the closure of the three-strand braid
$\sigma_1\sigma_2^{-1}\sigma_1\sigma_2^{-1}$, where $\sigma_i$ interchanges strands $i$ and $i+1$, with the strand entering from the right passing over the strand entering from the left; the braid is read from top to bottom. 

Let $J_N(q)$ denote the reduced Jones polynomial of $4_1$, colored by the $N$-dimensional irreducible representation of $\mathfrak{sl}_2$. It can be computed by the skein relations and is given by the Habiro-L\^e formula~\cite{Habiro2000,Le2003} (see also~\cite{Murakami2013}):
\begin{align}\label{eq:habiro}
 J_N(q)
=\sum_{k=0}^{N-1}\prod_{j=1}^k
 (q^N+q^{-N}-q^j-q^{-j}).
\end{align}
We study the large-$N$ asymptotics of this formula at fixed $x=2\pi\I+s$, where $s$ belongs to a compact interval $I\subset(0,\log\frac{3+\sqrt 5}{2})$. Set $\hbar_N=x/N$\footnote{Here and below we use $\hbar$ as a parameter independent from $x$, in formal expansions, and $\hbar_N$ as the parameter in the large-$N$
asymptotics at fixed $x$.} and $q_N=e^{\hbar_N}$.
We use $\hbar_N^{1/2}=N^{-1/2}\sqrt{x}$, with $\sqrt{x}$ principal
on $2\pi\I+I$ and analytically continued to $U$.
We are interested in an all-orders asymptotic expansion of the form
\begin{align}\label{eq:jones-asymptotic-wavefunction}
 J_N(e^{x/N})
 \sim \hbar_N^{-\frac 12}\Psi(x;\hbar_N),
\qquad 
 \Psi(x;\hbar)=
 \exp\bigg(\hbar^{-1}{S_0(x)}+S_1(x)
       +\sum_{r\geq1}\hbar^r S_{r+1}(x)\bigg).
\end{align}
The all-orders asymptotic expansion means that, for every $n\geq0$,
\begin{equation}\label{eq:jones-asymptotic-definition}
 \log\left(\hbar_N^{1/2}J_N(e^{x/N})
     \exp\left(-\hbar_N^{-1}{S_0(x)}-S_1(x)\right)\right)
 =\sum_{i=1}^{n}\hbar_N^i S_{i+1}(x)+O(\hbar_N^{n+1}),
\end{equation}
uniformly for $s\in I$, with the logarithm chosen to tend to zero. The leading terms $S_0(x)$ and $S_1(x)$ are given by the explicit Murakami formulas~\cite{Murakami2013,Murakami2024} derived by saddle calculation:
\begin{align}\label{eq:Murakami-leading}
	S_0(x)=\operatorname{Li}_2(e^{s-\varphi(s)})
	-\operatorname{Li}_2(e^{s+\varphi(s)})-s\varphi(s),\qquad
	S_1(x)=\log \frac{\I\sqrt{\pi}\sqrt{T(s)}}{2\sinh(s/2)},
\end{align}
where 
\begin{align}
	\varphi(s)=-\I\arccos(\cosh s-\tfrac12),\qquad
T(s)=\frac{2}{\sqrt{(2\cosh s+1)(2\cosh s-3)}}.
\end{align}
We use here the principal dilogarithm, the positive imaginary square root in the denominator of $T$, and the square root of $T$ with positive real part. The restriction $I\subset(0,\log\frac{3+\sqrt5}{2})$ excludes $s=0$, where the leading prefactor changes from order $N^{1/2}$ to $N^{3/2}$, and $\cosh s = \frac 32$, where the saddle points coalesce.

\begin{proposition} \label{prop:all-order-existence}
	An all-orders asymptotic expansion \eqref{eq:jones-asymptotic-wavefunction} exists for $(J_N(q))_{N\geq 1}$ given by~\eqref{eq:habiro}, for $x$ in an open neighborhood $U$ of $2\pi\I+I\subset\mathbb{C}$, uniformly on any compact subset $K\subset U$.
\end{proposition}
The proof of this proposition is the content of Section~\ref{sec:analytic-comparison}.

\begin{figure}[htbp]
    \centering
\begin{tikzpicture}[
    scale=1,
    line cap=round,
    line join=round,
    every node/.style={font=\scriptsize}
]

\begin{scope}[shift={(1.5,1.5)}]

  \draw[->] (-1.45,0) -- (2.65,0);
  \draw[->] (0,-1.45) -- (0,2.05);

  \draw[thin] (2.05,2.02) -- (2.05,1.55)
                         -- (2.65,1.55);
  \node at (2.35,1.79) {$q$};

  \draw[line width=0.8pt] (0,0) circle (1);

  \fill (1,0) circle (1.1pt);
  \node[right] at (1.15,-0.2) {$q=1$};

  \foreach \xx/\yy/\rr in {
      0.86/0.69/0.065,
      0.95/0.53/0.055,
      0.99/0.40/0.045
  }{
      \draw[
          line width=0.45pt,
          dash pattern=on 1pt off 1.2pt
      ] (\xx,\yy) circle (\rr);
      \fill (\xx,\yy) circle (1pt);
  }

  \draw[->, thin]
      (1.43,0.92)
      .. controls (1.52,0.57) and (1.32,0.22) ..
      (1.10,0.08);

  \node[right] at (1.52,0.65) {$N\to\infty$};

  \node at (0,-1.75) {$q=e^{x/N}$};

\end{scope}

\begin{scope}[shift={(6.4,0)}]

  \draw[->] (-0.35,0) -- (5.3,0);
  \draw[->] (0,-0.35) -- (0,3.65);

  \draw[thin] (4.65,3.53) -- (4.65,3.03)
                         -- (5.25,3.03);
  \node at (4.95,3.28) {$x$};

  \draw (-0.10,2) -- (0.10,2);
  \node[left] at (0,2) {$2\pi \I$};

  \draw (4.45,-0.10) -- (4.45,0.10);
  \node[below] at (4.45,0)
    {$\log\left(\frac{3+\sqrt{5}}{2}\right)$};

  \draw[line width=0.9pt] (0.95,2) -- (3.55,2);
  \draw (0.95,1.90) -- (0.95,2.10);
  \draw (3.55,1.90) -- (3.55,2.10);
  \node[below] at (1.55,2) {$I$};

  \draw (2.55,2) ellipse (0.40 and 0.27);
  \fill (2.55,2) circle (1.1pt);
  \node[above right] at (2.4,2.17) {$K$};

  \draw[
      line width=0.6pt,
      dash pattern=on 2pt off 1.6pt
  ] (2.25,2) ellipse (1.75 and 0.65);

  \node[above right] at (3.3,2.48) {$U$};

\end{scope}

\end{tikzpicture}
\caption{The compact interval $2\pi \I+I$ is contained in a fixed,
sufficiently small open neighborhood $U$ in the complex
$x$-plane. For every compact subset $K\subset U$, the colored
Jones polynomial admits an all-orders asymptotic expansion,
uniformly for $x\in K$. Under the map $q=e^{x/N}$, the fixed
neighborhood $U$ is mapped to a shrinking neighborhood
$e^{U/N}$ as $N\to\infty$. For every $x\in K$, the corresponding
points $q=e^{x/N}$ approach $1$ from the upper half-plane,
outside the unit disk, as $N\to\infty$.}

    \label{fig:asymptotic-domains}
\end{figure}
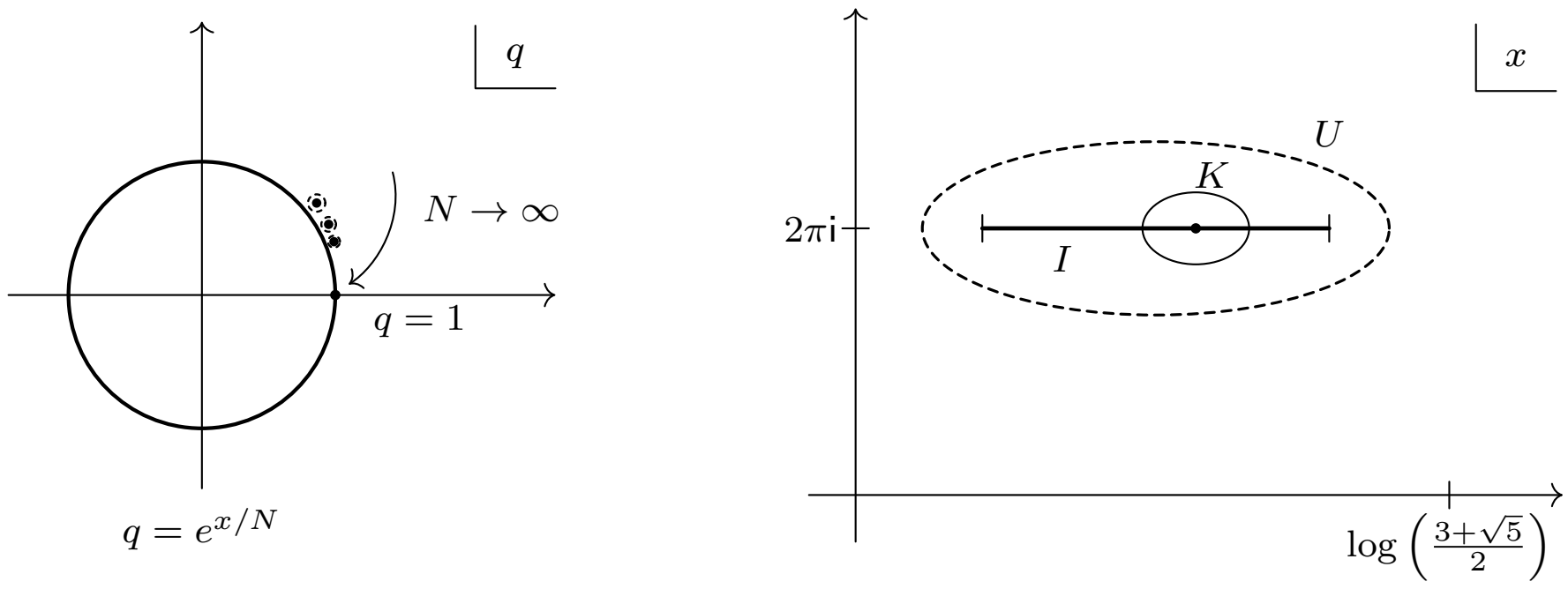

\subsection{The exact recurrence}\label{subsec:jones-recurrence}
The reduced zero-framed colored Jones polynomial~\eqref{eq:habiro} satisfies a linear recurrence in the color $N$~\cite{GaroufalidisLe2005} (see also a convenient presentation in~\cite{Zhu2017}). For $N\geq 3$ we have:
\begin{equation}\label{eq:jones-inhomogeneous}
 J_N(q)-b_N(q)J_{N-1}(q)+a_N(q)J_{N-2}(q)=e_N(q),
\end{equation}
where
\begin{align}\label{eq:jones-recurrence-coefficients}
 a_N(q)&\coloneqq \frac{(1-q^{N-2})(1-q^{2N-1})}
 {(1-q^N)(1-q^{2N-3})},
 \\ \notag 
 b_N(q)&\coloneqq \frac{q^{-2N+2}(1-q^{N-1})^2(1+q^{N-1}) (1-q^{N-1}-q^{2N-3}-q^{2N-1}-q^{3N-3}+q^{4N-4})}
 {(1-q^N)(1-q^{2N-3})}
,
 \\ \notag 
 e_N(q)& \coloneqq \frac{q^{-N+1}(1+q^{N-1})(1-q^{2N-1})}{1-q^N}.
\end{align}
The initial values are $J_1=1$ and $J_2=q^2-q+1-q^{-1}+q^{-2}$.

For a lattice function $f=(f_N)_{N\geq 1}$, $f_N\in \mathbb{Q}[q,q^{-1}]$, define
\begin{equation}\label{eq:jones-operators}
 (\widehat X f)_N=q^Nf_N,\qquad
 (\widehat Y f)_N=f_{N+1},\qquad
 \widehat Y\widehat X=q\widehat X\widehat Y.
\end{equation}
Then Equation~\eqref{eq:jones-inhomogeneous} implies, for $J=(J_N)_{N\geq 1}$,
\begin{align} \label{eq:widehat-A}
	\widehat{A} J =0, \qquad \widehat{A}  \coloneqq \widehat Y^3 (1-\tfrac{e_N}{e_{N-1}}\widehat Y^{-1})
	(1-b_N\widehat Y^{-1}+a_N\widehat Y^{-2}).
\end{align}

With $q=e^\hbar$ and $X=e^x=q^N$, the shift operator $\widehat{Y}$ becomes $\exp(\hbar \partial_x)$. Represent $\widehat A$ as
$\widehat A=\sum_{j=0}^3c_j(\widehat X,q)\widehat Y^j$. Its semi-classical limit $\hbar\to 0$ is given by
\begin{equation}\label{eq:jones-classical-symbol}
A(X,Y)=\sum_{j=0}^3c_j(X,1)Y^j
 =(Y-1)\left(Y^2-(X^2+X^{-2}-X-X^{-1}-2)Y+1\right)
\end{equation}
which is identical to the $A$-polynomial of the figure eight knot.

Let $a(e^x,\hbar)$ and $b(e^x,\hbar)$ be obtained
from~\eqref{eq:jones-recurrence-coefficients} by substituting
$q=e^\hbar$ and $q^N=e^x$. As we will show, the inhomogeneous term
$e_N(q)$ is exponentially small relative to the dominant
asymptotic contribution of the colored Jones polynomial
and therefore does not contribute to its all-orders
asymptotic expansion. Thus, we define $\Psi^{\mathsf{rec}}$
as a formal solution of the homogeneous second-order equation
\begin{align}
 \Psi^{\mathsf{rec}}(x;\hbar)
 -b(e^x,\hbar)\Psi^{\mathsf{rec}}(x-\hbar;\hbar)
 +a(e^x,\hbar)\Psi^{\mathsf{rec}}(x-2\hbar;\hbar)=0.
 \label{eq:rec-reduced-formal}
\end{align}
Solving it formally gives the WKB expansion~\cite{DGLZ2009,Zhu2017}
\begin{align}
 \Psi^{\mathsf{rec}}(x;\hbar)
 =\exp\left(\hbar^{-1}S^{\mathsf{rec}}_0(x)
 +S^{\mathsf{rec}}_1(x)
 +\sum_{r\geq1}\hbar^rS^{\mathsf{rec}}_{r+1}(x)\right).
 \label{eq:wkb-jones}
\end{align}
The leading multiplier $Y=\exp(\partial_xS_0^{\mathsf{rec}})$
satisfies the geometric factor of~\eqref{eq:jones-classical-symbol}.
For $k\geq1$, the coefficient of $\hbar^k$ expresses
$(1-Y^{-2})\partial_xS_k^{\mathsf{rec}}$ through earlier coefficients.
The factorization~\eqref{eq:widehat-A} also gives
$\widehat A\Psi^{\mathsf{rec}}=0$.

In the WKB approximation, choose $S^{\mathsf{rec}}_0=S_0$ and $S^{\mathsf{rec}}_1=S_1$, and consider the rest as holomorphic functions on $U$, that is, consider $ \Psi^{\mathsf{rec}}(x;\hbar)
\in e^{S_0(x)/\hbar+S_1(x)}
\bigl(1+\hbar\mathcal O(U)[[\hbar]]\bigr)$. Then Proposition~\ref{prop:all-order-existence} and formal uniqueness give the following corollary (cf.~\cite[Remark~2.4]{GaroufalidisGeronimo2006}):
\begin{corollary}\label{cor:psipsirec} As formal series, we have $\Psi(x;\hbar) = C(\hbar)\Psi^{\mathsf{rec}}(x;\hbar)$, where $C(\hbar) = 1+O(\hbar^1)$ is constant in $x$. 
\end{corollary}

In particular, this way we prove the first equality in Theorem~\ref{thm:main}.

\section{Recurrence vs all-orders asymptotic expansion}\label{sec:analytic-comparison}

\subsection{Structure of the argument}

The aim of this section is to prove
Proposition~\ref{prop:all-order-existence}, extending
Murakami's leading-order asymptotics of the colored Jones
polynomial on $2\pi \I+I$ to an all-orders expansion that
is locally uniform in an open complex neighborhood $U$
of $2\pi \I+I$. This means that the asymptotic expansion
holds uniformly on every compact subset $K\subset U$,
with coefficients holomorphic on $U$. We establish this
result by extending Murakami's saddle point analysis,
using the all-orders saddle point methods developed
in \cite{Ohtsuki2016}.

The local uniformity is essential for passing from
the exact inhomogeneous recurrence to the formal
homogeneous recurrence. Indeed, for $\hbar_N=x/N$,
the shifted arguments $x_j=x-j\hbar_N$ in \eqref{eq:rec-reduced-formal}, with
$j\in\{0,1,2\}$, remain in $U$ for sufficiently
large $N$, uniformly for $x$ in compact subsets
of $U$. Consequently, the same asymptotic expansion
applies to all shifted terms,
\begin{equation}
J_{N-j}(e^{\hbar_N})\sim
\hbar_N^{-1/2}
\exp\left(
\frac{S_0(x-j\hbar_N)}{\hbar_N}
+S_1(x-j\hbar_N)
+\sum_{r\geq 1}\hbar_N^r S_{r+1}(x-j\hbar_N)
\right).
\end{equation}
The holomorphic coefficients can then be expanded
around $x$, allowing the comparison of successive
orders in $\hbar_N$ in the recurrence.

The Gaussian saddle point expansion also yields
the structure of the higher-order coefficients,
given in \eqref{eq:analytic-gaussian-coefficients},
which will be used in
Section~\ref{sec:quantum-curve-proof}
to control their possible poles.
Finally, the inhomogeneous term $e_N$ in
\eqref{eq:jones-inhomogeneous} is exponentially small
relative to the dominant saddle contribution and
therefore does not contribute to the formal asymptotic
expansion. Consequently, $\Psi$ formally satisfies
the homogeneous second-order recurrence
\eqref{eq:rec-reduced-formal}, and formal uniqueness
of the WKB expansion gives
$d\log\Psi=d\log\Psi^{\mathsf{rec}}$.

\subsection{The analytic expansion}

\begin{proof}[Proof of Proposition~\ref{prop:all-order-existence}]

Fix $s_0\in(0,\log\frac{3+\sqrt{5}}{2})$ and set
$x_0=2\pi\I+s_0$. Let $K$ be a sufficiently small closed
disk in the $x$-plane around $x_0$. For $x\in K$, set
\begin{align}
s=x-2\pi\I,\qquad
\hbar_N=\frac{x}{N},\qquad
q_N=e^{\hbar_N}.
\end{align}
Since $\Re(x)=\Re(s)>0$ on $K$, we have $|q_N|>1$.
All estimates below will be uniform for $x\in K$,
with coefficients holomorphic on a neighborhood of $K$.

Following \cite{Murakami2024}, let the quantum dilogarithm \cite{MR1345554}
be defined by
\begin{equation}\label{eq:quantumdilogdefinition}
    T_N(v)=\frac14\int_{\mathbb R+\I 0}
    \frac{e^{(2v-1)u}}
         {u\,\sinh u\,\sinh\!\left(\frac{\hbar_N u}{2\pi\I}\right)}
    \,du,
\end{equation}
where the contour follows the real line from left to right and
passes above $0$ along a small semicircle. For $0<\Re(v)<1$,
the integral converges and defines a holomorphic function of
$v$ and $s$ with $\hbar_N=\frac{s+2\pi\I}{N}$.

The summands $B_{N,k}(s)$ in the Habiro--L\^e formula
$J_N(q_N)=\sum_{k=0}^{N-1} B_{N,k}(s)$ can be written as
\begin{equation}
B_{N,k}(s)
=\prod_{j=1}^k
\left(e^s+e^{-s}-q_N^j-q_N^{-j}\right) =e^{ks}\prod_{j=1}^k
\left(1-q_N^{-N-j}\right)
\left(1-q_N^{-N+j}\right).
\end{equation}

To write the expression in a symmetric form, we introduce
the variables
\begin{align}
\zeta_k=\frac{k+\frac12}{N},
\qquad
a_\pm(\zeta,s)=e^{-x(1\pm\zeta)}.
\end{align}
Next, we want to rewrite the summands by expressions in terms of the quantum dilogarithm to achieve analytic continuation.
For real $s$, applying the second identity of
\cite[Corollary~2.6]{Murakami2024} to the factors in 
the product and using \cite[Lemma~2.7]{Murakami2024}
gives
\begin{equation}
B_{N,k}(s)
=
\frac{1-e^{-4\pi^2N/x}}{2\sinh(s/2)}
e^{Nf_N(\zeta_k,s)},
\end{equation}
where
\begin{equation}
f_N(\zeta,s)
=
s\zeta+\frac1N\left(
T_N\!\left(2-\frac{x(1+\zeta)}{2\pi\I}\right)
-
T_N\!\left(1-\frac{x(1-\zeta)}{2\pi\I}\right)
\right).
\end{equation}
At the two arguments $v$ of $T_N$ in this formula,
$e^{2\pi\I v}$ equals $a_+(\zeta,s)$ and $a_-(\zeta,s)$,
respectively. This defines an analytic interpolation of the
summands wherever both arguments have real part in $(0,1)$.
For any fixed $\nu>0$ sufficiently small, this condition holds
at $\zeta=\zeta_k$ for $\nu\le k/N\le 1-\nu$ and all sufficiently
large $N$ (after shrinking $K$ if necessary). The identity for
these summands therefore extends to $x\in K$ by holomorphy.
Since $\Re(1/x)>0$ on $K$, we also have
$1-e^{-4\pi^2N/x}=1+O(e^{-cN})$ uniformly for some $c>0$.

\subsubsection*{Expansion of the phase.}
For $0<\Re(v)<1$, the quantum dilogarithm admits the following
asymptotic expansion (see for instance \cite{MR4394512}) as $N\to\infty$:
\begin{equation}
T_N(v)
=
\sum_{r=0}^{D}
\frac{B_{2r}(1/2)}{(2r)!}
\hbar_N^{2r-1}\operatorname{Li}_{2-2r}(e^{2\pi\I v})
+O(|\hbar_N|^{2D+1}).
\end{equation}
Here $\operatorname{Li}_2$ is taken on its principal branch.
This expansion can, for instance, be derived from the generating function of the
Bernoulli polynomials,
\begin{align}
\frac{t}{\sinh t}
=
\sum_{r=0}^{\infty}
\frac{B_{2r}(1/2)}{(2r)!}(2t)^{2r},
\end{align}
applied in the defining integral for $T_N(v)$ with
$t=\hbar_N u/(2\pi\I)$. Very explicit computation especially for the leading terms were performed in \cite{Murakami2024}, the higher order terms
are obtained by differentiating it with respect to $v$.

The estimate is uniform for $v$ in compact subsets of the strip
$0<\Re(v)<1$ and $x\in K$. Indeed, on $|u|\leq\sqrt{N}$ in \eqref{eq:quantumdilogdefinition} the
the higher order terms after the term of degree $2D$ is
$O(|t|^{2D+2})$. Its integral contributes
$O(|\hbar_N|^{2D+1})$, since $e^{(2v-1)u}/\sinh u$ decays
exponentially on the real axis. On $|u|>\sqrt{N}$ in \eqref{eq:quantumdilogdefinition} the same
exponential decay, together with
\begin{align}
\left|\sinh\!\left(\frac{\hbar_N u}{2\pi\I}\right)\right|
\geq |u|\,\Re\!\left(\frac{\hbar_N}{2\pi\I}\right)
\geq \frac{c|u|}{N}
\end{align}
for some $c>0$ uniform on $K$, makes the integral for $|u|>\sqrt{N}$
smaller than any power of $N^{-1}$. In particular, the expansion
applies locally uniformly in $(\zeta,s)$ wherever both arguments
of $T_N$ in the expression for $f_N$ have real part in $(0,1)$.

Substituting into the expression for $f_N$ gives
\begin{equation}
f_N(\zeta,s)
=
F_0(\zeta,s)
+\sum_{r=1}^{D}N^{-2r}F_r(\zeta,s)
+O(N^{-2D-2}),
\end{equation}
where
\begin{equation}\label{eq:F0}
F_0(\zeta,s)
=
s\zeta+
\frac{
\operatorname{Li}_2(a_+(\zeta,s))
-\operatorname{Li}_2(a_-(\zeta,s))
}{x},
\end{equation}
and, for $r\geq1$,
\begin{equation}
F_r(\zeta,s)
=
\frac{B_{2r}(1/2)x^{2r-1}}{(2r)!}
\left(
\operatorname{Li}_{2-2r}(a_+(\zeta,s))
-\operatorname{Li}_{2-2r}(a_-(\zeta,s))
\right).
\end{equation}
This expansion is well-known, but importantly is valid holomorphically on the chosen domain.
The asymptotic estimates remain valid
for all fixed-order derivatives with respect to $\zeta$ and $s$,
uniformly on compact subsets of the domain.

\subsubsection*{Saddle point approximation.}
Using the holomorphic interpolation by the quantum dilogarithm of the summands $B_{N,k}$
in the Habiro--L\^e formula, one can perform a saddle
point approximation à la Murakami. The idea
is to first derive, up to some negligible terms, a
contour integral formula and then perform
essentially Gaussian integrals.

For the leading phase $F_0(\zeta,s)$ in \eqref{eq:F0},
we have
\begin{equation}
\partial_\zeta F_0(\zeta,s)
=
\log\left(2\cosh s-2\cosh(x\zeta)\right),
\end{equation}
with the logarithmic branch determined by
\eqref{eq:F0}.

The saddle point appears at $\partial_\zeta F_0(\zeta,s)\vert_{\zeta=\zeta_*}
=0$ (see \cite{Murakami2013}) giving
\begin{align}
&
\varphi(s)=-\I\arccos(\cosh s-\tfrac12),
\qquad
\zeta_*=\frac{2\pi\I+\varphi(s)}{x},
\\ \notag 
&\Phi(s):=F_0(\zeta_*,s)=\frac{S_0(x)}{x},\\ \notag 
&
\left.\partial_\zeta^2F_0(\zeta,s)\right|_{\zeta=\zeta_*}
=
x\sqrt{(2\cosh s+1)(2\cosh s-3)}
\neq0,
\end{align}
where the square root has positive imaginary part at $s=s_0$
and is continued analytically to $K$.

In \cite{Murakami2024}, the real point $s=s_0$ is analysed, we want to extend this analytically in some complex neighborhood.
First, the boundary terms in the large $N$ limit of the
colored Jones polynomial can be estimated not to contribute.
For small enough fixed $\nu>0$, we have
\begin{align}\label{eq:analytic-endpoint-error}
e^{-N\Phi(s)}
\left(
\sum_{k/N<\nu}+\sum_{k/N>1-\nu}
\right)B_{N,k}(s)
=O(Ne^{-\epsilon N}),
\end{align}
uniformly for $x\in K$ (after shrinking $K$ if necessary).

Indeed, the factors
$2\cosh s-2\cosh(xt)$ have absolute value bounded by a
constant less than $1$ near the endpoints $t=0$ and $t=1$,
uniformly for $x\in K$. Near the left endpoint, this directly
bounds the summands, since $B_{N,0}=1$ and
$\Re\Phi(s_0)>0$ as in \cite[Lemma~5.2]{Murakami2024}.
Near the right endpoint, choose $\nu$ sufficiently small that
\begin{align}
\Re F_0(1-\nu,s_0)<\Re\Phi(s_0),
\end{align}
as in \cite[Lemma~6.1]{Murakami2024}. This strict inequality
persists on small enough $K$. The expansion of $f_N$ at
$k_0=\lfloor N(1-\nu)\rfloor$ therefore gives
\begin{align}
|B_{N,k_0}(s)|
=O\left(e^{N(\Re\Phi(s)-\epsilon)}\right).
\end{align}
The same bound holds for all $k>k_0$. This proves
\eqref{eq:analytic-endpoint-error}.

For the remaining terms in the colored Jones polynomial,
with $\nu\leq k/N\leq1-\nu$, we use the contours of
\cite[Proposition~5.1]{Murakami2024}, with $p=1$ and $m=0$.
At $s=s_0$, the upper and lower paths in \cite[Proposition~5.1, (4)--(5)]{Murakami2024} can be chosen compactly inside the domain
of the phase. On these paths,
\begin{align}
\Re\left(F_0(\zeta,s_0)-\Phi(s_0)\right)
-2\pi|\Im\zeta|\leq-\epsilon
\end{align}
for some $\epsilon>0$. By compactness, both arguments of $T_N$
remain in the strip $0<\Re(v)<1$ on a neighborhood of the
paths for $x\in K$, and the inequalities remain valid after
decreasing $\epsilon$. The expansion of $f_N$ and its
derivatives is therefore uniform there. Applying
\cite[Proposition~4.2 and Remark~4.4]{Ohtsuki2016} to
$f_N-\Phi$ gives
\begin{align}\label{eq:analytic-poisson}
\frac1N\sum_{\nu\leq k/N\leq1-\nu}
e^{N(f_N(\zeta_k,s)-\Phi(s))}
=
\int_{\Gamma(s)}
e^{N(f_N(\zeta,s)-\Phi(s))}\,d\zeta
+O(e^{-\epsilon N}).
\end{align}
Shifting the integration
endpoints by $O(1/N)$ contributes only to the error term,
by the strict inequalities at the endpoints.

The contour $\Gamma(s)$ is a descent path through $\zeta_*$.
At $s=s_0$,
\cite[Proposition~5.1 (1)--(3)]{Murakami2024} gives such a path with
$\Re(F_0-\Phi)\leq-\epsilon$ outside a small disk around
the saddle point. This bound persists on the compact part of the
path outside the disk. Inside the disk, since
$\Re\partial_\zeta^2F_0(\zeta_*,s_0)<0$ at $s=s_0$, we
can deform the path to pass through the moving saddle in a direction of quadratic decay, with
\begin{align}
\Re\left(F_0(\zeta,s)-\Phi(s)\right)
\leq-c|\zeta-\zeta_*|^2
\end{align}
near $\zeta_*$, uniformly for $x\in K$.

The saddle point expansion turns the integral into
Gaussian integrals evaluated at the saddle point
\begin{align}
\zeta_*=\frac{2\pi\I+\varphi(s)}{x}.
\end{align}
With the variable $z=e^{\varphi(s)}$, we find
\begin{align}
\left.\partial_\zeta^2F_0(\zeta,s)\right|_{\zeta=\zeta_*}
=
x\sqrt{(2\cosh s+1)(2\cosh s-3)}
=-x(z-z^{-1})\neq0.
\end{align}

To see how the Gaussian integrals arise, we introduce
$t=x\zeta-2\pi\I$ and $w=e^x=e^s$.
Since $2\cosh\varphi(s)=2\cosh s-1$, the variables
$z$ and $w$ satisfy the algebraic equation
\begin{align}\label{eq:zwalgebra}
    z+\frac1z-\bigg(w+\frac1w\bigg)+1=0.
\end{align}
Using identities of the dilogarithm, the leading phase becomes $xF_0(\zeta,s)=V(t)$, where
\begin{equation}
V(t)
=
\operatorname{Li}_2(we^{-t})
-\operatorname{Li}_2(we^t)-st.
\end{equation}
In particular,
\[
V(\varphi(s))=S_0(x),
\qquad
V''(\varphi(s))=-(z-z^{-1}).
\]
Similarly, the expansion of $f_N$ near the saddle is
\begin{equation}
Nf_N(\zeta,s)\sim
\frac{V(t)}{\hbar_N}
+\sum_{r\geq1}\hbar_N^{2r-1}\underbrace{\frac{B_{2r}(1/2)}{(2r)!}
\left(
\operatorname{Li}_{2-2r}(we^{-t})
-\operatorname{Li}_{2-2r}(we^t)
\right)}_{U_r(t)\coloneqq}.
\end{equation}

In the variable $t=x\zeta-2\pi\I$, the two arguments of $T_N$ are
$-(s+t)/(2\pi\I)$ and $1+(t-s)/(2\pi\I)$. Both have real part
in $(0,1)$ precisely when
\begin{align}
    -2\pi+|\Im s|<\Im t<-|\Im s|.
\end{align}
Thus $f_N$ and the coefficients $F_r$, including $F_0$, are
holomorphic there as functions of $(t,s)$. At $s=s_0$ this is the
strip $-2\pi<\Im t<0$. For fixed sufficiently small $\nu$, choose
$0<\delta<2\pi\nu$ and shrink $K$ about $x_0$. The truncated
contour in Figure~\ref{fig:saddle-contour} then stays in the common
strip $-2\pi+\delta<\Im t<-\delta$ as $x$ varies in $K$.
The saddle $t=\varphi(s)$ extends holomorphically from $s_0$.
For $s$ near $s_0$, the vertical part of the contour can be moved
to $\Re t=\Re\varphi(s)$, keeping the same endpoints in the
$\zeta$-plane.

\begin{figure}[htbp]
\centering
\resizebox{\linewidth}{!}{%
\begin{tikzpicture}[x=1cm,y=1cm,>=stealth,font=\small,
  contour/.style={draw=contourink,line width=1.3pt,line join=round},
  axis/.style={draw=black!60,line width=.48pt},
  boundary/.style={draw=stripink,densely dashed,line width=.5pt}]
  \definecolor{contourink}{RGB}{140,50,32}
  \definecolor{stripink}{RGB}{48,98,125}
  \definecolor{stripfill}{RGB}{226,239,246}
  \definecolor{decayfill}{RGB}{222,239,230}

  \node[anchor=west,font=\bfseries] at (-1.10,5.63)
    {Contour in the $t$-plane};
  \node[anchor=west,font=\bfseries] at (7.10,5.63)
    {Gaussian line in the $u$-plane};

  \fill[black!3] (-.93,0) rectangle (5.89,4.92);
  \fill[stripfill] (-.93,.18) rectangle (5.89,4.74);
  \draw[boundary] (-.93,.18)--(5.89,.18);
  \draw[boundary] (-.93,4.74)--(5.89,4.74);
  \draw[axis] (-.93,0)--(5.89,0);
  \draw[axis,->] (-.93,4.92)--(6.14,4.92)
    node[right] {$\Re t$};
  \draw[axis,->] (0,-.21)--(0,5.24)
    node[right] {$\Im t$};
  \node[above left,inner sep=2pt] at (0,4.92) {$0$};
  \node[below left,inner sep=2pt] at (0,0) {$-2\pi$};

  \coordinate (L) at (.33,.34);
  \coordinate (R) at (4.41,4.58);
  \coordinate (S) at (0,4.03);
  \draw[black!45,densely dotted,line width=.75pt] (L)--(R);
  \foreach \v in {.09,.18,.27,.36,.45,.54,.63,.72,.81,.90}
    {\path (L)--(R) coordinate[pos=\v] (Q);
     \fill[black!58] (Q) circle[radius=1.25pt];}
  \node[rotate=46,anchor=south,text=black!75,inner sep=3pt]
    at (2.05,2.12) {$\zeta\in[\nu,1-\nu]$};

  \draw[contour] (L)--(0,.34)--(0,4.58)--(R);
  \draw[contour,->] (0,1.35)--(0,2.13);
  \draw[contour,->] (2.18,4.58)--(3.06,4.58);
  \fill[black] (L) circle[radius=1.8pt];
  \fill[black] (R) circle[radius=1.8pt];
  \node[anchor=west,inner sep=1pt] at (.49,.40)
    {$x\nu-2\pi\I$};
  \draw[black!45,line width=.4pt] (4.46,4.53)--(4.72,4.27)--(4.72,3.87);
  \node[anchor=east,inner sep=1pt] at (5.80,3.64)
    {$x(1-\nu)-2\pi\I$};
  \draw[contourink,densely dashed] (S) ellipse (.30 and .25);
  \fill[contourink] (S) circle[radius=2.5pt];
  \node[anchor=west,text=contourink] at (.37,4.02)
    {$t=\varphi(s_0)$};
  \node[anchor=west,text=contourink] at (.09,3.10)
    {$x\Gamma(s_0)-2\pi\I$};

  \begin{scope}[shift={(9.52,2.65)}]
    \fill[decayfill] (0,0) rectangle (1.92,1.92);
    \fill[decayfill] (0,0) rectangle (-1.92,-1.92);
    \draw[axis,->] (-2.10,0)--(2.14,0) node[right] {$\Re u$};
    \draw[axis,->] (0,-2.05)--(0,2.12) node[above] {$\Im u$};
    \draw[contour] (-1.67,-1.80)--(1.67,1.80);
    \draw[contour,->] (.60,.647)--(1.17,1.261);
    \fill[contourink] (0,0) circle[radius=2.4pt];
    \node[below right,inner sep=2pt] at (0,0) {$0$};
    \node[text=contourink,anchor=west] at (1.30,1.18)
      {$\Gamma_{\mathrm G}$};
  \end{scope}
  \node at (9.56,.12)
    {$\displaystyle u=\frac{t-\varphi(s_0)}{\sqrt{\hbar_N}}$};

  \fill[stripfill] (-.88,-.72) rectangle (-.57,-.52);
  \draw[stripink,densely dashed] (-.88,-.52)--(-.57,-.52);
  \node[anchor=west] at (-.47,-.62) {common strip for $x\in K$};
  \fill[decayfill] (7.11,-.72) rectangle (7.43,-.52);
  \node[anchor=west] at (7.50,-.62)
    {$\Re((z-z^{-1})u^2)>0$};
\end{tikzpicture}%
}
\caption{The contour for real $s=s_0$, drawn schematically.
The dashed lines in the $t$-plane bound a strip containing the
contour truncated at $\nu$ and $1-\nu$. The dots represent the
original summands $B_{N,k}$. The contour passes vertically through
the saddle point in the $t$-plane. Under the rescaling, this local
segment tends to the oriented line
$\Gamma_{\mathrm G}=(i/\sqrt{x})\mathbb R$,
along which $\exp(-(z-z^{-1})u^2/2)$ decays.}
\label{fig:saddle-contour}
\end{figure}
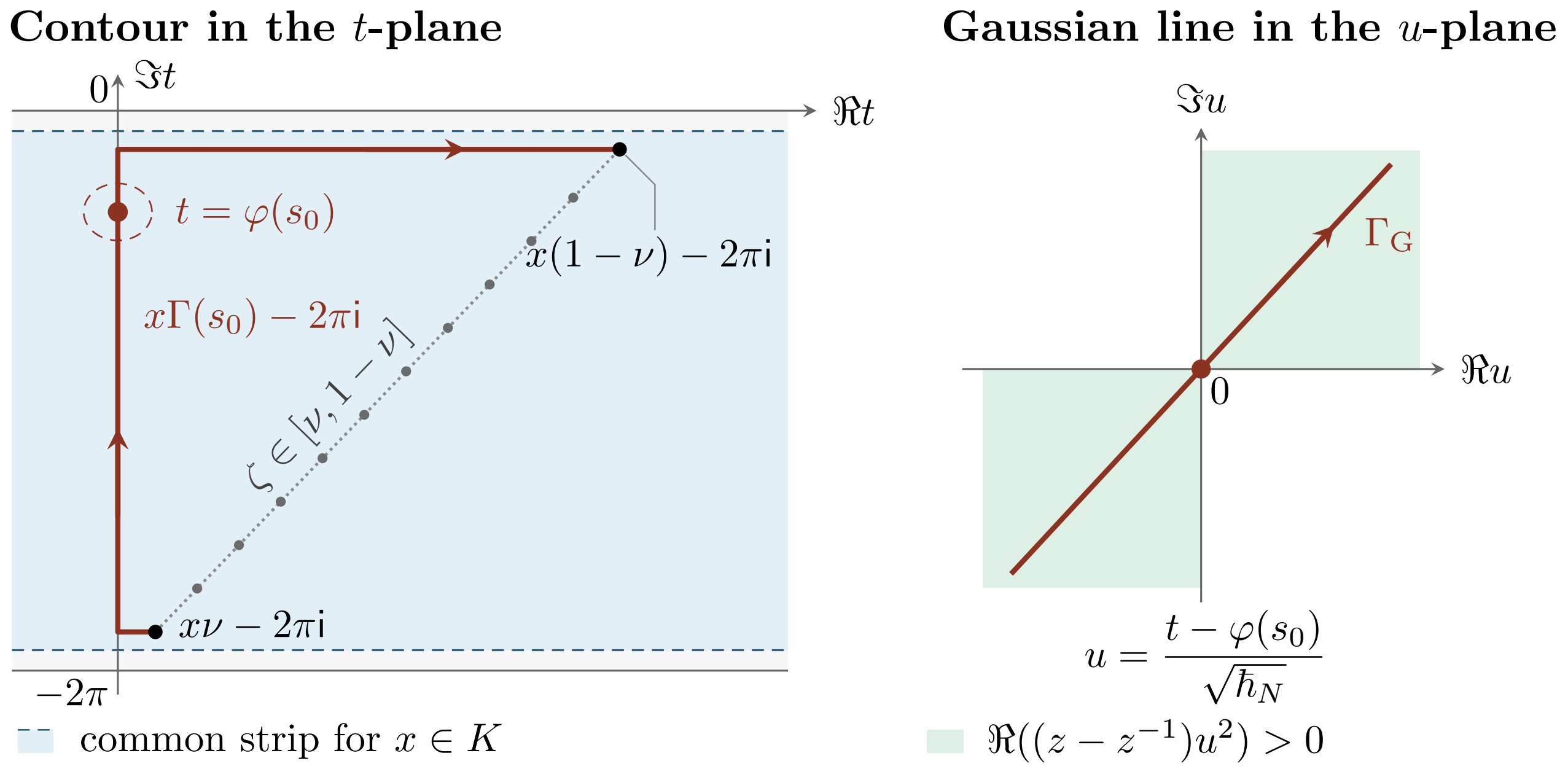

Now expanding around the saddle point with
$t=\varphi(s)+\sqrt{\hbar_N}\,u$ gives
\begin{equation} \label{eq:saddle-point-expansion}
\frac{V(t)}{\hbar_N}
=
\frac{S_0(x)}{\hbar_N}
-\frac{z-z^{-1}}{2}u^2
+\sum_{j\geq3}
\frac{V^{(j)}(\varphi(s))}{j!}
\hbar_N^{j/2-1}u^j.
\end{equation}
Consequently, after expanding the remaining exponential,
the odd Gaussian moments vanish, and at every fixed order
in $\hbar_N$ we obtain finitely many Gaussian integrals
of the form
\begin{equation}
c_{r,m}(s)
\int_{\Gamma_{\mathrm G}}
u^{2m}
e^{-\frac12(z-z^{-1})u^2}\,du,
\end{equation}
where $c_{r,m}(s)$ is independent of $u$ and is a
polynomial in the values and derivatives of $V$ and $U_r$
at the saddle point. The quadratic bound on $\Gamma(s)$ and the
uniform expansion of $e^{N(f_N-F_0)}$ near the saddle give
a uniform remainder at every fixed order.

Thus, inserting the expansion of $f_N$ into
\eqref{eq:analytic-poisson} and performing the Gaussian
integrals gives, since $N\,d\zeta=dt/\hbar_N$,
\begin{align} \label{eq:analytic-hbar-expansion}
e^{-S_0(x)/\hbar_N}\hbar_N^{1/2}J_N(e^{\hbar_N})
&=
\sum_{m=0}^{R}D_m(x)\hbar_N^m
+O(|\hbar_N|^{R+1}),
\qquad D_0=e^{S_1},
\end{align}
uniformly for $x\in K$. The orientation of the Gaussian
integral is fixed by the leading term in
\cite{Murakami2013}. The factor $1-e^{-4\pi^2N/x}$ in
the interpolation contributes only an exponentially small
error.

Passing to the associated formal series in $\hbar$, we see that at every fixed order in $\hbar$,
the coefficients $S_r$ are given by finitely many
terms arising from Gaussian integrals,
\begin{align}
\sum_{r\geq1}\hbar^r S_{r+1}(x)
=
\log\left(
1+\sum_{m\geq1}
\frac{D_m(x)}{D_0(x)}\hbar^m
\right).
\end{align}

Consequently, the expansion gives polynomials in
$(z-z^{-1})^{-1}$ and the polylogarithms of nonpositive
index at the saddle, that is,
$\operatorname{Li}_{-n}(wz)$ and
$\operatorname{Li}_{-n}(w/z)$ with $n\geq0$.
Indeed, the normalized Gaussian moment of order $2m$ is
$(2m-1)!!(z-z^{-1})^{-m}$, and the prefactor
$(2\sinh(s/2))^{-1}$ cancels in $D_m/D_0$.
Thus, we conclude for every $m\geq1$
\begin{align}
\frac{D_m}{D_0},\ S_{m+1}
\in\mathbb C\left[
(z-z^{-1})^{-1},\,
\operatorname{Li}_{-n}(wz),\,
\operatorname{Li}_{-n}(w/z)
:\ n\geq0
\right].
\label{eq:analytic-gaussian-coefficients}
\end{align}

Since $s_0$ was arbitrary, we can cover $2\pi\I+I$ by the
interiors of finitely many such disks $K_i$. By uniqueness
of the asymptotic coefficients, the functions $D_m$ agree
on overlaps and hence are holomorphic on the open neighborhood
$U$ of $2\pi\I+I$,
with locally uniform estimates. Furthermore, $z,w$ satisfy the algebraic relation
\eqref{eq:zwalgebra}, which defines an irreducible complex
algebraic curve. The map $x\mapsto(w,z)$ identifies $U$ with an open subset of this
curve. By \eqref{eq:analytic-gaussian-coefficients}, for every
$m\geq1$ the coefficients $D_m/D_0$ and $S_{m+1}$ on $U$
are rational expressions in $z,w$.
Consequently, these coefficients extend uniquely to meromorphic
functions on the smooth compact normalization of the curve
\eqref{eq:zwalgebra}.

This proves~\eqref{eq:jones-asymptotic-definition}
with the required complex locally uniform behaviour.
\end{proof}

\begin{remark}
The asymptotic expansion of the colored Jones polynomial
can also be obtained from the
state-integral models, as studied, for instance, in \cite{DGLZ2009,AK14,GK17}.
In this approach, the state integrals are expanded
around their saddle points, leading again to Gaussian
integrals and higher-order coefficients of the same
general form. The state-integral approach gives
the same algebraic structure of the higher-order
coefficients through Gaussian saddle point expansions.
The purpose of the present argument is mainly to
establish the locally uniform asymptotic expansion
directly from the colored Jones polynomial.
\end{remark}

\subsection{Passage to the recurrence}

We now use the locally uniform asymptotic expansion
established in the previous section to pass from the
exact inhomogeneous recurrence to the formal homogeneous
recurrence.

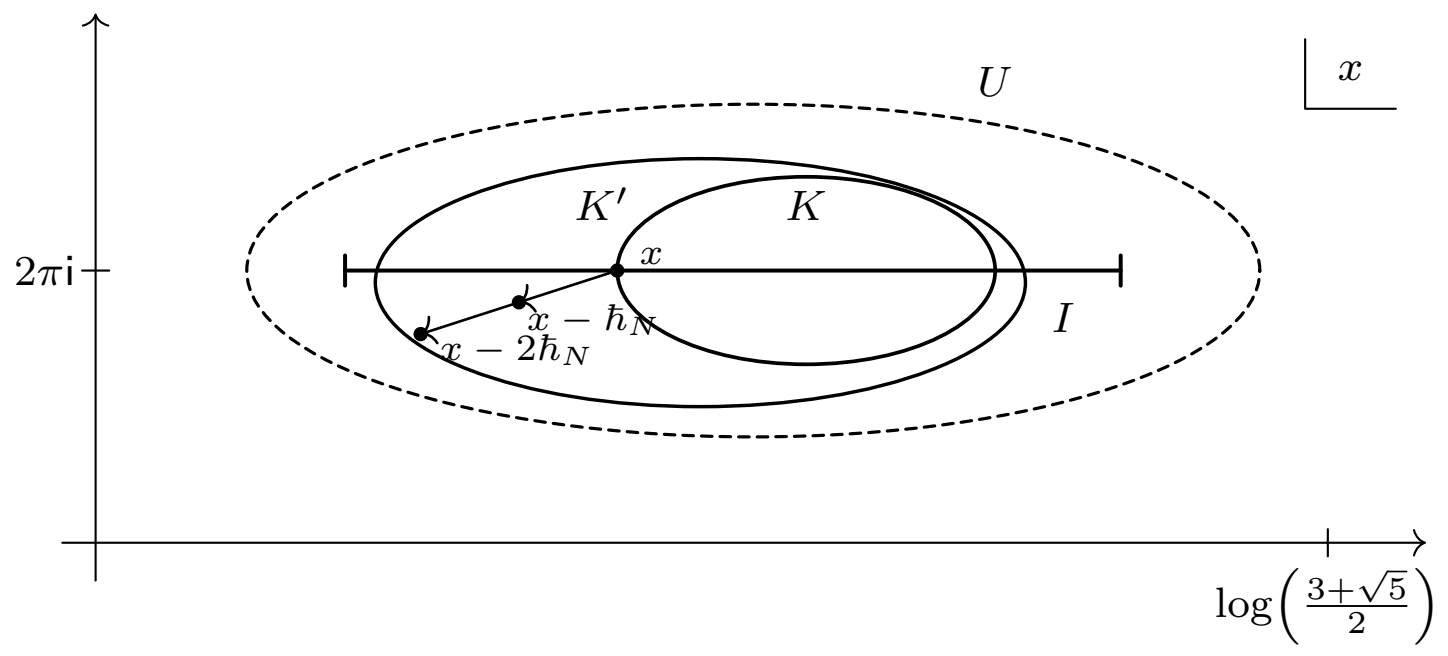
\begin{figure}[htbp]
    \centering
\begin{tikzpicture}[
  scale=1,
  line cap=round,
  line join=round,
  every node/.style={font=\scriptsize},
  shiftlabel/.style={font=\tiny, inner sep=1.0pt}
]
  \draw[->,thin] (-0.22,0) -- (8.8,0);
  \draw[->,thin] (0,-0.25) -- (0,3.5);

  \draw[thin] (8.00,3.33) -- (8.00,2.87) -- (8.60,2.87);
  \node at (8.30,3.12) {$x$};

  \draw[thin] (-0.09,1.8) -- (0.09,1.8);
  \node[left] at (0,1.8) {$2\pi\I$};
  \draw[thin] (8.15,-0.09)--(8.15,0.09);
  \node[below] at (8.15,-0.03) {$\log\!\left(\frac{3+\sqrt5}{2}\right)$};

  \draw[dash pattern=on 2pt off 1.5pt, line width=0.6pt]
       (4.35,1.8) ellipse (3.35 and 1.1);
  \node[above] at (5.95,2.82) {$U$};

  \draw[line width=0.75pt] (1.65,1.8) -- (6.78,1.8);
  \draw[line width=0.75pt] (1.65,1.7)--(1.65,1.9);
  \draw[line width=0.75pt] (6.78,1.7)--(6.78,1.9);
  \node[below] at (6.4,1.72) {$I$};

  \draw[line width=0.65pt] (4.00,1.72) ellipse (2.15 and 0.82);
  \node[above] at (3.35,2.) {$K'$};

  \draw[line width=0.7pt] (4.70,1.8) ellipse (1.25 and 0.62);
  \node[above] at (4.70,2) {$K$};

  \coordinate (p0) at (3.45,1.8);   
  \coordinate (p1) at (2.80,1.59);  
  \coordinate (p2) at (2.15,1.38);  

  \draw[->,line width=0.5pt] (p0) -- (p1);
  \draw[->,line width=0.5pt] (p1) -- (p2);

  \fill (p0) circle (1.35pt);
  \fill (p1) circle (1.35pt);
  \fill (p2) circle (1.35pt);

  \node[shiftlabel,above left] at (3.8,1.8) {$x$};
  \node[shiftlabel,below right] at (2.81,1.62) {$x-\hbar_N$};
  \node[shiftlabel,below right] at (2.23,1.42) {$x-2\hbar_N$};

\end{tikzpicture}
\caption{Local uniformity of the asymptotic expansion. Although $x\in K\subset K'\subset U$, the shifted points $x-j\hbar_N$, with $j=1,2$, may lie outside $K$. For sufficiently large $N$, however, they remain inside the slightly larger compact subset $K'\subset U$. Consequently, the same asymptotic expansion applies uniformly at the shifted arguments, with coefficient functions evaluated at $x-j\hbar_N$.}
\label{fig:xshift}
\end{figure}

Fix a compact subset $K\subset U$ and a slightly larger
compact subset $K'\subset U$, with $K$ contained in the
interior of $K'$. For $x\in K$ put $\hbar_N=x/N$ and
$x_j=x-j\hbar_N$, $j=0,1,2$. For sufficiently large $N$,
the shifted points lie in $K'$ and
\begin{align}
 \frac{x_j}{N-j}=\hbar_N,\qquad
 J_{N-j}(e^{\hbar_N})=J_{N-j}(e^{x_j/(N-j)}).
\end{align}
The importance of local uniformity for these shifted
arguments is illustrated in Figure~\ref{fig:xshift}.

Thus the expansions~\eqref{eq:analytic-hbar-expansion}
apply to all shifted terms in the inhomogeneous
recurrence~\eqref{eq:jones-inhomogeneous}. 
The coefficients of the recurrence, evaluated at
$q_N=e^{\hbar_N}$, are holomorphic in $x$ and uniformly
bounded on these compact sets for sufficiently large $N$.

We have
$\operatorname{Re}(S_0(x)/x)=\operatorname{Re}\Phi(s)>c>0$.
Consequently, after division by
$\hbar_N^{-1/2}e^{S_0(x)/\hbar_N}$, the right-hand side
satisfies
\begin{align}
 \hbar_N^{1/2}e^{-S_0(x)/\hbar_N}e_N(e^{\hbar_N})
 =O(N^{-1/2}e^{-cN}),
\end{align}
uniformly for $x$ in compact subsets of $U$.
It therefore has non-contributing formal asymptotic expansion.
Taylor expansion in the shifted arguments, using
the locally uniform property in
\eqref{eq:analytic-hbar-expansion}, gives identities between the coefficients of successive
powers of $\hbar_N$.
These identities state that
$\Psi(x,\hbar)$ satisfies the homogeneous equation
\eqref{eq:rec-reduced-formal} as a formal series in $\hbar$.

Finally, we compare $\Psi$ with the formal WKB
solution $\Psi^{\mathsf{rec}}$ of the same homogeneous
recurrence. Both have the same leading terms
$S_0$ and $S_1$. The higher-order WKB equations
determine the derivatives of the coefficients
uniquely at each order.
Thus the two formal expansions can differ only
by an $x$-independent formal series, and we conclude
\begin{align}
 \Psi(x;\hbar)
 &=C(\hbar)\Psi^{\mathsf{rec}}(x;\hbar),
 \qquad C(\hbar)\in1+\hbar\mathbb C[[\hbar]],\\
 d\log\Psi&=d\log\Psi^{\mathsf{rec}}.
 \label{eq:analytic-formal-identification}
\end{align}

\section{Spectral curve and Borot-Eynard wave function}\label{sec:A-polynomial}

\subsection{Equation and properties} The spectral curve $\Sigma$ is the smooth compact normalization of the one given by
\begin{equation}\label{eq:A-geom}
	A^{\mathsf{geom}}(X,Y)\coloneqq X^2Y^2-(X^4-X^3-2X^2-X+1)Y+X^2=0.
\end{equation}
It can be birationally equivalently represented as
\begin{align} \label{eq:normalized-curve}
	z+\frac1z-\left(w+\frac1w\right)+1=0,
\end{align}
with $X=w$ and $Y=(wz-1)/(w-z)$. The projection to $w$ presents the spectral curve as a double cover of $\mathbb C\mathrm P^1$ ramified over the four roots of
\begin{align} \label{eq:zeros-dx-branching}
	Q(w)\coloneqq (w^2-3w+1)(w^2+w+1)=0,   \qquad w = \frac{3\pm\sqrt5}{2},
	\frac{-1\pm\I\sqrt3}{2}.
\end{align}
Note that $\Sigma$ is of genus $1$, and $Y=1$ at $w = \frac{3\pm\sqrt5}{2}$ and  $Y=-1$ at $\frac{-1\pm\I\sqrt3}{2}$. In the presentation \eqref{eq:A-geom} there are two nodes, at $(X,Y)=(1,-1)$ and $(-1,1)$, whose preimages are, respectively,
\begin{align}\label{eq:node-preimages}
	(w,z)=\left(1,\frac{1\pm\I\sqrt3}{2}\right),\qquad
	(w,z)=\left(-1,\frac{-3\pm\sqrt5}{2}\right).
\end{align}
Let $\mathcal N$ denote the set of these four points. 

\begin{remark} Let us connect this parametrization of the spectral curve with the leading asymptotics of the colored Jones polynomial given in Section~\ref{sec:knot}, Equation~\eqref{eq:Murakami-leading}. 
	
	Over $x=2\pi\I+s$, $s\in I$, or, more generally, $x\in U$, we have $w=e^x$, $z=e^{\varphi(s)}$ (for $x\in U$ we analytically continue $\varphi(s)$ from $s\in I$), and $e^{S'_0(x)} = \frac{wz-1}{w-z}$. Thus $dS_0(x) = (\log Y +2\pi\I) d \log X$, where $\log$ denotes the principal branch of the logarithm. 
	Moreover,
	\begin{align}\label{eq:S1-on-curve}
		e^{S_1(x)} = \frac{\sqrt{2\pi}\,w}{(w-1) \sqrt[4]{Q(w)}},
		\qquad 
		S_1'(x)=-\frac1{w-1}-\frac{wQ'(w)}{4Q(w)}
		,
	\end{align}
	where the argument of $\sqrt[4]{\cdots}$ is close to $-\frac{\pi}{4}$ (more precisely, it is set to be equal to $-\frac{\pi}{4}$ for $s\in I$ and then analytically continued to $x\in U$). 

    For $x=2\pi\I+s$, $s\in I$, this expression for $e^{S_1(x)}$
is given in~\cite[Equation~(1.2)]{Murakami2013}. Its analytic extension for $x\in U$ amounts, in terms of Section~\ref{sec:analytic-comparison}, to applying the saddle point expansion~\eqref{eq:saddle-point-expansion}-\eqref{eq:analytic-hbar-expansion} to compute $D_0$, which gives Equation~\eqref{eq:S1-on-curve}.
\end{remark}

\subsubsection{Divisors and differentials}
The complement in $\Sigma$ of the affine curve in $(\mathbb C^*)^2$ given by~\eqref{eq:normalized-curve} consists
of four points $ \mathcal P=\{p_{00},p_{0\infty},p_{\infty0},p_{\infty\infty}\}$. In the $(w,z)$ coordinates, 
\begin{align}
	p_{00}=(0,0),\quad p_{0\infty}=(0,\infty),\quad
	p_{\infty0}=(\infty,0),\quad
	p_{\infty\infty}=(\infty,\infty).
\end{align}
The divisors of the coordinate functions and of $X$ and $Y$ are
\begin{align}\label{eq:divisors}
	(w)&=p_{00}+p_{0\infty}-p_{\infty0}-p_{\infty\infty},
	& (X) & =p_{00}+p_{0\infty}-p_{\infty0}-p_{\infty\infty}, 
	\\ \notag 
	(z)&=p_{00}-p_{0\infty}+p_{\infty0}-p_{\infty\infty},
	& (Y)&=-2p_{00}+2p_{0\infty}+2p_{\infty0}-2p_{\infty\infty}.
\end{align}

The differential
\begin{align}\label{eq:v8-holomorphic-differential}
	\eta_0=\frac{dw}{w(z-z^{-1})}
	=\frac{dz}{z(w-w^{-1})}
\end{align}
is holomorphic on $\Sigma$.
For local logarithmic lifts $x=\log X$ and $y=\log Y$,
direct differentiation gives
\begin{align}
	dx=\frac{dw}{w}=(z-z^{-1})\eta_0,\qquad
	dy=\frac{dY}{Y}=(2w-1+2w^{-1})\eta_0.
\end{align}
The zeros of $dx$ form the set (see Equation~\eqref{eq:zeros-dx-branching}) in $(w,z)$-coordinates
\begin{align}
	\mathcal R\coloneqq \{(a,1):a^2-3a+1=0\}
	\cup\{(a,-1):a^2+a+1=0\}.
\end{align}
They all are simple and $dy$ is holomorphic and nonzero
there. Both $dx$ and $dy$ have simple poles precisely at $\mathcal P$, cf.~Equation~\eqref{eq:divisors}.

Choose symplectic cycles $(\mathfrak A,\mathfrak B)$ and put
\begin{align}\label{eq:v8-periods}
	\eta=\frac{\eta_0}{\oint_{\mathfrak A}\eta_0},\qquad
	\mathcal T=\oint_{\mathfrak B}\eta.
\end{align}
These choices determine the normalized Bergman kernel $B$, the
Abel map $\mathcal A$, and the prime form $E$.

\subsubsection{Involutions}
The curve has two commuting involutions
\begin{align}\label{eq:v8-involutions}
	\sigma(w,z)=(w,z^{-1}),\qquad
	\iota(w,z)=(w^{-1},z).
\end{align}

The hyperelliptic involution $\sigma$ fixes
$\mathcal R$, $X\sigma = X$, $Y\sigma = Y^{-1}$, and it interchanges the points of $\mathcal N$ and of $\mathcal P$ with the same $w$ coordinate. The function $w$ identifies $\Sigma/\sigma$ with $\mathbb P^1$.
The reciprocal involution $\iota$ fixes $\mathcal N$, $X\iota = X^{-1}$, $Y\iota= Y^{-1}$, and it interchanges the points of  $\mathcal R$ and of $\mathcal P$ with the same $z$ coordinate. The function $z$ identifies $\Sigma/\iota$ with $\mathbb  P^1$.

Moreover,
$ \sigma^*\eta_0=\iota^*\eta_0=-\eta_0$, and $\sigma\iota$ is a translation on $\Sigma$  without fixed points, and we assume the choice of the symplectic cycles that makes this translation a shift by $\mathcal{T}/2$. 


\subsection{Borot-Eynard wave function} Apply topological recursion to the input data $(\Sigma, dx, dy, B, \mathcal{R})$. This produces a system of symmetric $n$-differentials $\omega^{(g)}_n$, $g\geq 0$, $n\geq 1$, with $\omega^{(0)}_1=ydx$ defined locally near $\mathcal{R}$, $\omega^{(0)}_2 = B$, and the remaining (stable) differentials produced by recursion and arranged into the generating series 
\begin{equation}\label{eq:stable}
	\omega_m=\sum_{\substack{g\geq0\\2g-2+m>0}}
	\hbar^{2g-2+m}\omega_m^{(g)}.
\end{equation}
Fix a theta function $\Theta(s)$. The non-perturbative disconnected $n|n$ half-differentials are defined as 
\begin{align}\label{eq:np-kernel}
	& 
	\Omega_n^{\mathsf{np},\bullet}(z^+_1,\dots,z^+_n,
	z^-_1,\dots,z^-_n \mid s)
	=
	\\ \notag & 
	\prod_{i=1}^n\frac{1}{E(z_i^+,z_i^-)}
	\prod_{i<j}\frac{E(z_i^+,z_j^+)E(z_i^-,z_j^-)}
	{E(z_i^+,z_j^-)E(z_i^-,z_j^+)} \exp\left(\frac 1\hbar \sum_{i=1}^n\int_{z_i^-}^{z_i^+} \omega_1^{(0)}\right)
	\\ \notag & 
	\frac{\exp \left(\sum_{m=1}^\infty \frac{1}{m!} \Big( \sum_{i=1}^n\int_{z_i^-}^{z_i^+}  +\frac{1}{2\pi\I}\oint_{\mathfrak B}\frac{\partial}{\partial s}\Big)^{\otimes m} \omega_m \right)
		\Theta(s+\mathcal A(\sum_{i=1}^n(z_i^+-z_i^-)))}
	{\exp \left(\sum_{m=1}^\infty \frac{1}{m!} \Big(\frac{1}{2\pi\I}\oint_{\mathfrak B}\frac{\partial}{\partial s}\Big)^{\otimes m} \omega_m \right) \Theta(s)}.
\end{align}

The specialization that we need to construct the Borot-Eynard wave function is
\begin{align}
	\Theta(s) 
	=\sum_{\ell \in\mathbb Z}
	\exp\left(
	\pi\I \mathcal T \ell^2
	+2\pi\I (s+\tfrac 12) \ell
	\right)
\end{align}
(that is, it is the function with the theta-characteristics $(0,\frac 12)$). Choose two pairs of points conjugated with respect to the involution $\iota$: $p, \iota(p), o, \iota(o) \in \Sigma$.  
Then the Borot-Eynard non-perturbative wave function is defined as
\begin{align} \label{eq:Psi-np-definition}
	\Psi^{\mathrm{np}} (p; o; \hbar) = \sqrt{\frac{ \Omega_2^{\mathsf{np},\bullet}(p, \iota(p), o, \iota(o) \mid 0) } {dx(p) dx(o)}
	}.
\end{align}

\begin{convention} \label{conv:Fixing-o}
It will be convenient to fix a particular $o$ such that $o\in\mathcal{R}$. To be concrete, let's assume that $o=((3+\sqrt5)/2,1)$ and $\iota(o)=((3-\sqrt5)/2,1)$ in $(w,z)$-coordinates. We will also always assume that paths from $o$ to $p$ and from $\iota(o)$ to $\iota(p)$ are connected by $\iota$. 

In this case, $dx$ has a simple zero at $o$ and at $\iota(o)$, hence some natural regularization will be required. The coefficients of strictly positive powers of $\hbar$ in
$d\log\Psi^{\mathrm{np}}$ are independent of the choice of
$o\in\mathcal R$, as proved
in~\cite[Proposition~6.2.2]{Potter2023}; see also below.
\end{convention}

\begin{remark}
	Note that $\mathcal A(p+\iota(p)-o-\iota(o))=0\in \mathbb C/(\mathbb Z+\mathcal T\mathbb Z)$, so this shift in the definition of $\Omega_2^{\mathsf{np},\bullet}$ can be ignored (in fact, with the assumption on paths conjugated by $\iota$ it is equal to $0$ in $\mathbb{C}$). Also note that 
	$dx(p) dx(o) = \sqrt{dx(p)} \sqrt {dx(\iota(p))}\sqrt{dx(o)} \sqrt {dx(\iota(o))}$ (for a consistent with $\iota$ choice of half-differentials), so it is indeed the standard normalization factor to convert $2|2$ half-differential into a function. Finally, $\Theta(s)$ that we use can be expressed in terms of the theta function $\Theta_*(s)$ with theta-characteristics $(\frac 12,\frac 12)$ as 
	$\Theta(s)=\I e^{\pi\I\mathcal T/4+\pi\I s}
	\Theta_*(s+\frac{\mathcal T}{2})$ (we refer to $\Theta_*(s)$ since it can be used to define the prime form).
\end{remark}

\begin{remark}
    We want to emphasize that, if one chooses compatible branches of
$\log Y$ along a path and its image under $\iota$, all closed cycle
periods of $\log Y\,d\log X$ on $\Sigma$ vanish, that is, 
    \[
        \oint_\gamma \log Y\,d\log X=0,
        \qquad \gamma\in H_1(\Sigma,\mathbb Z).
    \]
    With this choice, $\iota$ preserves $\log Y\,d\log X$.
    Since $\iota$ acts as $-1$ on $H_1(\Sigma,\mathbb Z)$, each period
    equals its own negative (see for instance \cite[Property~3.5]{BE2015}), and thus vanishes.
    This vanishing is related to a condition formulated from a physical
perspective in \cite{Gukov2005} and later referred to as the
"quantizability condition"; see also \cite{Gukov:2011qp}. In non-perturbative topological recursion,
    this vanishing accounts for the choice $s=0$
    in \eqref{eq:Psi-np-definition}, which together with the choice of theta characteristic will later be crucial for the projection onto the quotient curve.
\end{remark}

\subsection{Equation for the wave function}\label{sec:eqwave} Let $p$ be the point corresponding to the chosen branch of $w\colon \Sigma\to \mathbb{C}\mathrm{P}^1$ over $e^x$ for $x\in U$:
$w(p)=e^x$, $z(p)=e^{\varphi(x-2\pi\I)}$ 
(as before, the last expression denotes analytic continuation from $x\in2\pi\I+I$). Notice that 
$\log\Psi^{\mathrm{np}} (p;  \hbar)$ expands in $\hbar$ as a function of $x$
\begin{equation}\label{eq:BE-log-expansion}
	\log\Psi^{\mathsf{np}}(x;\hbar)
	=\hbar^{-1}S_0^{\mathsf{np}}(x)+S_1^{\mathsf{np}}(x)
	+\sum_{r\geq1}\hbar^r S_{r+1}^{\mathsf{np}}(x).
\end{equation}
Here we assume that $o$ is fixed and suppressed from the notation. As a function of $x$, $\Psi^{\mathsf{np}}(x;\hbar)
\in e^{S^\mathsf{np}_0(x)/\hbar+S^\mathsf{np}_1(x)}
\bigl(1+\hbar\mathcal O(U)[[\hbar]]\bigr)$.

\begin{lemma} We have $dS_0^{\mathsf{np}}(x) = ydx$, 
	$dS_1^{\mathsf{np}}(x) = \frac12 d\log\frac{dz}{dx} - \frac {dz}{z-z(o)}$, and for $r\geq 2$ holds that $S_r^{\mathsf{np}}(x)$ is given by a meromorphic function on $\Sigma$ with poles only at $\mathcal{R}$. 
\end{lemma}

\begin{proof}
	Since $\iota^*(y\,dx)=y\,dx$ (we assume the compatible choices of logarithms, namely, $y=S_0'(x)$ and $y\circ\iota=-y$), the $\hbar^{-1}$ term in~\eqref{eq:BE-log-expansion} is
	\begin{align}
		S_0^{\mathsf{np}}(x)=\frac12
		\left(\int_o^p y\,dx+\int_{\iota(o)}^{\iota(p)}y\,dx\right)
		=\int_o^p y\,dx, \qquad d S_0^{\mathsf{np}}(x) = ydx
	\end{align}
	
	The $\hbar^0$ term comes from the factor of prime forms in~\eqref{eq:np-kernel}. More precisely, 	$\exp(2S_1^{\mathsf{np}}(x))$ is obtained through $w(p) = e^x$ from 
	\begin{align} \label{eq:E-factor}
		\frac{E(p,\iota(p))E(o,\iota(o))}
		{E(p,o)E(\iota(p),\iota(o))E(p,\iota(o))E(o,\iota(p))},
	\end{align}
	further divided by  $dx(p)dx(o)$.
	Viewed as a bidifferential on $\Sigma\times\Sigma$,
\eqref{eq:E-factor} has simple zeros at the $\iota$-fixed points
in each variable and double poles along $p=o$ and $p=\iota(o)$.
These are exactly the zeros and poles of the pullback of
$dz(p)dz(o)/(z(p)-z(o))^2$. Their ratio is therefore a nonzero
constant $\tilde c$. Dividing by $dx(p)dx(o)$ and taking the
square root, we obtain 
	\begin{align}
		\exp(S_1^{\mathsf{np}}(x)) = c(o) \frac{\sqrt{dz/dx}}{z-z(o)},
	\end{align}
	where $c(o)$ is a constant that depends on $o$. Hence $dS_1^{\mathsf{np}}(x) = \frac12 d\log\frac{dz}{dx} - \frac {dz}{z-z(o)}$.
	
	For $r\geq 2$, the expressions that we get are polynomials in $(\int_o^p + \int_{\iota(o)}^{\iota(p)})^{\otimes k} \int_{\mathfrak{B}}^{m-k} \omega^{(g)}_m$, $m\geq 1$, $2g-2+m>0$, $0\le k\le m$, with coefficients given by polynomials in the derivatives of $\log\Theta(s)$ at $s=0$. Recall that for $2g-2+m>0$, the differentials $\omega^{(g)}_m$ have, in each variable, poles only at $\mathcal{R}$, with zero residues and vanishing $\mathfrak{A}$ periods. Note that the operator $\int_o^p + \int_{\iota(o)}^{\iota(p)}$ can be rewritten as $\int_o^p (\mathrm{Id}_\Sigma^*+\iota^*)$, and $(\mathrm{Id}_\Sigma^*+\iota^*)^{\otimes k}\int_{\mathfrak{B}}^{m-k}\omega^{(g)}_m$ has all periods in all its $k$ variables vanishing. Moreover, it still has poles only in the points of $\mathcal{R}$ with zero residues. Hence, $(\mathrm{Id}_\Sigma^*+\iota^*)^{\otimes k}\int_{\mathfrak{B}}^{m-k}\omega^{(g)}_m$ is exact in each variable. Thus its integral $\int_o^p$, applied in every variable, returns a globally defined meromorphic function.  
\end{proof}

\begin{remark}
	The disk terms $S_0^{\mathsf{np}}$ and $S_0$ differ by an additive
constant. On the chosen local branch
$x\in U$, $w=e^x$ and $z=e^{\varphi(s)}$, we have
	\begin{align}
		S_0^{\mathsf{np}}(x)-S_0(x) & = \text{constant};
		\\ \notag 
		S_1^{\mathsf{np}}(x)-S_1(x)
		& =
		\frac12\log\left(
		\frac{(w-1)^3(w+1)z}
		{w^2(z-z(o))^2}
		\right)+\text{constant},
	\end{align}
	where the constant depends on $o$. 
	Then, it is natural to define 
	\begin{align} \label{eq:tilde-Psi-definition}
		\widetilde{\Psi}^{\mathrm{np}}(x;\hbar)  \coloneqq e^{\hbar^{-1}S_0(x)-\hbar^{-1}S_0^{\mathsf{np}}(x) +S_1(x)-S_1^{\mathsf{np}}(x) }\Psi^{\mathsf{np}}(x;\hbar).
	\end{align}

This is the wave function $\widetilde{\Psi}^{\mathrm{np}}(x;\hbar)$, whose logarithmic derivative is identified with the logarithmic derivative of $\Psi^{\mathsf{rec}}$ in the second statement of Theorem~\ref{thm:main}. To this end, we identify $U$ with the chosen local branch of $\Sigma$ by
$w=e^x$ and $z=e^{\varphi(x-2\pi\I)}$, using analytic continuation from $x=2\pi\I+s$, $s\in I$ \emph{(and we continue to do so throughout the rest of the paper)}.
\end{remark}

An immediate corollary of the second equality in Theorem~\ref{thm:main} is the following proposition:

\begin{proposition} \label{prop:equation-for-tildepsi} We have
	\begin{align} \label{eq:operator-psi-nptilde}
		\left(1-b(e^x,\hbar)e^{-\hbar\partial_x}
		+a(e^x,\hbar)e^{-2\hbar\partial_x}\right)  \widetilde{\Psi}^{\mathrm{np}}(x;\hbar) =0,
	\end{align}
	where the coefficients $a(e^x,\hbar),b(e^x,\hbar)$ are obtained
	from~\eqref{eq:jones-recurrence-coefficients} by substituting
	$q=e^\hbar$ and $q^N=e^x$.
\end{proposition}

\begin{proof} 
Since $U$ is connected, every coefficient of
$\log\widetilde\Psi^{\mathsf{np}}-\log\Psi^{\mathsf{rec}}$
is constant in $x$. Their prescribed $S_0,S_1$ agree. Exponentiating therefore gives
\begin{align}
 \widetilde\Psi^{\mathsf{np}}(x;\hbar)
 =C(\hbar)\Psi^{\mathsf{rec}}(x;\hbar)
\end{align}
for some $C\in1+\hbar\mathbb C[[\hbar]]$. Equation~\eqref{eq:rec-reduced-formal} says that 
\begin{align}
\big(1-b(e^x,\hbar)e^{-\hbar\partial_x}
             +a(e^x,\hbar)e^{-2\hbar\partial_x}\big)\Psi^{\mathsf{rec}}=0.
\end{align}
The shift operators act on $x$ with $\hbar$ fixed, so they commute
with $C(\hbar)$, which implies that 
\begin{align}
    \big(1-b(e^x,\hbar)e^{-\hbar\partial_x}
             +a(e^x,\hbar)e^{-2\hbar\partial_x}\big)\widetilde{\Psi}^{\mathsf{np}}
             = C(\hbar)\big(1-b(e^x,\hbar)e^{-\hbar\partial_x}
             +a(e^x,\hbar)e^{-2\hbar\partial_x}\big) \Psi^{\mathsf{rec}}=0,
\end{align}
which proves~\eqref{eq:operator-psi-nptilde}.
\end{proof}

We also have the following corollary:
\begin{corollary} \label{cor:A-hat-Psi-np} We have $\widehat A \widetilde{\Psi}^{\mathrm{np}}(x;\hbar) =0$, for $\widehat A$ obtained from~\eqref{eq:widehat-A} by substituting
	$q=e^\hbar$ and $q^N=e^x$.	
\end{corollary}

\begin{proof}
    The factorization~\eqref{eq:widehat-A} writes $\widehat A$ as a left multiple of $\big(1-b(e^x,\hbar)e^{-\hbar\partial_x}
             +a(e^x,\hbar)e^{-2\hbar\partial_x}\big)$, hence also $\widehat A\widetilde\Psi^{\mathsf{np}}=0$.
\end{proof}


\section{Recurrence vs non-perturbative wave functions}
\label{sec:quantum-curve-proof}

We prove the second statement of Theorem~\ref{thm:main} by comparing the
logarithmic coefficients of $\widetilde\Psi^{\mathsf{np}}$ and
$\Psi^{\mathsf{rec}}$. The two expansions have the form
\begin{align}\label{eq:log-coefficients}
 \log\widetilde\Psi^{\mathsf{np}}
 =\frac{S_0}{\hbar}+S_1+
       \sum_{r\geq2}\hbar^{r-1}S_{r}^{\mathsf{np}},\qquad
 \log\Psi^{\mathsf{rec}}
 =\frac{S_0}{\hbar}+S_1+
       \sum_{r\geq2}\hbar^{r-1}S_{r}^{\mathsf{rec}}.
\end{align}
Here is a guide to the proof given below. 
Following the idea of~\cite{DijkgraafFujiManabe} (see also~\cite{BE2015}), we compare $dS_{r}^{\mathsf{np}}$ and $dS_{r}^{\mathsf{rec}}$ on the quotient $\mathsf C=\Sigma/\iota\simeq\mathbb P^1$.
Both families descend to meromorphic differentials
$\alpha_r^{\mathsf{np}}$ and $\alpha_r^{\mathsf{rec}}$ on
$\mathsf C$, with poles only at
$\mathcal R_{\mathsf C}=\mathcal R/\iota=\{1,-1\}$. In Section~\ref{subsec:RecurrenceGlobal}
we prove this on the recurrence side.
On the non-perturbative side, in Section~\ref{subsec:rational-reduction} we identify
the
projection of the connected non-perturbative differentials with topological
recursion on $\mathsf C$. The particular choice of the theta characteristics~\cite{BE2015}
gives the vanishing needed for the projected differentials
to satisfy the ordinary loop equations; uniqueness then gives
the identification.

In Section~\ref{subsec:alpha-equations}, we associate local
wave functions to both cases, with the same disk and
amplitude terms fixed by $S_0$ and $S_1$. For each case,
the two local branches of the local wave function satisfy a second-order differential
equation, a priori different, with regular coefficients at
the ramification points. In section~\ref{subsec:comparison} we prove the equality of the
$\alpha_r^{\mathsf{np}}$ and $\alpha_r^{\mathsf{rec}}$ by induction, using the regularity of their respective equations to exclude the polar parts. Then the pullback to $\Sigma$ gives equality of $dS_{r}^{\mathsf{np}}$ and $dS_{r}^{\mathsf{rec}}$, hence $d\log\widetilde\Psi^{\mathsf{np}}=d\log\Psi^{\mathsf{rec}}$
on $U$. The recurrence for $\Psi^{\mathsf{rec}}$
then gives Proposition~\ref{prop:equation-for-tildepsi}.

Throughout this section, the stable integrals defining
$\widetilde\Psi^{\mathsf{np}}$ use the centered prescription of
Section~\ref{subsubsec:endpoint-normalization}, at the points
$o=((3+\sqrt5)/2,1)$ and $\iota(o)=((3-\sqrt5)/2,1)$ fixed above.
This specifies the regularization used in
Proposition~\ref{prop:equation-for-tildepsi}. 

Various arguments below recycle a variety of ideas already present in the literature in local analysis near the simple ramification points and more generally, order two covering, in the translation between the properties of (local) wave function and differentials of (non-perturbative) topological recursion, such as~\cite{BBE,DijkgraafFujiManabe,BE-integrable,BE2015,MarchalOrantin}, combined with the KP  integrability methods developed in~\cite{ABDKS-KPxy,ABDKS2026,ABDKS-NP,ABDKS-Blobbed}. A related passage from equivariant topological recursion to ordinary
topological recursion on a quotient curve is proved
in~\cite{AlameddineHock2026}, together with
its effect on wave functions and quantum curves.

\subsection{Recurrence differentials: global analytic behavior} \label{subsec:RecurrenceGlobal}

Recall that we identify $U$ with the chosen local branch of $\Sigma$ by
$w=e^x$ and $z=e^{\varphi(x-2\pi\I)}$. For $r\geq2$, the differentials $dS_{r}^{\mathsf{rec}}$ are
initially defined on this chosen branch over $U$.

\begin{lemma}\label{lem:rec-meromorphy}
For $r\geq2$, the differentials $dS_{r}^{\mathsf{rec}}$ extend to
meromorphic differentials on $\Sigma$ with poles only at
$\mathcal R$. Moreover, 
\begin{align}\label{eq:rec-differential-invariance}
 \iota^*dS_{r}^{\mathsf{rec}}=dS_{r}^{\mathsf{rec}}.
\end{align}
\end{lemma}

We first reconstruct the differentials from the recurrence and prove
their $\iota$-invariance. The Gaussian expansion then shows that their
only possible poles are in $\mathcal R\cup\mathcal P$.
Finally, the recurrence excludes the poles in $\mathcal P$.

\subsubsection{Meromorphy}

The coefficients of~\eqref{eq:rec-reduced-formal}, with $w=e^x$,
are explicitly
\begin{align}
 a(w,\hbar)&=
 \frac{(1-we^{-2\hbar})(1-w^2e^{-\hbar})}
 {(1-w)(1-w^2e^{-3\hbar})},
 \\ \notag
 b(w,\hbar)&=
 \frac{e^{2\hbar}(1-we^{-\hbar})^2(1+we^{-\hbar}) \bigl(1-we^{-\hbar}-w^2e^{-3\hbar}-w^2e^{-\hbar}
 	-w^3e^{-3\hbar}+w^4e^{-4\hbar}\bigr)}
 {w^2(1-w)(1-w^2e^{-3\hbar})}
.
\end{align}
Recall that $S_0^{\mathsf{rec}}=S_0$ and $S_1^{\mathsf{rec}}=S_1$.
Divide~\eqref{eq:rec-reduced-formal} by
$\Psi^{\mathsf{rec}}(x;\hbar)$ and expand in $\hbar$.
For $j=1,2$, the contribution of the unknown coefficient is
\begin{align}
	\hbar^{r-1}\bigl(S_r^{\mathsf{rec}}(x-j\hbar)
	-S_r^{\mathsf{rec}}(x)\bigr)
	=-j\hbar^r\partial_xS_r^{\mathsf{rec}}(x)
	+O(\hbar^{r+1}).
\end{align}
The leading shifted ratios are $Y^{-j}$, while
$a(w,0)=1$ and $b(w,0)=Y+Y^{-1}$.
Thus the coefficient multiplying
$\partial_xS_r^{\mathsf{rec}}$ at order $\hbar^r$ is
\begin{align}
	b(w,0)Y^{-1}-2a(w,0)Y^{-2}
	=1-Y^{-2}=(Y-Y^{-1})Y^{-1}.
\end{align}
Moving the terms involving earlier coefficients to the
right-hand side gives, for $r\geq2$,
\begin{align}\label{eq:rec-coefficient-induction}
 (Y-Y^{-1}) Y^{-1}\partial_xS_r^{\mathsf{rec}}(x)
  =-[\hbar^r]\left(1-b(w,\hbar)
       \exp\left(\sum_{k=0}^{r-1}\hbar^{k-1}S_k^{\mathsf{rec}}(x-\hbar)-\sum_{k=0}^{r-1}\hbar^{k-1}S_k^{\mathsf{rec}}(x)\right) \right. 
       \\ \notag  \left. +a(w,\hbar)
       \exp\left(\sum_{k=0}^{r-1}\hbar^{k-1}S_k^{\mathsf{rec}}(x-2\hbar)-\sum_{k=0}^{r-1}\hbar^{k-1}S_k^{\mathsf{rec}}(x)\right)\right).
\end{align}
This is the triangular hierarchy of differential equations of
\cite[Section~2.2.2]{DGLZ2009}
in our notation. Starting from the explicit $S_0,S_1$, it determines
each derivative using only $Y$, derivatives of the locally defined function $y=S_0'(x)$,
for which $e^y=Y$, and derivatives
of the preceding coefficients. These are meromorphic on $\Sigma$, hence $\partial_xS_r^{\mathsf{rec}}(x)$ extends to a globally defined meromorphic function on $\Sigma$ as well.

\subsubsection{\texorpdfstring{$\iota$}{iota}-Invariance}
Invariance with respect to $\iota$ stated in Equation~\eqref{eq:rec-differential-invariance} follows directly from \begin{align}
	a(w,\hbar)a(e^{2\hbar}w^{-1},\hbar)=1; \qquad b(w,\hbar)=a(w,\hbar)b(e^{2\hbar}w^{-1},\hbar).
\end{align} Indeed, if a wave function $f(x,\hbar)$ satisfies Equation~\eqref{eq:rec-reduced-formal},
then, after local continuation to the $\iota$-reflected branch,
$g(x,\hbar)=f(-x,\hbar)$ satisfies the same equation: evaluate the
equation for $f$ at $-x+2\hbar$ and multiply it by $a(w,\hbar)$
(here $-x$ is the logarithmic coordinate on the $\iota$-reflected branch).
The reflected leading logarithmic derivative $-y\circ\iota$ differs
from $y$ by a locally constant multiple of $2\pi\I$.
This multiplies the wave function by a factor
$e^{2\pi\I kx/\hbar}$, $k\in\mathbb Z$, which is unchanged
under $x\mapsto x-\hbar$. Hence the equations determining the differentials
$dS_r^{\mathsf{rec}}$ on the $\iota$-reflected branch
are the same as on the original branch.

Moreover, for the $\iota$-reflected wave function $g(x,\hbar)$ we have
\begin{align}
	\frac{g(x+\hbar,\hbar)}{g(x,\hbar)}
	=(Y\circ\iota)^{-1}+O(\hbar)
	=Y+O(\hbar).
\end{align}
Thus the $\iota$-reflected wave function selects the same root $Y$ of the
semi-classical limit of~\eqref{eq:rec-reduced-formal}.
The one-loop differential is also preserved:
$\iota^*dS_1=dS_1$, as follows from its explicit formula.
Thus the $\iota$-reflected expansion has the same initial data for
the recursion~\eqref{eq:rec-coefficient-induction}.
Its uniqueness gives
$\iota^*dS_r^{\mathsf{rec}}=dS_r^{\mathsf{rec}}$ for $r\geq2$,
first locally, and then it extends globally
as an identity of meromorphic differentials on $\Sigma$.

\subsubsection{Possible poles}

To locate the poles, we use the saddle expansion of
Section~\ref{sec:analytic-comparison}. By~\eqref{eq:analytic-formal-identification}, for $r\geq2$,
the coefficient of $\hbar^{r-1}$ in the logarithm of
the saddle expansion of the contour integral~\eqref{eq:analytic-poisson} has differential $dS_r^{\mathsf{rec}}$. Furthermore, 
the Gaussian calculation
shows that these coefficients
are finite polynomials in $(z-z^{-1})^{-1}$ and the
polylogarithms with nonpositive index at $wz$ and $w/z$, see Equation~\eqref{eq:analytic-gaussian-coefficients}.
On the affine part of the curve $\Sigma$ with $w,z\ne0$ we have
\begin{align}
 (1-wz)(1-w/z)=w,\qquad
 \frac1{1-wz}=\frac{1-w/z}{w},\qquad
 \frac1{1-w/z}=\frac{1-wz}{w}.
\end{align}
Thus, since $\operatorname{Li}_{-n}(T)$ is polynomial in $T$ and $(1-T)^{-1}$
for $n\geq0$, every logarithmic coefficient $S_r^{\mathsf{rec}}$, $r\geq 2$, belongs to the
subring
\begin{align}
 \mathbb C[w^{\pm1},z^{\pm1},(z-z^{-1})^{-1}]
 \subset\mathbb C(\Sigma).
\end{align}
Its only possible poles on the affine part of the curve $\Sigma$ are therefore at $z=\pm1$, namely at $\mathcal R$. The complement of the affine part of $\Sigma$ consists of points in $\mathcal P$, thus the full set of possible poles is $\mathcal R\cup\mathcal P$, which holds for $S_r^{\mathsf{rec}}$ as well as their differentials $dS_r^{\mathsf{rec}}$. 

\subsubsection{Elimination of poles at \texorpdfstring{$\mathcal{P}$}{P}}
It remains to check and exclude the possible poles at $\mathcal{P}$. At the two points over $w=0$, $p_{0\infty}=(0,\infty)$ and $p_{00}=(0,0)$,
respectively,
\begin{align}
 Y=w^{\pm 2}(1+O(w)),\qquad
 \partial_x^2S_0=\pm 2+O(w),\qquad
 \partial_xS_1=1+O(w).
\end{align}
Assume $\partial_xS_k^{\mathsf{rec}}=O(w)$ for $2\leq k<r$.
This assumption is empty for $r=2$, and
$\partial_x=w\partial_w$ preserves $O(w)$.
Consequently, coefficientwise in $\hbar$,  $a(w,\hbar)=1+O(w)$,
$b(w,\hbar)=w^{-2}e^{2\hbar}(1+O(w))$, and 
\begin{align}
 \exp\left(\sum_{k=0}^{r-1}\hbar^{k-1}S_k^{\mathsf{rec}}(x-\hbar)-\sum_{k=0}^{r-1}\hbar^{k-1}S_k^{\mathsf{rec}}(x)\right)
 &=w^{\mp 2}e^{(\pm1-1)\hbar}(1+O(w)),
 \\ \notag  \exp\left(\sum_{k=0}^{r-1}\hbar^{k-1}S_k^{\mathsf{rec}}(x-2\hbar)-\sum_{k=0}^{r-1}\hbar^{k-1}S_k^{\mathsf{rec}}(x)\right)
 &=w^{\mp4}e^{(\pm4 -2)\hbar}(1+O(w)).
\end{align}
For $p_{0\infty}$, the two
$w^{-4}e^{2\hbar}$ terms cancel, leaving $O(w^{-3})$,
while $1-Y^{-2}=-w^{-4}(1+O(w))$. For $p_{00}$, the parentheses in
\eqref{eq:rec-coefficient-induction} are $O(w)$ and
$1-Y^{-2}$ is a unit. 
In both cases, $\partial_xS_r^{\mathsf{rec}}=O(w)$.
Thus $dS_r^{\mathsf{rec}}=(\partial_xS_r^{\mathsf{rec}})\,dw/w$
is holomorphic over $w=0$. Invariance under $\iota$ gives
holomorphy over $w=\infty$ and completes the proof.

\subsection{Reduction to topological recursion on the quotient}
\label{subsec:rational-reduction}

Consider the quotient by $\iota$:
\begin{align}\label{eq:quotient}
	\pi\colon\Sigma\longrightarrow\mathsf C=\Sigma/\iota\simeq\mathbb P^1_z,\qquad
	\pi(w,z)=z
\end{align}
The involution $\sigma$ projects to $\sigma_{\mathsf C}\colon z\mapsto z^{-1}$, thus we have the following coverings
\begin{equation}
\Sigma
\xrightarrow[\deg 2]{\ \pi\ }
\mathsf C\simeq\mathbb P^1_z
\xrightarrow[\deg 2]{\ x_{\mathsf C}\ }
\mathbb P^1_{x_{\mathsf C}}.
\end{equation}

We define the input data of topological recursion on $\mathsf C$ that would give the same local behavior at the points $\mathcal{R}_\mathsf{C}\coloneqq \mathcal{R}/\iota=\{\pm 1\}$.
Let 
\begin{align}
dx_{\mathsf C}\coloneqq \frac 12 d (z+z^{-1}),
\qquad 
 B_{\mathsf C}(z_1,z_2)=\frac{dz_1dz_2}{(z_1-z_2)^2},
\end{align}
and let $y_\mathsf{C}$ be the unique $\sigma_{\mathsf{C}}$-odd germ near $\mathcal{R}_\mathsf{C}$ such that $\pi^* y_\mathsf{C} dx_\mathsf{C} = \frac 12 (ydx - \sigma^* (ydx))$ near $\mathcal R$.

The map $\pi$ is unramified at $\mathcal R$, and
$\pi^*dx_{\mathsf C}/dx=(w-w^{-1})/2$ is nonzero there.
Since $dy$ is nonzero there, the odd germ $y_{\mathsf C}$ has
a simple zero at each point of $\mathcal R_{\mathsf C}$.

\begin{figure}
\begin{tikzpicture}[
  x=1cm,y=1cm,
  every node/.style={font=\scriptsize},
  strand/.style={line width=0.75pt,black},
  guide/.style={densely dashed,line width=0.42pt,gray!70},
  np/.style={circle,draw=black,fill=white,minimum size=5pt,inner sep=0pt},
  rp/.style={circle,draw=black,fill=black,minimum size=5pt,inner sep=0pt},
  mapping/.style={-{Latex[length=2.5mm]},line width=0.65pt},
  labelR/.style={draw=black,circle,inner sep=1.7pt,font=\footnotesize},
  labelN/.style={draw=black,circle,inner sep=1.7pt,font=\footnotesize}
]
\def\a{0}
\def\b{3.6}
\def\c{7.2}
\def\d{10.8}
\def\yt{6.70}
\def\yb{4.65}
\def\zupper{2.15}
\def\zlower{0.25}
\def\zbranch{1.20}
\def\xbase{-2.60}

\node[anchor=east,font=\normalsize] at (-0.70,7.92) {$\Sigma$};

\draw[strand] (\a,\yt) .. controls (1.05,7.35) and (2.55,7.35) .. (\b,\yt)
 .. controls (4.60,7.31) and (6.10,7.30) .. (\c,\yt)
 .. controls (8.40,7.33) and (9.73,7.31) .. (\d,\yt);
\draw[strand] (\a,\yb) .. controls (1.05,4.06) and (2.55,4.06) .. (\b,\yb)
 .. controls (4.55,4.05) and (6.18,4.04) .. (\c,\yb)
 .. controls (8.35,4.03) and (9.75,4.02) .. (\d,\yb);
\draw[strand,gray!70] (\a,\yt) .. controls (1.08,6.48) and (2.40,4.90) .. (\b,\yb);
\draw[strand,gray!70] (\a,\yb) .. controls (1.22,4.90) and (2.43,6.50) .. (\b,\yt);
\draw[strand,gray!70] (\b,\yt) .. controls (4.90,6.51) and (6.02,4.87) .. (\c,\yb);
\draw[strand,gray!70] (\b,\yb) .. controls (4.80,4.95) and (6.12,6.47) .. (\c,\yt);
\draw[strand,gray!70] (\c,\yt) .. controls (8.40,6.49) and (9.73,4.91) .. (\d,\yb);
\draw[strand,gray!70] (\c,\yb) .. controls (8.42,4.88) and (9.72,6.51) .. (\d,\yt);

\foreach \xx in {0,3.6,7.2,10.8}{\draw[guide] (\xx,4.43)--(\xx,2.59);}

\node[np] at (\a,\yt) {};
\node[np] at (\a,\yb) {};
\node[np] at (\c,\yt) {};
\node[np] at (\c,\yb) {};
\node[rp] at (\b,\yt) {};
\node[rp] at (\b,\yb) {};
\node[rp] at (\d,\yt) {};
\node[rp] at (\d,\yb) {};

\node[labelN] at (\a,7.74) {$\mathcal{N}$};
\node[labelR] at (\b,7.74) {$\mathcal{R}$};
\node[labelN] at (\c,7.74) {$\mathcal{N}$};
\node[labelR] at (\d,7.74) {$\mathcal{R}$};

\node[anchor=east] at (-0.13,\yt) {$w=-1$};
\node[anchor=east] at (-0.13,\yb) {$w=-1$};
\node[anchor=south] at (\b+0.02,6.82) {$w=\frac{-1+i\sqrt3}{2}$};
\node[anchor=north] at (\b+0.06,4.51) {$w=\frac{-1-i\sqrt3}{2}$};
\node[anchor=south] at (\c,6.82) {$w=1$};
\node[anchor=north] at (\c,4.52) {$w=1$};
\node[anchor=south] at (\d+0.10,6.82) {$w=\frac{3+\sqrt5}{2}$};
\node[anchor=north] at (\d+0.10,4.52) {$w=\frac{3-\sqrt5}{2}$};

\draw[<->,densely dotted,gray!70,thin] (\b-0.42,6.40)--(\b-0.42,4.93);
\node[anchor=east] at (\b-0.45,5.62) {$\iota$};
\draw[<->,densely dotted,gray!70,thin] (\d-0.42,6.40)--(\d-0.42,4.93);
\node[anchor=east] at (\d-0.45,5.62) {$\iota$};

\draw[mapping] (12.45,5.46)--(12.45,2.40);
\node[anchor=west,align=left] at (12.56,3.55) {$\pi(w,z)=z$};

\node[font=\normalsize,anchor=east] at (-0.7,2.08) {$\mathsf C\simeq\mathbb P^1_z$};
\draw[strand] (\a,\zupper) .. controls (1.3,2.12) and (2.30,1.17) .. (\b,\zbranch)
 .. controls (4.90,1.23) and (5.96,2.15) .. (\c,\zupper)
 .. controls (8.55,2.13) and (9.64,1.23) .. (\d,\zbranch);
\draw[strand] (\a,\zlower) .. controls (1.30,0.26) and (2.34,1.23) .. (\b,\zbranch)
 .. controls (4.88,1.19) and (5.98,0.23) .. (\c,\zlower)
 .. controls (8.46,0.25) and (9.67,1.17) .. (\d,\zbranch);

\node[np] at (\a,\zupper) {};
\node[np] at (\a,\zlower) {};
\node[rp] at (\b,\zbranch) {};
\node[np] at (\c,\zupper) {};
\node[np] at (\c,\zlower) {};
\node[rp] at (\d,\zbranch) {};
\node[anchor=south] at (\a,2.31) {$z=\frac{-3+\sqrt5}{2}$};
\node[anchor=north] at (\a,0.07) {$z=\frac{-3-\sqrt5}{2}$};
\node[anchor=north] at (\b,1.02) {$z=-1$};
\node[anchor=south] at (\c,2.31) {$z=\frac{1+i\sqrt3}{2}$};
\node[anchor=north] at (\c,0.07) {$z=\frac{1-i\sqrt3}{2}$};
\node[anchor=north] at (\d,1.02) {$z=1$};

\foreach \xx in {0,3.6,7.2,10.8}{\draw[guide] (\xx,-0.68)--(\xx,-2.17);}
\draw[mapping] (12.45,0.63)--(12.45,-2.22);
\node[anchor=west] at (12.55,-0.69) {$x_{\mathsf C}(z)=\frac12(z+z^{-1})$};

\node[font=\normalsize,anchor=east] at (-0.7,-2.52) {$\mathbb P^1_{x_{\mathsf C}}$};
\draw[-{Latex[length=2.3mm]},line width=0.7pt] (-0.40,\xbase)--(11.55,\xbase);
\foreach \xx in {0,3.6,7.2,10.8}{\draw[thin] (\xx,-2.45)--(\xx,-2.75);}
\node[np] at (\a,\xbase) {};
\node[rp] at (\b,\xbase) {};
\node[np] at (\c,\xbase) {};
\node[rp] at (\d,\xbase) {};
\node[anchor=north] at (\a,-2.84) {$-\frac32$};
\node[anchor=north] at (\b,-2.84) {$-1$};
\node[anchor=north] at (\c,-2.84) {$\frac12$};
\node[anchor=north] at (\d,-2.84) {$1$};
\node[anchor=west] at (11.61,\xbase) {$x_{\mathsf C}$};
\node[anchor=south,font=\tiny] at (\a,-2.38) {node};
\node[anchor=south,font=\tiny] at (\b,-2.38) {ram.};
\node[anchor=south,font=\tiny] at (\c,-2.38) {node};
\node[anchor=south,font=\tiny] at (\d,-2.38) {ram.};
\end{tikzpicture}
\caption{The two successive double coverings
$\Sigma\to \mathsf C\to\mathbb P^1_{x_\mathsf C}$.
The four ramification points $\mathcal{R}$ of the original
projection to the $w$-plane are identified in pairs
under the involution $\iota$, giving
$\mathcal{R}_\mathsf C=\{-1,1\}$ on ${\mathsf C}$.
In contrast, the four preimages of the nodes are
fixed by $\iota$ and remain distinct on $\mathsf C$,
but are mapped in pairs to the same $x_\mathsf C$-values.
The points $z=\pm1$ are the ramification points
of the projection $x_\mathsf C$.}
\label{fig:spectral-projections}
\end{figure}
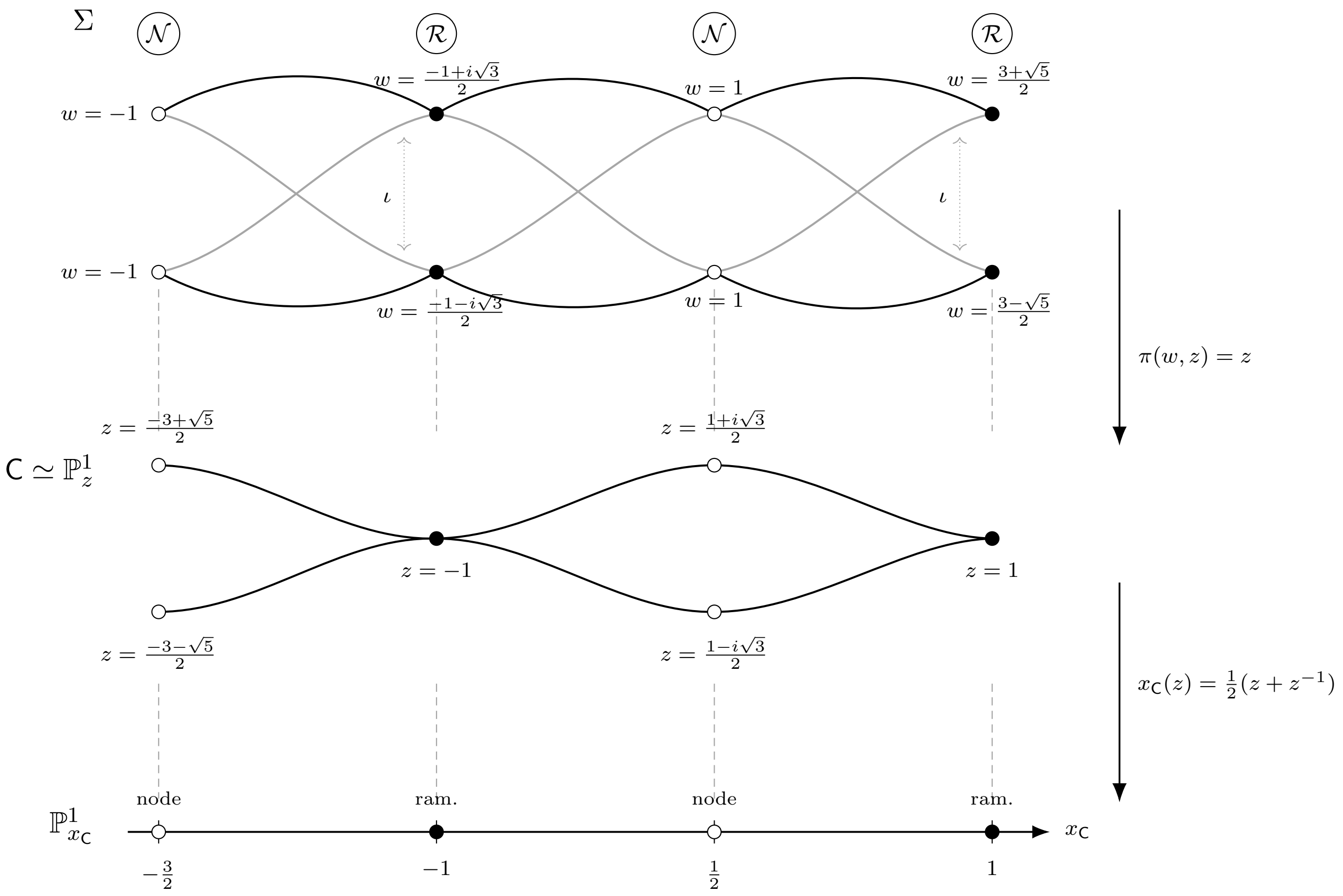

These input data allows us to apply topological recursion and obtain a system of differentials $\omega^{\mathsf C, (g)}_n$. Note that for $2g-2+n>0$, the resulting differentials $\omega^{\mathsf C, (g)}_n$ are odd with respect to $\sigma_\mathsf{C}$ in each variable. As before, we arrange them into the generating series in $\hbar$,
\begin{equation}\label{eq:stable-C}
	\omega^{\mathsf C}_n=\sum_{\substack{g\geq0\\2g-2+n>0}}
	\hbar^{2g-2+n}\omega_n^{\mathsf C,(g)}, \qquad n\geq 1
\end{equation}

\begin{lemma}\label{lem:descent}
We have:
\begin{align}  \label{eq:stable-descent}
 &d\log\widetilde\Psi^{\mathsf{np}}(x;\hbar)
 -\hbar^{-1}dS_0(x)-dS_1(x)
 =
 d\left(
 \sum_{n\geq1}
 \frac{1}{n!}
 \left(\int^{\pi(x)}\right)^{\otimes n}
 \omega_n^{\mathsf C}
 \right).
\end{align}
Here $\int^{\pi(x)}$ evaluates the unique
$\sigma_{\mathsf C}$-odd meromorphic primitive in the
corresponding variable at $\pi(x)$. All $n$ slots are integrated,
and $d$ is taken with respect to $x$, with $\hbar$ fixed.
\end{lemma}

We first discuss the connected non-perturbative differentials.
Our choice of theta characteristic gives an additional zero of the
kernel, which removes an extra term from their projected loop equations.
The projected differentials then satisfy the ordinary loop equations
on $\mathsf C$, with initial bidifferential $B_{\mathsf C}$.
Uniqueness identifies them with topological recursion on $\mathsf C$
and eliminates the theta functions from this projected system.
With this input, an alternative standard formula for the $2|2$ kernel in terms of the integrals of non-perturbative differentials reduces to the right hand side of~\eqref{eq:stable-descent}.

\subsubsection{Connected non-perturbative differentials and their symmetries}

Let $\omega^{\mathrm{np}}_n$ denote the connected non-perturbative differentials
associated with~\eqref{eq:np-kernel}, with the disk term omitted
and the Bergman term retained for $n=2$; see
\cite[Lemma~3.2]{ABDKS-NP}. Their positive coefficients in $\hbar$
are meromorphic, with poles only at $\mathcal R$ in each variable.
Their constant  terms in $\hbar$ are the so-called Krichever differentials, which differ from $\delta_{n,2}B$ by a holomorphic differential; see \cite[Lemma~3.4]{ABDKS-NP}.

Consider $\sigma\iota$, which acts as the translation by $\mathcal{T}/2$. The stable differentials $\omega_n^{(g)}$
are odd with respect to $\sigma$  in each variable and invariant under simultaneous
application of $\iota$ to all variables. 
Consequently,
\begin{align}
 (\sigma\iota)^{*\otimes n}\omega_n^{(g)}=(-1)^n\omega_n^{(g)}.
\end{align}
Since $\Theta$ is an even function, the definition~\eqref{eq:np-kernel} implies
simultaneous $\iota$-invariance of $\omega^{\mathrm{np}}_n$.

\subsubsection{A zero of the kernel}

We prove the following vanishing, which will remove the additional
connected term from the projected quadratic loop equation:
for every $n\geq0$,
\begin{align}  \label{eq:extra-handle-zero}
\left(\prod_{i=1}^n(\mathrm{Id}_\Sigma^*+\iota^*)\big|_{q_i}\right)
 \omega^{\mathrm{np}}_{n+2}(p,p',q_1,\ldots,q_n)
 \big|_{p'=\sigma\iota(p)}=0.
\end{align}
For $n=0$, it simply states that
$\omega^{\mathrm{np}}_2(p,\sigma\iota(p))=0$.

We first prove a kernel identity from which this follows. Let $K_D(p,p')$ be the modified kernel associated to a divisor $D=\sum_{i=1}^{k}(v_i^+-v_i^-)$, as in \cite[Section~2.8]{ABDKS-Blobbed}:
\begin{align} \label{eq:def-modified-K}
 K_D(p,p')\coloneqq 
 \frac{\Omega_{k+1}^{\mathsf{np},\bullet}
 (p,v_1^+,\ldots,v_k^+;p',v_1^-,\ldots,v_k^-\mid0)}
 {\Omega_k^{\mathsf{np},\bullet}
 (v_1^+,\ldots,v_k^+;v_1^-,\ldots,v_k^-\mid0)}.
\end{align}
We are interested in the case of even $k$ such that $\iota(v^\pm_{2i}) = v^{\pm}_{2i-1}$, $i=1,\dots,\frac k2$, and the paths connecting the $\iota$-conjugated pairs in $\sum_{i=1}^k\int_{v_i^-}^{v_i^+}$ are $\iota$-conjugated  as well. We claim that, for every such $D$,
\begin{align} \label{eq:K-D-vanishing}
 K_D(p,p')|_{p'=\sigma\iota(p)}=0. 
\end{align}
\begin{remark}
    We use the same conventions as in \cite[Section~2.8]{ABDKS-Blobbed}, that is, we assume that the endpoints are initially generic, so the denominator in the definition of $K_D$~\eqref{eq:def-modified-K} has nonzero leading coefficient after removing its disk exponential. In the case $k=0$ the denominator in~\eqref{eq:def-modified-K} is set to $1$.
\end{remark}

Choose a path $\gamma$ from $\sigma\iota(p)$ to $p$
so that
$\gamma+\sigma\iota(\gamma)$ represents $\mathfrak B$.
Consequently, on meromorphic differentials with vanishing residues which are odd under $\sigma$,
\begin{align}
 \left(\int_\gamma-\frac12\oint_{\mathfrak B}\right)
 (\sigma\iota)^*
 =-\left(\int_\gamma-\frac12\oint_{\mathfrak B}\right),
 \qquad \left(\sum_{i=1}^k\int_{v_i^-}^{v_i^+}\right)(\sigma\iota)^* = -\left(\sum_{i=1}^k\int_{v_i^-}^{v_i^+}\right).
\end{align}
Indeed, $\sigma\iota$ replaces the first half of the $\mathfrak B$-cycle by
the second half and preserves the full period. 
The operator $\sum_{i=1}^k\int_{v_i^-}^{v_i^+}$ also changes sign on stable differentials: $\iota$ exchanges the two intervals in each pair, while $\sigma$ changes the sign of the differential. 

Now recall~\eqref{eq:np-kernel} and apply it to $\Omega_{k+1}^{\mathsf{np},\bullet}
 (p,v_1^+,\ldots,v_k^+;p',v_1^-,\ldots,v_k^-\mid0)|_{p'=\sigma\iota(p)}$. In the numerator, the theta factor is
$\Theta(s+\mathcal T/2)$, evaluated after differentiation at $s=0$.
The integration operator acting on each slot can be written as
\begin{align}
 \int_\gamma+\sum_{i=1}^k\int_{v_i^-}^{v_i^+}+\frac{1}{2\pi\I}\oint_{\mathfrak B}\partial_s
 =\underbrace{\left(\int_\gamma-\frac12\oint_{\mathfrak B}+\sum_{i=1}^k\int_{v_i^-}^{v_i^+}\right)}_{I_1\coloneqq} + \underbrace{\left(
 \frac{1}{2\pi\I}\oint_{\mathfrak B}(\partial_s+\pi\I) \right)}_{I_2\coloneqq }.
\end{align}
Consider a term with $j$ period integrals $I_2$ and $m-j$ integration operators $I_1$ applied to a stable
$m$-differential. The operator $I_2$ is preserved by $\sigma\iota$. Applying $\sigma\iota$ in all $m$ variables
multiplies the differential by $(-1)^m$ and the product of integration
operators $I_1$ by $(-1)^{m-j}$. Thus this term equals $(-1)^j$
times itself and vanishes for odd $j$. Only even powers of $\partial_s+\pi\I$ remain. On the other hand, 
$e^{\pi\I s}\Theta(\mathcal T/2+s)$ is odd in $s$. Therefore,
\begin{align}
 \left.(\partial_s+\pi\I)^{2j}
 \Theta(\mathcal T/2+s)\right|_{s=0}
 =\left.\partial_s^{2j}
 \bigl(e^{\pi\I s}\Theta(\mathcal T/2+s)\bigr)\right|_{s=0}
 =0,\qquad j\geq0.
\end{align}
Thus every remaining term in the numerator vanishes. This proves
\eqref{eq:K-D-vanishing}.

The KP integrability of the non-perturbative system \cite[Theorem~3.14 and Corollary~3.15]{ABDKS-NP}, the modified two-point identity \cite[Section~2.8]{ABDKS-Blobbed}, and the corresponding connected expansion \cite[Lemma~3.2]{ABDKS-NP} all together give
\begin{align} \label{eq:omega-additional-n-pts}
 \sum_{m\geq0}\frac{1}{m!}\left(\sum_{i=1}^k\int_{v_i^-}^{v_i^+}\right)^{\otimes m}
 \omega^{\mathrm{np}}_{m+2}(p,p',\cdot)
 =-K_D(p,p')K_D(p',p).
\end{align}
The disk factors cancel in the product on the right. The sum
is understood as a formal expansion in the lengths of the
additional intervals. Setting $p'=\sigma\iota(p)$
makes its right-hand side zero by~\eqref{eq:K-D-vanishing}.

We now recover~\eqref{eq:extra-handle-zero}  by differentiating~\eqref{eq:omega-additional-n-pts} with respect to the additional endpoints assuming $p'=\sigma\iota(p)$. To this end, take $n$ independently varying small intervals near
$q_1,\dots,q_n$, e.g. $v_{2i-1}^- = q_i$, $v_{2i-1}^+ = q_i+\epsilon_i$ (the addition here is understood in some local coordinate), and their $\iota$-images.
In this case, the coefficient linear in $\epsilon_i$  in~\eqref{eq:omega-additional-n-pts}
applies $(\mathrm{Id}_\Sigma^*+\iota^*)_{q_i}$ in the
corresponding slot of $\omega^{\mathrm{np}}_{m+2}$. Consider the coefficient of $\epsilon_1\cdots\epsilon_n$. Then
only $m=n$ contributes. Symmetry of
$\omega^{\mathrm{np}}_{n+2}$ gives $n!$ equal terms, cancelling
the factor $1/n!$. This coefficient is precisely the left-hand
side of~\eqref{eq:extra-handle-zero}, so the required
vanishing follows.

\subsubsection{Projection to the quotient}

For $n\geq1$, define the quotient differentials $\omega_n^{\mathsf C,\mathrm{np}}$ by
\begin{align} \label{eq:projector-identity}
 (\pi^{\times n})^*\omega_n^{\mathsf C,\mathrm{np}} (p_1,\dots,p_n)
&=\frac12\prod_{i=1}^n(\mathrm{Id}_\Sigma^*+\iota^*)_{p_i}
 \omega_n^{\mathrm{np}} (p_1,\dots,p_n)
=\prod_{i=2}^n(\mathrm{Id}_\Sigma^*+\iota^*)_{p_i}
 \omega_n^{\mathrm{np}} (p_1,\dots,p_n).
\end{align}
The first expression is invariant under $\iota$ in every
variable, so it defines a meromorphic differential on
$\mathsf C^n$. The second equality follows from simultaneous
$\iota$-invariance of $\omega_n^{\mathrm{np}}$.
By symmetry, any one variable, not necessarily the first one, can be left out of the product
in this second expression.

Our goal now is to collect and prove the properties of $\omega_n^{\mathsf C,\mathrm{np}}$ that would allow to identify them with  $\delta_{n,2}B_{\mathsf C}+\omega_n^{\mathsf C}$ at the next step of the argument. To this end, we need the following conventions.

\begin{convention} \label{convention:centered-disk}
Recall the local disk differential:
\begin{align}
 \pi^*(y_{\mathsf C}\,dx_{\mathsf C})
 =\frac12\bigl(y\,dx-\sigma^*(y\,dx)\bigr),
 \qquad
 \pi^*dx_{\mathsf C}=\frac{w-w^{-1}}2\,dx.
 \label{eq:rational-ordinate}
\end{align}
To write the loop equations uniformly, we include the disk
terms by setting, locally near the ramification points,
\begin{align} 
 \widehat\omega_n^{\mathrm{np}}
 =\omega_n^{\mathrm{np}}
 +\frac{\delta_{n,1}}{2\hbar}
 \bigl(y\,dx-\sigma^*(y\,dx)\bigr),
\qquad 
 \widehat\omega_n^{\mathsf C,\mathrm{np}}
 =\omega_n^{\mathsf C,\mathrm{np}}
 +\frac{\delta_{n,1}}{\hbar}y_{\mathsf C}\,dx_{\mathsf C}.
\end{align}
Equation~\eqref{eq:projector-identity} also holds for these
series, since the centered disk differential is
$\iota$-invariant. 
\end{convention}

\begin{convention}
Below, $z_{\llbracket n \rrbracket} \coloneqq \{z_i\}_{i\in \{1,\dots, n\}}$ denotes the tuple of remaining
arguments, the sum of $I\sqcup J$ ranges over ordered partitions $\{1,\dots, n\}=I\sqcup J$ including empty parts, $z_{I} \coloneqq \{z_i\}_{i\in I}$, $z_{J} \coloneqq \{z_i\}_{i\in J}$.
\end{convention}

We prove that the positive coefficients of
$\omega_n^{\mathsf C,\mathrm{np}}$ have poles only at
$\mathcal R_{\mathsf C}$, that
$[\hbar^0]\omega_n^{\mathsf C,\mathrm{np}}=\delta_{n,2}B_{\mathsf C}$,
and that the series with the disk included satisfy the ordinary
loop equations on $\mathsf C$. More explicitly, for every
$n\geq0$, the linear expression
\begin{align}
 \widehat\omega_{n+1}^{\mathsf C,\mathrm{np}}(z,z_{\llbracket n \rrbracket})
 +\widehat\omega_{n+1}^{\mathsf C,\mathrm{np}}
 (\sigma_{\mathsf C}(z),z_{\llbracket n \rrbracket})
\end{align}
is holomorphic at $z=\pm1$, and the quadratic expression
\begin{align}
 &\widehat\omega_{n+2}^{\mathsf C,\mathrm{np}}
 (z,\sigma_{\mathsf C}(z),z_{\llbracket n \rrbracket})
 +
 \sum_{I\sqcup J}
 \widehat\omega_{|I|+1}^{\mathsf C,\mathrm{np}}(z,z_I)
 \widehat\omega_{|J|+1}^{\mathsf C,\mathrm{np}}
 (\sigma_{\mathsf C}(z),z_J)
\end{align}
is holomorphic at $z=\pm1$ with a zero of order at least two.

The pole statement follows from that for
$\omega_n^{\mathrm{np}}$---it is easy to check via local computation that descent procedure~\eqref{eq:projector-identity} introduces no additional
poles at the fixed points of $\iota$. The constant coefficient for $n=2$ is determined by
\begin{align}
 B+(\mathrm{Id}_\Sigma\times\iota)^*B
 =(\pi\times\pi)^*B_{\mathsf C}.
 \label{eq:bergman-descent}
\end{align}
Indeed, the left-hand side descends in both variables and has
the same diagonal principal part as $B_{\mathsf C}$ on the
quotient. Their difference is holomorphic in each variable on
$\mathbb P^1$, hence zero. The remaining $\hbar^0$ terms
are holomorphic and are annihilated by
$\mathrm{Id}_\Sigma^*+\iota^*$.

For the loop equations, we use
\cite[Theorem~3.7]{ABDKS-NP}, whose proof applies unchanged
to our theta characteristic. The local replacement of $y$
by $(y-y\circ\sigma)/2$ is a constant shift, since
$Y\circ\sigma=Y^{-1}$. It preserves the loop equations:
the change in the quadratic expression is a multiple of $dx$
times the linear trace, together with a multiple of $dx^2$.
Both have a double zero, because a holomorphic
$\sigma$-invariant one-form vanishes at a fixed point of
$\sigma$. Thus the hatted differentials on $\Sigma$ satisfy
the loop equations with the centered disk shifts introduced in Convention~\ref{convention:centered-disk}.

In order to prove the linear loop equations on $\mathsf C$, apply $\prod_{i=1}^n(\mathrm{Id}_\Sigma^*+\iota^*)_{q_i}$, $z_i=z(q_i)$,
to the loop equations on $\Sigma$, leaving the distinguished variable $p$
fixed. By the second expression
in~\eqref{eq:projector-identity}, the projected linear loop 
equation on $\Sigma$ is precisely the pullback of the linear equation on
$\mathsf C$.

For the quadratic loop equations, first consider their connected
term. Expanding the two additional factors in the variables
$p,p'$ gives
\begin{align}
 &\left.
 \bigl((\pi^{\times(n+2)})^*
 \omega_{n+2}^{\mathsf C,\mathrm{np}}\bigr)(p,p',q_1,\dots,q_n)
 \right|_{p'=\sigma(p)}
 \label{eq:projected-connected-term}
 \\ \notag
 &=
 \prod_{i=1}^n(\mathrm{Id}_\Sigma^*+\iota^*)_{q_i}
 \Bigl(\omega_{n+2}^{\mathrm{np}}(p,\sigma(p),q_1,\dots,q_n)
 +\omega_{n+2}^{\mathrm{np}}(p,\sigma\iota(p),q_1,\dots,q_n)\Bigr).
\end{align}
To see the factor in this formula, the four terms obtained
from the $p,p'$ factors have coefficient $1/2$.
After applying the operators in $q_1,\dots,q_n$, simultaneous
$\iota$-invariance identifies the term with both $p,p'$
transformed with the term with neither transformed.
It also identifies the two terms with exactly one of them
transformed. The two equal pairs cancel the factor $1/2$.
The second summand on the right of
\eqref{eq:projected-connected-term} vanishes
by~\eqref{eq:extra-handle-zero}. Thus the projected
connected term is exactly the connected term required on
$\mathsf C$. 

For a product indexed by $I\sqcup J$, each remaining variable occurs in only one of the two factors. Consequently,
\begin{align}
 &\prod_{i=1}^n (\mathrm{Id}_\Sigma^*+\iota^*)_{q_i}
 \Bigl(\widehat\omega_{|I|+1}^{\mathrm{np}}(p,q_I)
 \widehat\omega_{|J|+1}^{\mathrm{np}}(\sigma(p),q_J)\Bigr)
 \\ \notag
 & =
 \left(\prod_{i\in I}(\mathrm{Id}_\Sigma^*+\iota^*)_{q_i}
 \widehat\omega_{|I|+1}^{\mathrm{np}}(p,q_I)\right)
 \left(\prod_{i\in J}(\mathrm{Id}_\Sigma^*+\iota^*)_{q_i}
 \widehat\omega_{|J|+1}^{\mathrm{np}}(\sigma(p),q_J)\right).
\end{align}
By~\eqref{eq:projector-identity}, each parenthesis is the
pullback of the corresponding quotient differential, with
its distinguished variable left out of the product.
There is therefore no additional factor of $1/2$.
This also covers the empty parts, hence all disk terms.

We have identified every term of the projected quadratic
equation with the pullback of its quotient counterpart.
Finally, $\pi$ is unramified at $\mathcal R$, so holomorphy
and the order of vanishing are preserved under this local
pullback. The quadratic expression on $\mathsf C$ therefore
has the required double zero, completing the proof of the
stated loop equations.

\subsubsection{Identification with topological recursion}

We prove that, for every $n\geq1$,
\begin{align}  \label{eq:projected-tr}
 \omega_n^{\mathsf C,\mathrm{np}}
 =\delta_{n,2}B_{\mathsf C}+\omega_n^{\mathsf C}
\end{align}
We use the standard characterization of topological recursion
by the loop equations and the projection property; see
\cite[Proposition~2.7]{BEO},
\cite[Theorem~2.1 and Section~2.4]{BS}, and, for instance, an exposition in~\cite[Section~1.1]{BDKS-TR}.

The preceding subsection gives the loop equations and global
meromorphy, with poles of the positive coefficients only at
$\mathcal R_{\mathsf C}$. Since $\mathsf C\simeq\mathbb P^1$,
this also gives the projection property. Indeed, the linear loop
equation implies zero residues at the ramification points, and the principal parts there determine each coefficient in every
variable.

The usual uniqueness argument applies directly in the degree
of $\hbar$, simultaneously for all $n$, so it does not require
a topological expansion in advance. Indeed, suppose two solutions with
the same disk and constant coefficients first differ at
degree $\hbar^d$, $d\geq1$. The linear loop equation makes the
$\sigma_{\mathsf C}$-invariant part of this difference
holomorphic. In the quadratic loop equation at degree $d-1$, all
terms involving lower coefficients cancel. What remains is
$-2$ times the product of $y_{\mathsf C}\,dx_{\mathsf C}$ with the
anti-invariant part of the difference. The disk differential
has a zero of order exactly two, and the quadratic equation
requires this product to be holomorphic with a double zero.
Thus the anti-invariant part is holomorphic as well.
The projection property implies that the difference of the two solutions is equal to zero.

The ordinary topological recursion series on the right-hand
side of~\eqref{eq:projected-tr} has the same disk,
constant coefficients, loop equations and pole restriction.
Uniqueness therefore proves~\eqref{eq:projected-tr},
including the a priori not evident topological expansion of the quotient
differentials~\eqref{eq:projector-identity}.

\subsubsection{Revisited formula for the wave function} Equation~\eqref{eq:projected-tr} allows us to revisit the equation for $d\log\widetilde\Psi^{\mathsf{np}}$. Initially it was defined by Equations~\eqref{eq:Psi-np-definition} and~\eqref{eq:tilde-Psi-definition}. An alternative formula follows from \cite[Lemma~3.2]{ABDKS-NP}: for every $r\geq1$,
\begin{align} \label{eq:Alternative-tilde-Psi}
 [\hbar^r]d\log\widetilde\Psi^{\mathsf{np}}
 =\frac12\,d\left(
 \sum_{n\geq1}\frac1{n!}
 \left(\int_o^p + \int_{\iota(o)}^{\iota(p)}\right)^{\otimes n}[\hbar^r]\omega_n^{\mathrm{np}}
 \right).
\end{align}
With the convention that the paths of integration are identified by $\iota$, we have 
\begin{align}
    \int_o^p + \int_{\iota(o)}^{\iota(p)} = \left(\int_o^p\right)(\mathrm{Id}_\Sigma^*+\iota^*).
\end{align}
Then, recall~\eqref{eq:projector-identity}, which gives
\begin{align}
 (\pi^{\times n})^*\omega_n^{\mathsf C,\mathrm{np}}
 =\frac12(\mathrm{Id}_\Sigma^*+\iota^*)^{\otimes n}
 \omega_n^{\mathrm{np}}.
\end{align}
Using this, and, for the next equality, Equation~\eqref{eq:projected-tr}, we obtain, for $r\geq 1$,
\begin{align} \label{eq:from-np-to-C}
    \frac12
 \left(\int_o^p + \int_{\iota(o)}^{\iota(p)}\right)^{\otimes n}[\hbar^r] \omega_n^{\mathrm{np}}
= \left( \int_o^p \right)^{\otimes n} (\pi^{\times n})^*[\hbar^r]\omega_n^{\mathsf C,\mathrm{np}}
= \left(\int_{\pi(o)}^{\pi(p)} \right)^{\otimes n}[\hbar^r]\omega_n^{\mathsf C}.
\end{align}
Note that the factor $\frac 12$ that came from the square root in Equation~\eqref{eq:Psi-np-definition}, is thus absorbed. Since in our notation we can identify $\int^{\pi(p)}$ and $\int^{\pi(x)}$, substituting~\eqref{eq:from-np-to-C} into~\eqref{eq:Alternative-tilde-Psi} indeed gives Equation~\eqref{eq:stable-descent}, 
provided the regularized definite integrals agree with the odd primitives in Lemma~\ref{lem:descent}. To conclude the proof of Lemma~\ref{lem:descent}, we discuss this remaining bit of convention in the next paragraph.

\subsubsection{Normalization and primitives}
\label{subsubsec:endpoint-normalization}


Recall Convention~\ref{conv:Fixing-o}. Both lower endpoints, $o$ and $\iota(o)$, project to $z=1$. On $\mathsf C$, use a local coordinate $t$ near $z=1$ for which $t\circ\sigma_{\mathsf C}=-t$: $t=(z-1)/(z+1)$. On $\Sigma$, use the lifts of this local coordinate with respect to the projection $\pi$. 

Then, for a residue-free one-form $\alpha$ with local primitive
$F(t)=\sum_{k\geq-M}c_k t^k$, set
\begin{align}
 \int_o^p\alpha\coloneqq F(p)-c_0,
\end{align}
where $F$ is continued along the chosen path, and similarly at $\iota(o)$. Apply this
prescription separately for every integration variable in Equation~\eqref{eq:np-kernel}. This regularization can be pulled through Equation~\eqref{eq:from-np-to-C}.

By~\eqref{eq:projected-tr}, each positive coefficient of
$\omega_n^{\mathsf C,\mathrm{np}}$ is residue-free and
$\sigma_{\mathsf C}$-odd in each variable. On
$\mathsf C\simeq\mathbb P^1$ it therefore has a unique odd
meromorphic primitive in each slot. The Laurent series at
$z=1$ is odd with respect to $\sigma_{\mathsf C}$, so its constant coefficient is zero. Thus the prescription above gives precisely $\int^{\pi(x)}$ as we defined it.




\subsection{Local differential equations with regular coefficients}
\label{subsec:alpha-equations}

We prove that, near each ramification point of $x_{\mathsf C}$,
the logarithmic differentials of both the recurrence and the non-perturbative wave functions
come from a second-order differential equation with regular
coefficients. The equations may be different a priori; their regularity
is the property used in Section~\ref{subsec:comparison}, which identifies them finally.

By Lemma~\ref{lem:rec-meromorphy}, there are unique meromorphic
one-forms $\alpha_{r}^{\mathsf{rec}}$ on $\mathsf C$, $r\geq2$,
such that
\begin{align}
 \pi^*\alpha_{r}^{\mathsf{rec}}=dS_{r}^{\mathsf{rec}}.
 \label{eq:rec-quotient-differentials}
\end{align}
On the non-perturbative side, using Lemma~\ref{lem:descent}, set for $r\geq2$
\begin{align}
 \alpha_{r}^{\mathsf{np}}(z)
 =d\left(
 \sum_{\substack{g\geq0,\ n\geq1\\2g-2+n=r-1}}
 \frac{1}{n!}
 \left(\int^z\right)^{\otimes n}\omega_n^{\mathsf C, (g)}
 \right).
 \label{eq:np-quotient-differentials}
\end{align}
The same lemma gives
\begin{align}
 \pi^*\alpha_{r}^{\mathsf{np}}=dS_{r}^{\mathsf{np}}.
 \label{eq:np-quotient-pullback}
\end{align}

Furthermore, in the coordinate $z$ the explicit formula for $S_1$ given in Equation~\eqref{eq:S1-on-curve} implies 
\begin{align} \label{eq:S-1-C-definition}
    dS_1 = \pi^* d \underbrace{\log \frac{\sqrt{dz/dx_{\mathsf C}}}
 {\sqrt{(z-e^{\pi\I/3})(z-e^{-\pi\I/3})}}}_{S_1^{\mathsf C}\coloneqq}
\end{align}
(this can be checked explicitly using $w+w^{-1}=z+z^{-1}+1$). Recall also that we defined locally $y_{\mathsf C}$ near the points of $\mathcal R_{\mathsf C}$. 
This allows to define near either point $\rho\in\mathcal R_{\mathsf C}$ local
formal wave functions $\Psi^{\mathsf C,\star}$,
$\star\in\{\mathsf{rec},\mathsf{np}\}$, by
\begin{align}
 d\log\Psi^{\mathsf C,\star}
 =\hbar^{-1}y_{\mathsf C}\,dx_{\mathsf C}
  +d S_1^{\mathsf C}
  +\sum_{r\geq2}\hbar^{r-1}\alpha_{r}^{\star}.
 \label{eq:quotient-waves}
\end{align}
Primitives and square roots are chosen on a simply connected
sector of a punctured neighborhood of $\rho$. Locally near any preimage of $\rho$ in $\mathcal{R}$ under $\pi \colon \Sigma\to\mathsf{C}$, we have
\begin{align} \label{eq:local-wave-pullback}
\pi^*d\log\Psi^{\mathsf C,\mathsf{rec}}
=d\log\Psi^{\mathsf{rec}}
-\frac{y+y\circ\sigma}{2\hbar}dx,
\qquad 
\pi^*d\log\Psi^{\mathsf C,\mathsf{np}}
=d\log\widetilde\Psi^{\mathsf{np}}
-\frac{y+y\circ\sigma}{2\hbar}dx.
\end{align}

Choose a new local coordinate $t=\sqrt{x_{\mathsf C}-x_{\mathsf C}(\rho)}$,
$\sigma_{\mathsf C}(t)=-t$. 
For either wave function, put
$u^\star_\pm=\hbar\partial_{x_{\mathsf{C}}} \log\Psi^{\mathsf{C},\star}(\pm t)$ and $\Delta^\star=u^\star_+-u^\star_-$. Note that 
$\Delta^\star=2y_{\mathsf C}(t)+O(\hbar)$ is invertible in
$\mathbb C((t))[[\hbar]]$. 
For each $\star\in\{\mathsf{rec},\mathsf{np}\}$, there is a unique possible monic
second-order equation satisfied by both branches $\Psi^{\mathsf C,\star}(\pm t)$; its
coefficients are
\begin{align}
 \mathfrak a^\star&=-(u^\star_++u^\star_-)-\hbar\partial_{x_{\mathsf{C}}}\log\Delta^\star,
 \label{eq:reconstructed-ode}
 \\ \notag
 \mathfrak b^\star&=u^\star_+u^\star_-+u^\star_+\hbar\partial_{x_{\mathsf{C}}}\log\Delta^\star-\hbar\partial_{x_{\mathsf{C}}}u^\star_+.
\end{align}
Indeed, requiring both branches $\Psi^{\mathsf C,\star}(\pm t)$ to satisfy $((\hbar\partial_{x_{\mathsf{C}}})^2+\mathfrak a^\star \hbar\partial_{x_{\mathsf{C}}}+\mathfrak b^\star)\Psi^{\mathsf C,\star}=0$, and dividing each equation by its corresponding branch, gives
\begin{align} \label{eq:linearsystem-a-b}
\hbar\partial_{x_{\mathsf{C}}}u^\star_\pm+(u^\star_\pm)^2+\mathfrak a^\star u^\star_\pm+\mathfrak b^\star=0,
\end{align}
since $u^\star_\pm=\hbar\partial_{x_{\mathsf{C}}}\log\Psi^{\mathsf C,\star}(\pm t)$ and hence $(\hbar\partial_{x_{\mathsf{C}}})^2\Psi^{\mathsf C,\star}(\pm t)/\Psi^{\mathsf C,\star}(\pm t)=\hbar\partial_{x_{\mathsf{C}}}u^\star_\pm+(u^\star_\pm)^2$. Then we solve~\eqref{eq:linearsystem-a-b} as a linear system for the unknowns $\mathfrak a^\star$ and $\mathfrak b^\star$. 

\begin{lemma}\label{lem:rec-equation}
For each $\star\in\{\mathsf{rec},\mathsf{np}\}$ and for each $\rho\in\mathcal R_{\mathsf C}$, the two local branches
of $\Psi^{\mathsf C,\star}$ satisfy
\begin{align} \label{eq:regular-equations-branches}
 ((\hbar\partial_{x_{\mathsf{C}}})^2+\mathfrak a^\star \hbar\partial_{x_{\mathsf{C}}}+\mathfrak b^\star)\Psi^{\mathsf C,\star}(\pm t)=0,
 \qquad \mathfrak a^\star,\mathfrak b^\star\in\mathbb C[[t^2,\hbar]],
\end{align}
with the coefficients given by~\eqref{eq:reconstructed-ode}.
\end{lemma}

This lemma is proved in the following subsections. 

\subsubsection{From the recurrence to a differential equation}\label{sec:recdif}

In this Section, we prove Lemma~\ref{lem:rec-equation} for $\star=\mathsf{rec}$. We fix a point $\rho\in\mathcal R_{\mathsf C}$, and without loss of generality let us assume that $z(\rho)=1$, thus $x_{\mathsf C}(\rho)=1$. Let $\tilde\rho\in\pi^{-1}(\rho)\subset\Sigma$ be one of the preimages of $\rho$ on $\Sigma$. We have to use Equation~\eqref{eq:rec-reduced-formal}, and thus, as the first step of the proof, we derive a local equation for the branches of $d\log \Psi^{\mathsf{rec}}$ at $\tilde\rho$. Then Equation~\eqref{eq:local-wave-pullback} allows us to descend the derived equation to the branches of $d\log \Psi^{\mathsf C,\mathsf{rec}}$ at $\rho$. 

In order to derive an equation for the branches of $\Psi^{\mathsf{rec}}$ at $\tilde\rho$, we work on a small disk in the $x$-plane centered at $x(\tilde \rho)$. There, we rewrite Equation~\eqref{eq:rec-reduced-formal} in matrix form as 
\begin{align} \label{eq:Shift-Matrix}
e^{\hbar \partial_x}\begin{pmatrix}\phi(x)\\\psi(x)\end{pmatrix}
= \underbrace{\begin{pmatrix}0&1\\-a(e^{x+\hbar},\hbar)&b(e^{x+\hbar},\hbar)\end{pmatrix}}_{M(x,\hbar)\coloneqq}
\begin{pmatrix}\phi(x)\\\psi(x)\end{pmatrix},
\end{align}
for unknown functions $\psi(x)$ and $\phi(x)$ (note that $\phi(x) = \psi(x-\hbar)$). Note that the coefficients of $M(x,\hbar)$ are holomorphic on the disk. The
eigenvalues of $M(x,0)$ are $Y$ and $Y^{-1}$; at the center of the disk they coincide and are equal to $1$. Our immediate goal is to represent the shift~\eqref{eq:Shift-Matrix} as a solution of an ODE. Indeed, there exists a unique matrix $C(x,\hbar) = \sum_{k=0}^\infty \hbar^k C_k(x)$, holomorphic on the disk, such that exponentiating the flow given by  {}the ODE
\begin{align} \label{eq:ODE-Matrix}
{\hbar \partial_x}\begin{pmatrix}\phi(x)\\\psi(x)\end{pmatrix}
= C(x,\hbar)
\begin{pmatrix}\phi(x)\\\psi(x)\end{pmatrix},
\end{align}
we obtain expression~\eqref{eq:Shift-Matrix} for the operator $\exp(\hbar \partial_x)$. The coefficients $C_k(x)$ can be recursively uniquely reconstructed by this property starting from a choice of the holomorphic matrix logarithm $C_0(x)=\log M(x,0)$; $C_0(x)$ we choose such that its eigenvalues are precisely the chosen local branches of $y$ and $y\circ\sigma$. 

Notice that at the center of the disk 
\begin{align}
M(x,0)|_{x=x(\tilde \rho)} = \begin{pmatrix}0  &1\\-1& 2\end{pmatrix}
\qquad \text{and thus} \qquad 
C_0(x)|_{x=x(\tilde \rho)} \in \begin{pmatrix} -1 & 1\\-1 &  1\end{pmatrix}+\mathrm{Id} \cdot 2\pi\I\mathbb{Z}
\end{align}
Thus, $(C_0(x))_{21}|_{x=x(\tilde \rho)}=-1$, and hence $C(x,\hbar)_{21}$ is invertible on a sufficiently small disk, as a formal power series in $\hbar$ with coefficients holomorphic in $x$. This allows, using~\eqref{eq:ODE-Matrix}, to express the function $\phi(x)$ as 
\begin{align}
    \phi(x) = \frac{1}{C(x,\hbar)_{21}} \big(\hbar \partial_x -C(x,\hbar)_{22}\big) \psi(x).
\end{align}
Thus, we can rewrite the first order ODE~\eqref{eq:ODE-Matrix} as the second order ODE for a single function $\psi(x)$, with the coefficients being formal in $\hbar$ and holomorphic in $x$. Explicitly, the equation reads
\begin{align} \label{eq:final-eq-branches-psi}
 & \bigg( (\hbar\partial_x)^2
 -\left(\operatorname{tr}C(x,\hbar)
       +\hbar\frac{\partial_xC(x,\hbar)_{21}}{C(x,\hbar)_{21}}\right)
       \hbar\partial_x
\\ \notag & 
+\det C(x,\hbar)-\hbar\partial_xC(x,\hbar)_{22}
       +\hbar C(x,\hbar)_{22}\frac{\partial_xC(x,\hbar)_{21}}{C(x,\hbar)_{21}}\bigg) \psi(x) = 0.
\end{align}
Away from $x(\tilde\rho)$, its two formal asymptotic solutions
have leading logarithmic derivatives
$\hbar\partial_x\log\psi=y+O(\hbar)$ and
$y\circ\sigma+O(\hbar)$, respectively, since these are the
eigenvalues of $C_0(x)$. 
By construction, these solutions satisfy the original recurrence given in Equation~\eqref{eq:rec-reduced-formal}. As explained after Equation~\eqref{eq:wkb-jones}, their logarithmic
derivatives are uniquely determined by their leading terms. Therefore, their logarithmic $\hbar \partial_x$-derivatives
coincide with the two local branches of
$\hbar \partial_x \log\Psi^{\mathsf{rec}}$.

Now, in order to descend the equation to the branches of $\Psi^{\mathsf C,\mathsf{rec}}$ at $\rho$, recall 
that by Equation~\eqref{eq:local-wave-pullback} the pullbacks of the two branches of
$\Psi^{\mathsf C,\mathsf{rec}}$ are obtained from the branches of $\Psi^{\mathsf{rec}}$ by multiplication by 
$\exp(-\frac {x(y + y\circ\sigma)}{2\hbar})$ (up to factors independent of $x$). Here $\frac{y+y\circ\sigma}2$ is constant.
Thus, in order to obtain an equation for the two branches of
$\Psi^{\mathsf C,\mathsf{rec}}$, we first have to replace in Equation~\eqref{eq:final-eq-branches-psi} $\hbar\partial_x$ by
$\hbar\partial_x+\frac{y+y\circ\sigma}2$ (which amounts to conjugating it by 
$\exp(\frac {x(y + y\circ\sigma)}{2\hbar})$) and then rewrite it in terms of $\hbar\partial_{x_{\mathsf C}} = \frac{dx}{dx_{\mathsf C}} \hbar \partial_x$ using that $\frac{dx_{\mathsf C}}{dx}\not=0$.
 This, after dividing by $\big(\frac{dx_{\mathsf C}}{dx}\big)^2$, gives the required equation for the branches of $\Psi^{\mathsf C,\mathsf{rec}}$ and its coefficients are given by 
\eqref{eq:reconstructed-ode} by uniqueness.

\subsubsection{Towards an equation for the non-perturbative wave function}\label{sec:difeqnp}

Now we prove Lemma~\ref{lem:rec-equation} for $\star=\mathsf{np}$. The plan of the proof is as follows. First of all, it will be convenient to replace in the definition~\eqref{eq:np-quotient-differentials} the odd primitive by an actual endpoint integral, as in~\cite{MarchalOrantin}. This allows to interpret the non-perturbative wave function $\Psi^{\mathsf C,\mathsf{np}}$ as a deformed Baker-Akhiezer kernel. 

This interpretation has an advantage, since it allows to use the methods of KP integrability~\cite{BE-integrable,ABDKS-KPxy,ABDKS2026,ABDKS-NP}, in particular, the determinantal identities for the deformed Baker-Akhiezer kernels~\cite[Section 2.8]{ABDKS-Blobbed}. This way we find, following the ideas initially suggested in~\cite[Section 8]{BE-integrable}, the counterterms that regularize, in a uniform way via the sums over local sheets, both $\Psi^{\mathsf C,\mathsf{np}}(\pm t)$ and their $\hbar \partial_{x_{\mathsf C}}$-derivatives. 

We further use the regularizing counterterms to derive a uniform second order equation for $\Psi^{\mathsf C,\mathsf{np}}(\pm t)$ in a matrix form, which proves Lemma~\ref{lem:rec-equation} for $\star = \mathsf{np}$.

\subsubsection{Kernel interpretation for the non-perturbative wave function}

Recall Equation~\eqref{eq:def-modified-K}. It gives a definition of the deformed Baker-Akhiezer kernel $K_D(p,p')$. As in~\cite[Section 2.8]{ABDKS-Blobbed}, by taking the limits of the points in the divisor $D$ where the points collide one can consider divisors of degree zero with integer coefficients. We need a variation of this concept for the divisors of the type $D=\sum_{i} c_i(\gamma^+_i-\gamma^-_i)$ for half-integer coefficients $c_i$ and initially distinct finite points $\gamma_i^\pm\in\mathsf C\setminus\mathcal R_{\mathsf C}$. Since $\mathsf C$ is of genus $0$, with a global rational coordinate $z$, we have:
\begin{align}  \label{eq:source-kernel}
 K_D(p,p')
 &=e^{\hbar^{-1}\int_{p'}^p y_{\mathsf C}\,dx_{\mathsf C}}
 \frac{\sqrt{dz(p)\,dz(p')}}{z(p)-z(p')}
 \prod_i \left(\frac{z(p)-z(\gamma^+_i)}{z(p)-z(\gamma^-_i)}\right)^{c_i}
 \prod_i \left(\frac{z(p')-z(\gamma^+_i)}{z(p')-z(\gamma^-_i)}\right)^{-c_i} \times
 \\ \notag 
 & \quad 
 \frac{
    \exp\left(
 \sum_{\substack{g\geq0,\ n\geq1\\2g-2+n>0}}
 \frac{\hbar^{2g-2+n}}{n!}
 \big(\int_{p'}^p + \sum_i c_i \int_{\gamma_i^-}^{\gamma_i^+}\big)^{\otimes n}\omega_n^{\mathsf C,(g)}\right)
 }{    
 \exp\left(
 \sum_{\substack{g\geq0,\ n\geq1\\2g-2+n>0}}
 \frac{\hbar^{2g-2+n}}{n!}
 \big(\sum_i c_i \int_{\gamma_i^-}^{\gamma_i^+}\big)^{\otimes n}\omega_n^{\mathsf C,(g)}\right)}
.
\end{align}
These deformed kernels are KP integrable because
$\mathsf C$ has genus zero; see~\cite{ABDKS2026,ABDKS-NP} for the integrability statement and involved determinant identities. Note that the determinantal identities, initially derived for integral weights $c_i$ (cf.~\cite[Section 2.8]{ABDKS-Blobbed}), also hold more generally, in particular, for half-integer weights $c_i$. Indeed, integer $c_i$'s are obtained in \emph{op.~cit.} by repeating the corresponding endpoint pairs and taking their coincident limits. Now take a closer look at the determinantal identities (see e.g.~\cite[Proposition 1.3]{ABDKS-NP}). Note that from the shape of the determinantal identities, one can cancel the disk factors (the first exponential in~\eqref{eq:source-kernel}), the half-differentials, and the $c_i$-dependent prime form factors in the first line of~\eqref{eq:source-kernel}, and then expand the remaining identity in a formal power series in $\hbar$. Each coefficient in $\hbar$ is a polynomial in $c_i$'s that vanishes on a Zariski dense subset of possible values of $c_i$'s. Hence, these polynomials vanish for arbitrary values of  $c_i$'s.

We are interested in $K_D(p,p')$ for a very particular divisor $D_{\mathsf{np}}\coloneqq \frac{1}{2} (\beta - \sigma_{\mathsf C}(\beta))$, with $\beta\in \mathsf C$ such that $z(\beta) = e^{\pi\I/3}$ and $z(\sigma_{\mathsf C}(\beta)) = e^{-\pi\I/3}$, and we also intend to take the limit $p'\to \beta$.
The reason for this choice is that the odd primitive $\int^{z}$, $z=z(p)$ for $p\in \mathsf C$, used in the definition of $\Psi^{\mathsf C,\mathsf{np}}$ \eqref{eq:np-quotient-differentials} can then be represented as (cf.~\cite[Section~4.2]{MarchalOrantin})
\begin{align}
    \int^z = \int^{p}_{\beta} + \frac 12 \int^\beta_{\sigma_{\mathsf C}(\beta)}.
\end{align}
The latter formula holds, of course, for any $\beta\in \mathsf C$, $\beta\notin\mathcal R_{\mathsf C}$, but this particular choice is aligned with Equation~\eqref{eq:S-1-C-definition} for $S_1^{\mathsf C}$, and as a result the following deformed kernel formula for the wave function $\Psi^{\mathsf C,\mathsf{np}}$ will automatically have the right amplitude factor. By explicit computation, for a suitable choice of the multiplicative constant for the integral of the local wave function, otherwise given only by its logarithmic differential
in~\eqref{eq:quotient-waves}, we have:
\begin{align} \label{eq:Psi-np-C-kernel}
\Psi^{\mathsf C,\mathsf{np}} & = e^{\hbar^{-1}\int^{z} y_{\mathsf C}\,dx_{\mathsf C}}
 e^{S_1^{\mathsf C}(z)}
 \frac{
    \exp\left(
 \sum_{\substack{g\geq0,\ n\geq1\\2g-2+n>0}}
 \frac{\hbar^{2g-2+n}}{n!}
 \big(\int^z\big)^{\otimes n}\omega_n^{\mathsf C,(g)}\right)
 }{    
 \exp\left(
 \sum_{\substack{g\geq0,\ n\geq1\\2g-2+n>0}}
 \frac{\hbar^{2g-2+n}}{n!}
 \big(\frac 12\int_{\sigma_{\mathsf C}(\beta)}^{\beta}\big)^{\otimes n}\omega_n^{\mathsf C,(g)}\right)}
 \\ \notag & = \lim_{p'\to\beta} K_{D_{\mathsf{np}}}(p,p') \cdot  \frac{e^{\hbar^{-1}\int^{p'} y_{\mathsf C}\,dx_{\mathsf C}}}{\sqrt{dx_{\mathsf C}(p)}\sqrt{dz(p')}} \left(\frac{z(p')-e^{\pi\I/3}}{z(p')-e^{-\pi\I/3}}\right)^{1/2},
\end{align}
with $z=z(p)$. Note that the expression to which we apply $\lim_{p'\to\beta}$ is indeed regular at $p'\to \beta$. 

There is a useful normalization of the kernel by its disk factors,
half-differentials, and divisor factors. This normalization is prompted
by equation~\eqref{eq:Psi-np-C-kernel}, where the second argument is held
fixed, and it turns out to be very convenient in the computations below. We underline an argument to indicate the
normalization, and we will call it a "fixed" variable. This normalization is defined by:
\begin{align}  \label{eq:kernel-fixed-moving}
 K_D(p,p')
 &=\sqrt{dx_{\mathsf C}(p)}\sqrt{dz(p')} e^{-\hbar^{-1}\int^{p'} y_{\mathsf C}\,dx_{\mathsf C}}
 \prod_i \left(\frac{z(p')-z(\gamma^+_i)}{z(p')-z(\gamma^-_i)}\right)^{-c_i} K_D(p,\underline{p'})
 \\ \notag 
 & = \sqrt{dx_{\mathsf C}(p')}\sqrt{dz(p)} e^{\hbar^{-1}\int^{p} y_{\mathsf C}\,dx_{\mathsf C}}
 \prod_i \left(\frac{z(p)-z(\gamma^+_i)}{z(p)-z(\gamma^-_i)}\right)^{c_i} K_D(\underline{p},{p'})
 \\ \notag
 & = \sqrt{dz(p)}\sqrt{dz(p')}e^{\hbar^{-1}\int_{p'}^p y_{\mathsf C}\,dx_{\mathsf C}} \prod_i \left(\frac{z(p)-z(\gamma^+_i)}{z(p)-z(\gamma^-_i)}\right)^{c_i} \prod_i \left(\frac{z(p')-z(\gamma^+_i)}{z(p')-z(\gamma^-_i)}\right)^{-c_i} K_D(\underline{p},\underline{p'}),
\end{align}
and thus $\Psi^{\mathsf C,\mathsf{np}}(p) = K_{D_{\mathsf{np}}}(p,\underline{\beta})$.

\subsubsection{Regularity of products of kernels}

Choose distinct regular finite endpoints $\xi,\zeta\in\mathsf C$, and an arbitrary degree zero divisor $D$ with support disjoint from $\mathcal R_{\mathsf C}$ and, initially, also disjoint from $\{\xi,\zeta\}$. Consider the following expression
\begin{align} \label{eq:reg-branch-sum}
K_D(e^{\frac{u\hbar}{2}\partial_{x_{\mathsf C}} }z,\underline{\xi})K_D(\underline{\zeta},e^{-\frac{u\hbar}{2}\partial_{x_{\mathsf C}}}z)+K_D(e^{\frac{u\hbar}{2}\partial_{x_{\mathsf C}} }\sigma_{\mathsf C}(z),\underline{\xi})K_D(\underline{\zeta},e^{-\frac{u\hbar}{2}\partial_{x_{\mathsf C}}}\sigma_{\mathsf C}(z)),
\end{align}
$z=z(p)$ and $\sigma_{\mathsf C}(z)= z^{-1}$ is the coordinate of the point $\sigma_{\mathsf C}(p)$ (recall that $\sigma_{\mathsf C}$ is the global deck transformation of $x_{\mathsf C}$ on $\mathsf C$). Let us prove that it is regular at $p\to \rho\in\mathcal R_{\mathsf C}$, or, in coordinate $z$, at $z\to \pm 1$, as a function of $z$, $\hbar$, and $u$. In other words, it belongs to $\mathbb C[[t^2,\hbar,u]]$, where $t$ is a local coordinate at $\rho$ such that $x_{\mathsf C}(z) = x_{\mathsf C}(\rho) + t^2 $.

In order to analyze this expression, we use the toolkit of KP integrability developed in \cite{ABDKS-KPxy,ABDKS2026,ABDKS-NP,ABDKS-Blobbed}. The two-by-two determinantal identity (whose leading term in $\hbar$ is a version of the Fay identity) gives, for any $p_1,p_2\in\mathsf C$,
\begin{align}\label{eq:source-kernel-identity}
 K_D(p_1,\underline\xi)K_D(\underline\zeta,p_2)
 &={K_D(\underline\zeta,\underline\xi)}\left( \frac{K_D(p_1,p_2)}{\sqrt{dx_{\mathsf C}(p_1)dx_{\mathsf C}(p_2)}} -\frac {K_{D+\zeta-\xi}(p_1,p_2)} {\sqrt{dx_{\mathsf C}(p_1)dx_{\mathsf C}(p_2)}} \right).
\end{align}
Thus, \eqref{eq:reg-branch-sum} is equal to, in the coordinate $t$,
\begin{align} \label{eq:reg-branch-sum-2}
    & \frac{K_D(\underline\zeta,\underline\xi)}{{dx_{\mathsf C}(z)}} \bigg( {K_D(e^{\frac{u\hbar}{2}\partial_{x_{\mathsf C}} }z,e^{-\frac{u\hbar}{2}\partial_{x_{\mathsf C}} }z)} - {K_{D+\zeta-\xi}(e^{\frac{u\hbar}{2}\partial_{x_{\mathsf C}} }z,e^{-\frac{u\hbar}{2}\partial_{x_{\mathsf C}} }z)} \\ \notag & + 
     {K_D(e^{\frac{u\hbar}{2}\partial_{x_{\mathsf C}} }\sigma_{\mathsf C}(z),e^{-\frac{u\hbar}{2}\partial_{x_{\mathsf C}} }\sigma_{\mathsf C}(z))} - {K_{D+\zeta-\xi}(e^{\frac{u\hbar}{2}\partial_{x_{\mathsf C}} }\sigma_{\mathsf C}(z),e^{-\frac{u\hbar}{2}\partial_{x_{\mathsf C}} }\sigma_{\mathsf C}(z))} \bigg)
\end{align} 
(note that in this formula we use that $\sigma_{\mathsf C}$ is the global deck transformation for $x_{\mathsf C}$, which implies  $\sigma_{\mathsf C}^*dx_{\mathsf C}=dx_{\mathsf C}$).
Let
$\mathcal W_m(z,u;q_1,\ldots,q_m)$ be the extended connected 
differential~\cite[Equations (4)-(5)]{ABDKS-NP}.
Then, for $D=\sum_{i} c_i(\gamma^+_i-\gamma^-_i)$, we have
\begin{align} \label{eq:curly-W}
 \frac{1}{dx_{\mathsf C}(z)} K_D(e^{\frac{u\hbar}{2}\partial_{x_{\mathsf C}} }z,e^{-\frac{u\hbar}{2}\partial_{x_{\mathsf C}} }z)
 =\frac{e^{u y_{\mathsf C}(z)}}{dx_{\mathsf C}(z)}
 \sum_{m\geq0}\frac{1}{m!}
 \left(\sum_i c_i \int_{\gamma_i^-}^{\gamma_i^+}\right)^{\otimes m}\mathcal W_m(z,u;\cdots).
\end{align}
The $\mathcal W_m(z,u;\cdots)$ differentials on the right hand side of this expression satisfy the extended loop equations~\cite[Definition~1.9 and Lemma~1.11]{ABDKS-NP}, which in this form state that 
\begin{align}
   \frac{e^{u y_{\mathsf C}(z)}}{dx_{\mathsf C}(z)} \mathcal W_m(z,u;\cdots) + \frac{e^{u y_{\mathsf C}(\sigma_{\mathsf C}(z))}}{dx_{\mathsf C}(z)}\mathcal W_m(\sigma_{\mathsf C}(z),u;\cdots) 
\end{align}
has Laurent expansion in $u$ with coefficients regular in $z$ at $z\to \pm 1$ (or, equivalently, $p\to \rho\in\mathcal R_{\mathsf C}$). Therefore,
\begin{align}
    & \frac{1}{dx_{\mathsf C}(z)} \bigg( e^{u y_{\mathsf C}(z)}
 \sum_{m\geq0}\frac{1}{m!}
 \left(\sum_i c_i \int_{\gamma_i^-}^{\gamma_i^+}\right)^{\otimes m}\mathcal W_m(z,u;\cdots) + 
 \\ \notag & 
 e^{u y_{\mathsf C}(\sigma_{\mathsf C}(z))}
 \sum_{m\geq0}\frac{1}{m!}
 \left(\sum_i c_i \int_{\gamma_i^-}^{\gamma_i^+}\right)^{\otimes m}\mathcal W_m(\sigma_{\mathsf C}(z),u;\cdots) \bigg)
\end{align}
is also regular in $z$ at $z\to \pm 1$. 
Also, the right hand side of~\eqref{eq:curly-W} expands, as a function of $\hbar$, as $e^{uy_{\mathsf C}(z)}/(u\hbar)+O(1)$, where the regular part is also regular in $u$. Combining these two observations together (the first one establishes regularity in $z$ at $z\to \pm 1$ and the second one takes care of cancellation of negative powers of $u$ and $\hbar$), and using that $K_D(\underline\zeta,\underline\xi)$ doesn't depend on $z$ and $u$ and is regular in $\hbar$, we see that \eqref{eq:reg-branch-sum-2} (and thus \eqref{eq:reg-branch-sum}) is regular in all its variables at $p\to \rho\in\mathcal R_{\mathsf C}$.

Here we initially assumed that in \eqref{eq:reg-branch-sum} the points $\xi$ and $\zeta$ are disjoint from $D$. At the next step, we use the regularity of this expression for $D=D_{\mathsf{np}}$ and taking the limit $\xi\to\beta$; note that in this limit the same regularity statement still holds. 

\subsubsection{Construction of the regular differential equation}\label{sec:difeqnpend}

We construct the equation for $\Psi^{\mathsf C,\mathsf{np}}(\pm t)$, with the method that roughly follows~\cite[Section 5]{MarchalOrantin}. 

Let $t$ be a local coordinate near $\rho \in \mathcal R_{\mathsf C}$, $x_{\mathsf C}(t) = x_{\mathsf C}(\rho) + t^2$. 
Recall that $\Psi^{\mathsf C,\mathsf{np}}(\pm t) = K_{D_{\mathsf{np}}}(\pm t,\underline{\beta})$. Choose an arbitrary finite point $\zeta\in \mathsf C$, $\zeta\notin\{\beta, \sigma_{\mathsf C}(\beta)\}\cup \mathcal{R}_{\mathsf C}$ (this choice won't play any role), and consider 
$K_{D_{\mathsf{np}}}(\underline{\zeta},\pm t)$.
As we proved above,
\begin{align} \label{eq:regularexpression}
\underbrace{\Psi^{\mathsf C,\mathsf{np}}(e^{\frac{u\hbar}{2}\partial_{x_{\mathsf C}}}t)K_{D_{\mathsf{np}}}(\underline{\zeta},e^{-\frac{u\hbar}{2}\partial_{x_{\mathsf C}}}t) + \Psi^{\mathsf C,\mathsf{np}}(-e^{\frac{u\hbar}{2}\partial_{x_{\mathsf C}}}t)K_{D_{\mathsf{np}}}(\underline{\zeta},-e^{-\frac{u\hbar}{2}\partial_{x_{\mathsf C}}}t)}_{L(t^2,u,\hbar)\coloneqq}\in\mathbb C[[t^2,\hbar,u]];
\end{align}
note that here, in the coordinate $t$, $\partial_{x_{\mathsf C}} = (2t)^{-1}\partial_t$. 

Consider the following two matrices:
\begin{align}
 U=
 \begin{pmatrix}\Psi^{\mathsf C,\mathsf{np}}(t)&\Psi^{\mathsf C,\mathsf{np}}(-t)
 \\ \hbar \partial_{x_{\mathsf C}} \Psi^{\mathsf C,\mathsf{np}}(t) & \hbar \partial_{x_{\mathsf C}} \Psi^{\mathsf C,\mathsf{np}}(-t) \end{pmatrix},
 \qquad
 V=
 \begin{pmatrix}K_{D_{\mathsf{np}}}(\underline{\zeta}, t)& \hbar \partial_{x_{\mathsf C}}K_{D_{\mathsf{np}}}(\underline{\zeta},t)\\ K_{D_{\mathsf{np}}}(\underline{\zeta},-t)& \hbar \partial_{x_{\mathsf C}} K_{D_{\mathsf{np}}}(\underline{\zeta},-t) \end{pmatrix}.
\end{align}
Using derivatives of \eqref{eq:regularexpression}, it is easy to show that the entries of $UV$ and $(\hbar\partial_{x_{\mathsf C}}U)V$ belong to $\mathbb C[[t^2,\hbar]]$. Indeed, for $a,b\in\mathbb Z_{\geq 0}$,
\begin{align}
 & \left(\tfrac \hbar 2\partial_{x_{\mathsf C}}+\partial_u\right)^a
 \left(\tfrac \hbar2\partial_{x_{\mathsf C}}-\partial_u\right)^b
 L(t^2,u,\hbar)\Big|_{u=0} = 
 \\ \notag 
 & 
 \big((\hbar\partial_{x_{\mathsf C}})^a \Psi^{\mathsf C,\mathsf{np}}(t)\big) \big((\hbar\partial_{x_{\mathsf C}})^b K_{D_{\mathsf{np}}}(\underline{\zeta},t)\big) 
 + 
  \big((\hbar\partial_{x_{\mathsf C}})^a \Psi^{\mathsf C,\mathsf{np}}(-t)\big) \big((\hbar\partial_{x_{\mathsf C}})^b K_{D_{\mathsf{np}}}(\underline{\zeta},-t)\big) \in \mathbb C[[t^2,\hbar]],
\end{align}
and we obtain the entries of $UV$ and $(\hbar\partial_{x_{\mathsf C}}U)V$ taking $a=0,1,2$ and $b=0,1$. 
Note also that $UV$ is invertible  over $\mathbb C[[t^2,\hbar]]$. Indeed, from the definition of $K_{D_{\mathsf{np}}}(z,\underline{\beta})$ and $K_{D_{\mathsf{np}}}(\underline{\zeta},z)$ one can explicitly compute 
\begin{align}
 [\hbar^0]\det(UV)
 =c(t)c(-t)\left(\frac{y_{\mathsf C}(t)}{t}\right)^2,
 \qquad c(t)\coloneqq\frac{z'(t)}{(z(t)-z(\beta))(z(\zeta)-z(t))}.
\end{align}
Thus $[\hbar^0]\det(UV)|_{t=0} = c(0)^2y_{\mathsf C}'(0)^2\ne0$.

These properties imply that  the entries of $(\hbar\partial_{x_{\mathsf C}}U)U^{-1}=(\hbar\partial_{x_{\mathsf C}}U)V(UV)^{-1}$ belong to $\mathbb{C}[[t^2,\hbar]]$.
The first row of $(\hbar\partial_{x_{\mathsf C}}U)U^{-1}$ is $(0,1)$ and we denote the entries of the second row by $(-\mathfrak b^{\mathsf{np}}, -\mathfrak a^{\mathsf{np}})$. Then the second row of the tautological identity
\begin{align}
\Big(\hbar \partial_{x_{\mathsf C}} - ((\hbar\partial_{x_{\mathsf C}}U)U^{-1})\Big) U =0 
\end{align}
gives the desired equations~\eqref{eq:regular-equations-branches} for $\star=\mathsf{np}$.

\subsection{Comparison of the wave functions}
\label{subsec:comparison}

The goal of this section is to complete the proof of the second statement of Theorem~\ref{thm:main}. To this end, we first compare the two families of meromorphic differentials on
$\mathsf C$, $\alpha_{r}^{\mathsf{rec}}$ and $\alpha_{r}^{\mathsf{np}}$, defined in~\eqref{eq:rec-quotient-differentials}
and~\eqref{eq:np-quotient-differentials}, respectively.

\begin{lemma}\label{lem:quotient-comparison}
For every $r\geq2$, 
 $\alpha_{r}^{\mathsf{np}}=\alpha_{r}^{\mathsf{rec}}$.
\end{lemma}

\begin{proof} We prove this by induction on $r$. The strategy is as follows. At each step, the regularity of the two local equations proved in Lemma~\ref{lem:rec-equation} will exclude poles of the difference $\alpha_{r}^{\mathsf{np}}-\alpha_{r}^{\mathsf{rec}}$ at $\mathcal R_{\mathsf C}$. Then, since there are no other possible poles, and the only holomorphic one-form on $\mathsf C\cong\mathbb P^1$ is zero, we conclude that $\alpha_{r}^{\mathsf{np}}=\alpha_{r}^{\mathsf{rec}}$.

Some of the technical steps below follow rather closely the computations and ideas in~\cite{IwakiSaenz,BBE}.

\subsubsection*{The first possible difference}

Assume that $\alpha_{k}^{\mathsf{np}}=\alpha_{k}^{\mathsf{rec}}$
for $2\leq k<r$, and put
$\alpha=\alpha_{r}^{\mathsf{np}}-\alpha_{r}^{\mathsf{rec}}$.
For $r=2$ the induction assumption is empty.
By Lemmas~\ref{lem:rec-meromorphy} and~\ref{lem:descent},
$\alpha$ is a meromorphic one-form on $\mathsf C$ with poles
only at $\mathcal R_{\mathsf C}=\{1,-1\}$.

Fix $\rho\in\mathcal R_{\mathsf C}$ and use the local coordinate
$t=\sqrt{x_{\mathsf C}-x_{\mathsf C}(\rho)}$, so that
$\sigma_{\mathsf C}(t)=-t$ and
$\hbar\partial_{x_{\mathsf C}}=\hbar(2t)^{-1}\partial_t$.
In this coordinate, write
\begin{align}
 \alpha=f(t)\,dt,\qquad \text{and} \qquad 
 y_{\mathsf C}(t)=t\,v(t^2),\quad v(0)\ne0.
\end{align}
The function $f$ is meromorphic at $t=0$, and $v$ is
holomorphic and nonvanishing at $t=0$. Our goal will be to show that $f$ is \emph{holomorphic} at $t=0$.

Use $u^\star_\pm$, $\Delta^\star$, $\mathfrak a^\star$, and
$\mathfrak b^\star$ as defined in Section~\ref{subsec:alpha-equations}.
The disk and amplitude terms in~\eqref{eq:quotient-waves}
are the same, and the induction assumption gives
\begin{align}
 u_\pm^{\mathsf{np}}&=\pm t\,v(t^2)+O(\hbar),
  & u_+^{\mathsf{np}}-u_+^{\mathsf{rec}}
 &=\hbar^r\frac{f(t)}{2t}+O(\hbar^{r+1}),
 \\ \notag 
 u_\pm^{\mathsf{rec}}& =\pm t\,v(t^2)+O(\hbar),
 & u_-^{\mathsf{np}}-u_-^{\mathsf{rec}}
 &=-\hbar^r\frac{f(-t)}{2t}+O(\hbar^{r+1}).
\end{align}
Indeed, the first difference of the logarithmic differentials $d\log\Psi^{\mathsf C,\star}$
is $\hbar^{r-1}\alpha$; multiplying by $\hbar(dx_{\mathsf C})^{-1}$ gives the difference of $u_+^\star$'s. On the other sheet,
$\sigma_{\mathsf C}^*\alpha=-f(-t)\,dt$, which gives the difference of $u_-^\star$'s.

\subsubsection*{Elimination of the poles}
In order to show that $f$ is \emph{holomorphic} at $t=0$, we compare the coefficients of $\hbar^r$ in the differences of $\mathfrak{a}^\star$'s and $\mathfrak{b}^\star$'s in
\eqref{eq:reconstructed-ode}. We have 
\begin{align}
 \mathfrak a^{\mathsf{np}} - \mathfrak a^{\mathsf{rec}}&=-(u^{\mathsf{np}}_++u^{\mathsf{np}}_-)+(u^{\mathsf{rec}}_++u^{\mathsf{rec}}_-)-\hbar\partial_{x_{\mathsf{C}}}\log(\Delta^{\mathsf{np}}/\Delta^{\mathsf{rec}}),
 \\ \notag
 \mathfrak b^{\mathsf{np}} - \mathfrak b^{\mathsf{rec}}&=u^{\mathsf{np}}_+u^{\mathsf{np}}_-
 -u^{\mathsf{rec}}_+u^{\mathsf{rec}}_-
 +(u^{\mathsf{np}}_+\hbar\partial_{x_{\mathsf{C}}}\log\Delta^{\mathsf{np}} - u^{\mathsf{rec}}_+\hbar\partial_{x_{\mathsf{C}}}\log\Delta^{\mathsf{rec}})
 -\hbar\partial_{x_{\mathsf{C}}}(u^{\mathsf{np}}_+-u^{\mathsf{rec}}_+).
\end{align}
Each $\Delta^\star$ has nonzero
leading coefficient $2t\,v(t^2)$, so division by it does not
introduce negative powers of $\hbar$. In particular,
$\Delta^{\mathsf{np}}/\Delta^{\mathsf{rec}}=1+O(\hbar^r)$,
and applying $\hbar\partial_{x_{\mathsf{C}}}\log$ to this ratio gives $O(\hbar^{r+1})$.
The other differences of terms involving $\hbar\partial_{x_{\mathsf{C}}}$ also start
at order $\hbar^{r+1}$, since there is an additional
factor $\hbar$.
Thus, only $-(u_+^\star+u_-^\star)$ and $u_+^\star u_-^\star$ contribute to the coefficient of $\hbar^r$. Substitution gives
\begin{align}
 [\hbar^r]
 (\mathfrak a^{\mathsf{np}}-\mathfrak a^{\mathsf{rec}})
 &=-\frac{f(t)-f(-t)}{2t},
 \label{eq:first-difference}
 \\ \notag
 [\hbar^r]
 (\mathfrak b^{\mathsf{np}}-\mathfrak b^{\mathsf{rec}})
 &=-v(t^2)\frac{f(t)+f(-t)}2.
\end{align}
By Lemma~\ref{lem:rec-equation},
both left-hand sides of~\eqref{eq:first-difference} belong
to $\mathbb C[[t^2]]$. As $v(0)\ne0$, it follows that
\begin{align}
 \frac{f(t)-f(-t)}2\in t\mathbb C[[t^2]],\qquad
 \frac{f(t)+f(-t)}2\in\mathbb C[[t^2]].
\end{align}
Thus both the odd and the even parts of $f$ are holomorphic,
hence $\alpha$ has no pole at $\rho$. Applying the same argument
at both points of $\mathcal R_{\mathsf C}$ makes $\alpha$
a global holomorphic one-form on $\mathbb P^1$. Hence
$\alpha=\alpha_{r}^{\mathsf{np}}-\alpha_{r}^{\mathsf{rec}}=0$, which completes the step of induction and the proof of
Lemma~\ref{lem:quotient-comparison}.
\end{proof}

As a direct corollary of Lemma~\ref{lem:quotient-comparison}, we obtain the following statement:

\begin{corollary}\label{cor:dsnpdsrec}
For every $r\geq2$, we have
$dS_{r}^{\mathsf{np}}=dS_{r}^{\mathsf{rec}}$ on $\Sigma$. Consequently,
$d\log\widetilde\Psi^{\mathsf{np}}(x;\hbar)
 =d\log\Psi^{\mathsf{rec}}(x;\hbar)$
on $U$; thus the second equality of Theorem~\ref{thm:main} holds.
\end{corollary}

\begin{proof}
Just pull back the identity of Lemma~\ref{lem:quotient-comparison} by $\pi$.
Equations~\eqref{eq:rec-quotient-differentials}
and~\eqref{eq:np-quotient-pullback} give the asserted
comparison of $dS_{r}$. The terms of degrees $-1$ and $0$
in the logarithmic expansions agree by the definition of
$\widetilde\Psi^{\mathsf{np}}$, proving equality of the full
logarithmic differentials.
\end{proof}

\appendix

\section{Local topological recursion on the quotient curve}

Here we want to illustrate how the first few multidifferentials
$\omega_n^{\mathsf C,(g)}$ obtained from local topological
recursion around the ramification points $\mathcal{R}_{\mathsf C}$
give the higher-order coefficients of the asymptotic expansion
established in the previous sections.
Note that $y_{\mathsf C}$ is defined only locally around the
ramification points, whereas $x_{\mathsf C}$ and $B_{\mathsf C}$ are defined
globally on $\mathsf C$, that is,
\begin{align}
    x_{\mathsf C}=&\frac12\left(z+\frac{1}{z}\right),
    \quad B_{\mathsf C}=\frac{dz_1dz_2}{(z_1-z_2)^2},\\ \notag
    y_{\mathsf C}=&\frac{2\log\left[
\frac{\rho}{2}\left(
z+z^{-1}
+(z-z^{-1})
\left(
z-z^{-1}
+\sqrt{(z+z^{-1}-1)(z+z^{-1}+3)}
\right)
\right)
\right]}{\sqrt{(z+z^{-1}-1)(z+z^{-1}+3)}}
\end{align}
for $\rho\in\mathcal{R}_{\mathsf C}=\{\pm1\}$, where the
logarithm is chosen to vanish at $z=\rho$.
Near the two ramification points, the local expansions
of $y_{\mathsf C}$ are given by
\begin{align}
y_{\mathsf C}(z)
&=2(z-1)-(z-1)^2-\frac23(z-1)^3
+\frac32(z-1)^4-(z-1)^5
+O((z-1)^6),
&& z\to1,\\
y_{\mathsf C}(z)
&=-2(z+1)-(z+1)^2-2(z+1)^3
-\frac52(z+1)^4-\frac{59}{15}(z+1)^5
+O((z+1)^6),
&& z\to-1.
\end{align}
Applying topological recursion \cite{EynOra} to this
local spectral curve, the first few multidifferentials
are given by
\begin{align}
\omega_3^{\mathsf C,(0)}(z_1,z_2,z_3)
&=
-\frac{dz_1\,dz_2\,dz_3}{2}
\left(
\frac{1}{\prod_{i=1}^3(z_i-1)^2}
+
\frac{1}{\prod_{i=1}^3(z_i+1)^2}
\right).
\\
\omega_1^{\mathsf C,(1)}(z)
&=
\left(
-\frac{z^2+z+1}{48(z-1)^4}
+\frac{z^2+3z+1}{16(z+1)^4}
\right)dz.
\\
\omega_4^{\mathsf C,(0)}(z_1,z_2,z_3,z_4)
&=
\frac{\prod_{i=1}^4 dz_i}{16}
\Bigg[
\frac{
1+3\displaystyle\sum_{i=1}^4
\left(\frac{z_i+1}{z_i-1}\right)^2
}{
\displaystyle\prod_{i=1}^4(z_i-1)^2
}
+
\frac{
-15+3\displaystyle\sum_{i=1}^4
\left(\frac{z_i-1}{z_i+1}\right)^2
}{
\displaystyle\prod_{i=1}^4(z_i+1)^2
}
\\ \nonumber
&\quad+
\sum_{1\le i<j\le4}
\frac{1}{
(z_i-1)^2(z_j-1)^2
\displaystyle\prod_{k\notin\{i,j\}}(z_k+1)^2
}
\Bigg].
\\
\omega_2^{\mathsf C,(1)}(z_1,z_2)
&=
\frac{dz_1\,dz_2}{512}
\Bigg[
\frac{
5\displaystyle\sum_{i=1}^2
\left(\frac{z_i+1}{z_i-1}\right)^4
+
3\displaystyle\prod_{i=1}^2
\left(\frac{z_i+1}{z_i-1}\right)^2
}{
(z_1-1)^2(z_2-1)^2
}+
\frac{
2\displaystyle\sum_{i=1}^2
\left(\frac{z_i+1}{z_i-1}\right)^2-21
}{
(z_1-1)^2(z_2-1)^2
}
\\ \nonumber
&\qquad+
\frac{
5\displaystyle\sum_{i=1}^2
\left(\frac{z_i-1}{z_i+1}\right)^4
+
3\displaystyle\prod_{i=1}^2
\left(\frac{z_i-1}{z_i+1}\right)^2
}{
(z_1+1)^2(z_2+1)^2
}
+
\frac{
-30\displaystyle\sum_{i=1}^2
\left(\frac{z_i-1}{z_i+1}\right)^2+43
}{
(z_1+1)^2(z_2+1)^2
}
\nonumber\\ \notag 
&\qquad+
\frac{4}{(z_1-1)^2(z_2+1)^2}
+
\frac{4}{(z_1+1)^2(z_2-1)^2}
\Bigg].
\end{align}

Next, we integrate the multidifferentials using the
symmetric integration prescription introduced in
Section~\ref{subsubsec:endpoint-normalization}. We define
\begin{align}
S_k^{(g,n)}(z)
:=
\frac{1}{n!}\left(\frac12\right)^n
\left.
\int_{z_1^{-1}}^{z_1}\cdots
\int_{z_n^{-1}}^{z_n}
\omega_n^{\mathsf C,(g)}(\zeta_1,\ldots,\zeta_n)
\right|_{z_1=\cdots=z_n=z},
\qquad k=2g-1+n.
\end{align}
Here the integrations are performed independently
in each variable before setting all variables equal.
Direct integration yields the following expressions.

For $S_2$, the two contributions are
\begin{align}
S_2^{(1,1)}(z)&=
\frac{
z^6-z^5-z^4+6z^3-z^2-z+1
}{
24(z^2-1)^3
},
\\[2mm]
S_2^{(0,3)}(z)
&=
\frac{
z(z^2+3)(3z^2+1)
}{
24(z^2-1)^3
}. 
\end{align}
Hence
\begin{align}
S_2(z)=\frac{z^6+2z^5-z^4+16z^3-z^2+2z+1}{24(z^2-1)^3}.
\end{align}

Similarly, for $S_3$, we obtain
\begin{align}
S_3^{(1,2)}(z)
&=
-\frac{z}{48(z^2-1)^6}
\Big(
z^{10}+z^9-19z^8-18z^7
+18z^6-78z^5
+18z^4-18z^3-19z^2+z+1
\Big),
\\[2mm]
S_3^{(0,4)}(z)
&=
\frac{z}{48(z^2-1)^6}
\Big(
z^{10}+z^9+5z^8+30z^7
-6z^6+66z^5
-6z^4+30z^3+5z^2+z+1
\Big).
\end{align}
Hence
\begin{align}
S_3(z)=\frac{z^3(z^6+2z^5-z^4+6z^3-z^2+2z+1)}{2(z^2-1)^6}.
\end{align}

These results agree with the perturbative coefficients
computed from the quantum $A$-polynomial in
\cite[Table~2]{DGLZ2009}, which were also obtained
from the state-integral expansion in
\cite[Section~4.2.2]{DGLZ2009}.
The identification of variables is given by
\begin{align}
    z+z^{-1}=m^2+m^{-2}-1, \qquad 
    \Delta(m)=i\sqrt{-m^{-4}+2m^{-2}+1+2m^2-m^4}
    =-(z-z^{-1}),
\end{align}
where the square-root branch is chosen to agree
with the geometric branch (the other sign is what they call the conjugate branch).
Since our convention is $q=e^\hbar$, whereas
\cite{DGLZ2009} use $q=e^{2\hbar_{\mathrm{DGLZ}}}$,
we have $\hbar_{\mathrm{DGLZ}}=\hbar/2$.
Consequently, the corresponding coefficients satisfy
\begin{align}
S_2(z)&=\frac12 S_2^{\mathrm{DGLZ}}(m),\\
S_3(z)&=\frac14 S_3^{\mathrm{DGLZ}}(m).
\end{align}
More generally, the rescaling of the contribution is
$\left(\frac12\right)^{2g-2+n}
=\left(\frac12\right)^{-\chi}$,
with $\chi=2-2g-n$.

\printbibliography

\end{document}